\documentclass[prl,aps,nofootinbib,twocolumn,noshowpacs,noshowkeys,superscriptaddress]{revtex4-1}
\usepackage{comment}

\usepackage{amsfonts}
\usepackage{amsmath}
\usepackage{amssymb}
\usepackage{graphicx}
\usepackage{textcomp}%
\usepackage{graphicx}% Include figure files
\usepackage{dcolumn}% Align table columns on decimal point
\usepackage{bm}% bold math

\usepackage{color}

\newcommand{\e}[1]{\mathrm{e}^{#1}}
\newcommand{\abs}[1]{\left\vert #1 \right\vert}
\newcommand\VecTwoD[2]{
\begin{pmatrix}
#1\\
#2
\end{pmatrix}
}
\newcommand\MatTwoD[4]{
\begin{pmatrix}
#1 & #2\\
#3 & #4
\end{pmatrix}
}

\newcommand\VecFourD[4]{
\begin{pmatrix}
#1\\
#2\\
#3\\
#4\\
\end{pmatrix}
}

\begin{document}

\title{Cavity nano-optomechanics with suspended subwavelength-sized nanowires}

\def\neel{Institut N\'{e}el, Universit\'{e} Grenoble Alpes - CNRS:UPR2940,\\ 38042 Grenoble, France}

\def\lkb{Laboratoire Kastler Brossel, ENS-Universit\'e PSL, CNRS, Sorbonne Universit\'e, Coll\`{e}ge de France,\\ 75005, Paris, France}

\author{Antoine Reigue}
\affiliation{\neel}
\author{Francesco Fogliano}
\affiliation{\neel}
\author{Philip Heringlake}
\affiliation{\neel}
\author{Laure Mercier de L\'epinay}
\affiliation{\neel}
\author{Benjamin Besga}
\affiliation{\neel}
\author{Jakob Reichel}
\affiliation{\lkb}
\author{Benjamin Pigeau}
\affiliation{\neel}
\author{Olivier Arcizet}
\affiliation{\neel}
\email{olivier.arcizet@neel.cnrs.fr}

\begin{abstract}
In the field of cavity nano-optomechanics, the nanoresonator-in-the-middle approach consists in inserting a sub-wavelength sized deformable resonator, here a nanowire, in the small mode volume of a fiber microcavity. Internal resonances in the nanowire enhance the light nanowire interaction which provide giant coupling strengthes -sufficient to enter the single photon regime of cavity optomechanics- at the condition to precisely position the nanowire within the cavity field. Here we expose a theoretical description that combines an analytical formulation of the Mie-scattering of the intracavity light by the nanowire and an input-output formalism describing the dynamics of the intracavity optical eigenmodes. We investigate both facets of the optomechanical interaction describing the position dependent parametric and dissipative optomechanical coupling  strengths, as well as the optomechanical force field experienced by the nanowire. We find a quantitative agreement with recent experimental realization.
We discuss the specific phenomenology of the optomechanical interaction which acquires a vectorial character since the nanowire can identically vibrate along both transverse directions: the optomechanical force field presents a non-zero rotational, while anomalous positive cavity shifts are expected. Taking advantage of the large Kerr-like non linearity, this work opens perspectives in the field of quantum optics with nanoresonator with for instance broadband squeezing of the outgoing cavity fields close to the single photon level.
\end{abstract}

\maketitle
\tableofcontents

\section{\label{sec:Intro}Introduction}

The field of optomechanics investigates the parametric coupling between the motional degrees of freedom of a resonator and a light field. The interaction first describes the impact of the resonator motion on the optical field which is usually enhanced using a high finesse optical cavity, and reciprocally, the intra-cavity field exerts an optical force on the resonator. This constitutes both facets of the optomechanical interaction. Such systems have been subject to impressive developments in the last decades~\cite{Aspelmeyer2014} enabling ultra sensitive optical readout of the resonator displacement or the manipulation of the mechanical system through optical forces. In that context, quantum non demolition (QND) measurements of the intra-cavity shot noise~\cite{Purdy2013, Lecocq2015}, generation of non-classical states of both resonator and electromagnetic fields~\cite{Brooks2012, Purdy2013a, Palomaki2013, Lecocq2015, Pirkkalainen2015, Wollman2015, Riedinger2016}, or ground state cooling of the macroscopic oscillator~\cite{Teufel2011, chan2011laser} were demonstrated.

\begin{figure}[b!]
\begin{center}
\includegraphics[width=0.99\linewidth]{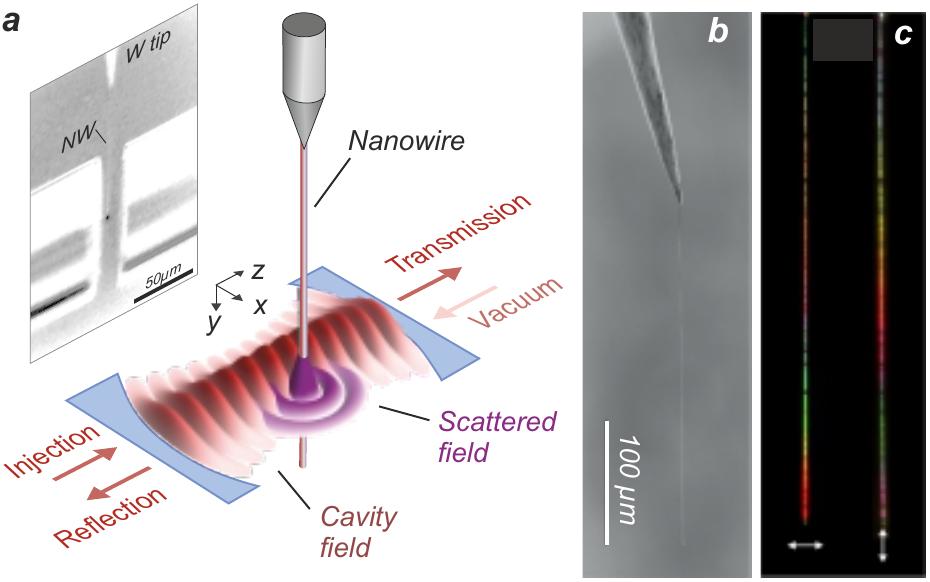}
\caption{\textbf{(a)} Optomechanical system consisting of a suspended nanowire inserted in the middle of a high finesse fiber based Fabry-P\'erot micro-cavity. Details on an experimental realization, shown in the photograph, can be found in ref.~\cite{Fogliano2021}. \textbf{(b)} SEM  image of a suspended nanowire attached to a sharp tungsten tip. \textbf{(c)} White light images of a conical nanowire obtained for perpendicular and parallel light polarisations (white arrow). Its reflected color is governed by  internal Mie resonances, which will have a large impact on its optomechanical coupling to the cavity field.}
\label{Fig:Intro:NW_and_cavity_intro}
\end{center}
\end{figure}

However, the optomechanical coupling strength per photon remained generally relatively weak and in most experiments it was necessary to strongly pump the cavity mode to observe or exploit the optical back-action. It leads to a situation where the optomechanical interaction can be seen as linearized around a mean intra-cavity photon number, then preventing the observation of its intrinsic Kerr-like non linearity~\cite{Reynaud1989} (due to the intensity dependence of the optical phase accumulated in the cavity) whose relative photon number sensitivity is enhanced at low photon number.

To circumvent this issue and guarantee both large field enhancement and ultra-high sensitivity mechanical resonator, a natural strategy consists in decoupling the optical and mechanical components of the system. It has emerged in the pioneering membrane in the middle experiments \cite{Thompson2008} and, in the last decade, several other strategies have been implemented to insert a mechanical resonator (ultra thin membrane, nanorod, carbon nanotube or nanogram-scale "trampoline") in the middle of an optical cavity~\cite{Thompson2008, Favero2009, Sankey2010, Purdy2013, Stapfner2013, Joeckel2015, Peterson2016, Reinhardt2016, Kampel2017, Rossi2018, Ruelle2022}.

Nevertheless, nanostring oscillator coupled to toroidal optical microresonator~\cite{Anetsberger2009,Anetsberger2010, wilson2015measurement, sudhir2017appearance}, photonic crystal nanobeam~\cite{leijssen2015strong, leijssen2017nonlinear} or levitated particles~\cite{delic2020cooling} have also provided experimental platforms with the required characteristics. These realizations have demonstrated ultra high displacement sensitivities~\cite{Anetsberger2009, Arcizet2006c,leijssen2015strong, wilson2015measurement, Kampel2017}, an ability to cool down the oscillator close to its mechanical ground state~\cite{wilson2015measurement, Peterson2016, Rossi2018, delic2020cooling}, or the possibility to observe correlations between imprecision (shot noise) and back action (radiation pressure force) of a meter laser~\cite{sudhir2017appearance}. They have also been used to perform QND measurement of the intra-cavity photon fluctuations~\cite{Purdy2013} and the great tunability of the optomechanical interaction also opened the road to QND measurements of the mechanical phonon number~\cite{Thompson2008, Sankey2010, leijssen2017nonlinear}. Still, optomechanical interaction at the single photon scale remains out of range, preventing the observation of original and unexplored phenomena~\cite{bose1997preparation, mancini1997ponderomotive, rabl2011photon, nunnenkamp2011single, nunnenkamp2012cooling, he2012quantum, hong2013open, nation2013nonclassical, kockum2019ultrastrong, forn2019ultrastrong} among which one of the long standing goal of optomechanics, the single photon blockade regime~\cite{rabl2011photon} where a single intra-cavity photon shifts the optical resonance by more than its linewidth. It is worth mentioning that up to now, only atom-based analogue optomechanical experiments~\cite{murch2008observation, brennecke2008cavity} could access such a regime, while recent developments with highly deformable photonic crystals~\cite{leijssen2015strong, leijssen2017nonlinear} or trampoline resonators~\cite{Reinhardt2016} appear to be promising in that perspective. More recently, nanowires in the middle of a fiber optical cavity have shown their potential allowing to perform optomechanical measurements close to the single photon regime~\cite{Fogliano2021}, and we present here the extended associated theoretical description of the system.
\

Moreover, the use of sub-wavelength-sized mechanical resonators enriches the phenomenology of the optomechanical interaction, and a quantitative modeling allowing to overpass the one dimensional approximation~\cite{jayich2008dispersive, favero2008cavity} is still lacking for those multidimensional systems. The computation of the optical forces also requires a precise knowledge of the intra-cavity field surrounding the resonators making essential to go beyond the dipolar approximation when calculating the field diffused by the scatterer which was in general treated in the Rayleigh regime for cavity optomechanics~\cite{romero2010toward, chang2010cavity, gieseler2012subkelvin, kiesel2013cavity, neukirch2015nano, bhattacharya2015rotational, delic2020levitated} (at the exception of ref.~\cite{welker2021nanofiber}).

In this paper we model the optomechanical system made of a suspended nanowire inserted in a high finesse Fabry-P\'erot micro-cavity~\cite{Fogliano2021} as shown Fig.~\ref{Fig:Intro:NW_and_cavity_intro}. The ultra high force sensitivity of the nanowires~\cite{nichol2008displacement, gil2010nanomechanical, arcizet2011single, siria2012electron, gloppe2014bidimensional, pigeau2015observation, de2017universal, rossi2017vectorial, de2018eigenmode, rossi2019magnetic}, their sub-wavelength sized diameter and their strong interaction with light~\cite{Bohren1998}, associated to the small mode volume of the fiber Fabry-P\'erot micro-cavity~\cite{colombe2007strong, hunger2010fiber}, provides large optomechanical coupling strengths and an ideal configuration to finely study the two facets of the optomechanical interaction. Combining Mie theory~\cite{Bohren1998} to properly describe the scattering of the light by the nanowire and a transfer matrix formalism~\cite{karrai1988magnetooptical, deutsch1995photonic, jayich2008dispersive, Reinhardt2016} to describe the propagation of the intra-cavity fields, we expose here the mean field solutions of the model giving access to the outgoing and intra-cavity fields for varying nanowire geometries and positions within the micro-cavity mode. We first present a description taking into account the ensemble of the cavity modes cross coupled by the nanowire scattering and then restrict our study to the situation where only the fundamental cavity mode is efficiently pumped, the scattering towards the other modes being treated as a loss channel. Our work highlights the existence of configurations where the single photon optomechanical coupling strength becomes larger than the vibrational frequency of the nanowire by several order of magnitude, leading to a situation where a single photon in the cavity displaces the nanowire by much more than its zero point fluctuations. The recoil exerted by a single intra-cavity photon is even shown at cryogenic temperature to overpass the residual thermal position fluctuations of the nanowire, thus motivating the development of such experimental configuration~\cite{Fogliano2021a}.

Additionally, this system allows to spatially map the intra-cavity field through the modifications of the output fields induced by the nanowire but also through the optical force applied on the nanowire, which is derived from the Maxwell's stress tensor formalism. This analysis reveals the existence of trapping and anti-trapping locations within the intra-cavity field positions and predicts the emergence of a non-conservative, or rotational structure of the intracavity force field which could be analysed in a future work using the 2D force field sensor capacity of the nanowires~\cite{Fogliano2021, de2018eigenmode}. Moreover, it is important to stress that our model presents a very good quantitative agreement with the experimental work of ref.~\cite{Fogliano2021} for both optomechanical interaction strength and optical forces.

Finally, we show that the system possesses a near unity single photon parametric cooperativity opening the road to the observation of Kerr-like non linearity and broadband optomechanical squeezing at the ultra low intra-cavity photon level~\cite{Reynaud1989, Fabre1994}. Furthermore, QND measurement of the intra-cavity field intensity fluctuations~\cite{Purdy2013} below one photon seems achievable, opening the road towards the investigation of a possible deviation to the mean field approximation, where the fluctuations in the cavity mode have a comparable impact on the oscillator dynamics as the mean field. The input-output formalism implemented with the Mie formalism, should also be suitable to investigate the Casimir forces in this original configuration.

The paper is organized as follow, section~\ref{sec:NW_in_cavity} details the model used to treat the interaction between the intra-cavity field and the nanowire, leading to the determination of the spatial dependence of the cross coupling coefficients from one cavity mode towards another, induced by the nanowire scattering. Section~\ref{sec:optomecha_coupling_TM00} presents the results of the transfer matrix formalism applied on the complete system highlighting the strong optomechanical interaction achieved. The impact of the temperature is discussed as well as the possibility to reach regimes of strong optomechanical interaction at the single photon scale. Section~\ref{sec:2D_charac} discusses the two dimensions characterization of the intra-cavity fields and computes the vectorial optomechanical force applied on the nanowire. Finally, section~\ref{sec:conclu_and_perpectives} summarizes the main results of this work and presents the perspectives.

\section{Modeling a nanowire in the middle of an optical cavity}
\label{sec:NW_in_cavity}

In this Section we derive the transfer matrices of the different elements of the system (cavity mirrors, propagation in the cavity, nanowire) that will be used in Section~\ref{sec:optomecha_coupling_TM00} for the complete description of the nanowire in the middle (NIM) configuration. For that purpose, the light-nanowire interaction will be consider with a free space field whose characteristics are the ones of the intra-cavity field. In particular, the geometrical parameters of the intra-cavity field (Rayleigh length, waist, curvature radius) will be fully determined by the cavity parameters (length, mirrors geometry, laser wavelength).

\subsection{Generalities about light-nanowire interaction}
\label{sec:NW_in_cavity:light_nanowire_interaction}

We consider the optomechanical system depicted in Fig.~\ref{Fig:Intro:NW_and_cavity_intro} consisting of a suspended nanowire inserted in the middle of a high finesse fiber based Fabry-P\'erot micro-cavity (for experimental details see ref.~\cite{Fogliano2021}). The nanowire is modelled as an infinite lossless dielectric (non-magnetic) cylinder of radius $R_\mathrm{nw}$ and refractive index $n$ (around 2.7) collinear to the $y$ axis and located at $\textbf{r}_0 = x_0 \, \textbf{e}_x + z_0 \, \textbf{e}_z$. This relatively large refractive index causes that for the radius employed, ranging from 25 to $250 \, \mathrm{nm}$, several arches of the EM field can be localized inside the nanowire (internal resonances), producing a multipolar scattering regime, which spectrally and geometrically structures and possibly allows to enhance the light-nanowire interaction. The light scattering by the nanowire is treated in the framework of the Mie formalism (see Appendix~\ref{Appendix:Optical_prop_SiC_NW} and~\cite{Bohren1998}) using the cylindrical coordinates $(r, \varphi, y)$ centred at the nanowire position $\textbf{r}_0$ (Fig.~\ref{Fig:NW_in_cavity:Mie_scattering}(a)). This cylindrical description makes that our model will be suited for a nanowire sufficiently inserted in the optical mode (in practice the extremity effects disappear for vertical insertions larger than $\sim 2 \, \mu \mathrm{m}$). For plane wave incidence (wavevector $\textbf{k}_i$), we consider the two possible orthogonal polarization states of the incident light: electric field polarized parallel or perpendicular to the incidence plane $(\textbf{k}_i, \textbf{e}_y)$. For an amplitude $E_0$ at the nanowire position and an incident wavevector in the $(xz)$ plane, the scattered fields for both polarizations are
\begin{subequations}
\label{sec:NW_in_cavity:scattered_fields_single_orthogonal_incidence}
\begin{align}
& \textbf{E}_{\mathrm{scat}}^{\parallel}(\textbf{r}) {=} E_0 \sum_{l = - \infty}^{+ \infty} (-i)^l \, \e{-i l \phi_i} b_l^\parallel H_l^{(1)}(kr) \, \e{i l \varphi} \, \textbf{e}_y, \\
& \textbf{E}_{\mathrm{scat}}^{\perp}(\textbf{r}) {=} E_0 \, i \sum_{l = - \infty}^{+ \infty} (-i)^l \, \e{-i l \phi_i} a_l^\perp \bigg[ il \frac{H_l^{(1)}(kr)}{kr} \, \textbf{e}_r \nonumber \\
& \hspace*{4cm} - H_l^{(1)'}(kr) \, \textbf{e}_\varphi \bigg] \, \e{i l \varphi},
\end{align}
\end{subequations}
where the coefficients $a_l^\perp$ and $b_l^\parallel$ are imposed by the boundary conditions and given in Appendix~\ref{Appendix:Optical_prop_SiC_NW:Mie_formalism_infinite_cylinder}. They depend on the nanowire properties (radius and refractive index) as well as on the incident wavelength. In these expressions, $\phi_i$ characterizes the direction of $\textbf{k}_i$ in the $(xz)$ plane, $H_l^{(1)}$ is the Hankel function of the first kind and the prime denotes a derivative with respect to the argument. Here we have chosen the incident polarization vectors to be $\textbf{e}_P^{\parallel} = - \textbf{e}_y$ and $\textbf{e}_P^{\perp} = \textbf{e}_x$.

From the knowledge of the scattered electric and magnetic fields we obtain (Appendix~\ref{Appendix:additional_result_mie_scattering:emission_diagram}) for a single incident wavevector $\textbf{k}_i$ the angular dependence of the emission diagram as shown Fig.~\ref{Fig:NW_in_cavity:Mie_scattering}(cd) for different nanowire radii and for both polarizations. Moreover, for a given half optical collection angle $\theta_\mathrm{col}$ we define the reflected, transmitted and scattered 1D cross-sections caracterizing the amount of light scattered in the corresponding channels. These quantities are given Fig.~\ref{Fig:NW_in_cavity:Mie_scattering}(cd) in the case of a numerical aperture $\mathrm{NA} = \sin \theta_\mathrm{col} = 0.15$ corresponding to the situation described in the following and for comparison, for a $\mathrm{NA} = 0.7$ corresponding to the situation where a microscope objective is used to collect the scattered light. As expected from the Mie resonances, the cross-sections strongly depends on the nanowire radius and we will see that the choice of the ratio $R_\mathrm{NW}/\lambda$ will have a strong impact on the optomechanical coupling between the nanowire and the cavity mode.

\subsection{Optical cavity field}
\label{sec:NW_in_cavity:optical_cavity_field}

The optical cavity is assumed to be made of two spherical mirrors of curvature radii $R_c$ located at $z = \pm L_\mathrm{cav}/2$. They are treated as lossless beam splitters of reflection and transmission coefficients $R_{L,R}$ and $T_{L,R}$ (intensity related) satisfying $R_{L,R} + T_{L,R} = 1$ where the index $L,R$ stands for the left and right mirrors. For $R_{L,R} = 0.994$ (corresponding to the experimental situation of ref.~\cite{Fogliano2021} where $\lambda = 770 \, \mathrm{nm}$), the finesse of the empty cavity is $\mathcal{F}_0 = 522$ and the settling time of the intra-cavity field is of the order of $10 \, \mathrm{ps}$ which justifies the use of the slowly varying envelope approximation in the following. A position in the cavity volume is characterized by it Cartesian coordinates $(x, y, z)$ using an origin located at the center of the cavity. The cavity is pumped from the left side with a fixed laser wavelength $\lambda$ and for the geometry considered ($R_c = 28 \, \mu \mathrm{m}$), the waist and Rayleigh length of the intra-cavity field are $w_0 = 1.7 \, \mu \mathrm{m}$ and $z_R = 11.5 \, \mu \mathrm{m}$, corresponding to a weakly diverging beam for a cavity length of the order of $10 \, \mu \mathrm{m}$. For this reason, we work in the framework of the paraxial approximation allowing to treat easily the polarization effects. However, we will see that a more accurate description of the beam polarization structure is necessary to obtain a vectorial expression of the optical force applied by the intra-cavity field on the nanowire. Finally, we employ a mean field description sufficient to characterize the two facets of the optomechanical coupling. We note that the same input-output formalism can also be used to describe the field fluctuations, but this will be the topic of future work.

\subsubsection{Hermite-Gaussian modes}
\label{sec:NW_in_cavity:Hermite_Gaussian_cavity_modes}

We assume a spherical profile of the cavity mirrors leading to Gaussian cavity modes. These modes are given in the paraxial approximation by the set of Hermite-Gaussian beams~\cite{jones2015optical}, defined for propagation along $\pm z$ as, $\textbf{E}^{(\pm, p)}_{n_x, n_y}(\textbf{r}) = E^{(\pm)}_{n_x, n_y}(\textbf{r}) \, \textbf{e}_P^{(p)}$ with
\begin{subequations}
\label{Eq:NW_in_cavity:General_Gaussian_beam_definition}
\begin{align}
%& \textbf{E}^{(\pm)}_{n_x, n_y}(\textbf{r}) = E^{(\pm)}_{n_x, n_y}(\textbf{r}) \, \textbf{e}_P^{(p)}, \\
& E^{(\pm)}_{n_x, n_y}(\textbf{r}) = \rho_{n_x, n_y}(\textbf{r}) \, \e{\pm i \varphi_{n_x, n_y}(\textbf{r})}, \\
& \rho_{n_x, n_y}(\textbf{r}) {=} \mathcal{A}_0^{n_x, n_y} \frac{w_0}{w(z)} \e{{-}\dfrac{\textbf{r}_\perp ^2}{w^2(z)}} \, H_{n_x} \left( \frac{\sqrt{2} x}{w(z)} \right) H_{n_y} \left( \frac{\sqrt{2} y}{w(z)} \right), \label{Eq:NW_in_cavity:General_Gaussian_beam_definition_amplitude} \\
& \varphi_{n_x, n_y}(\textbf{r}) = k z - (1 + n_x + n_y) \Psi(z) + k \frac{\textbf{r}_\perp^2}{2 R(z)},
\end{align}
\end{subequations}
where $\rho_{n_x, n_y}$ and $\varphi_{n_x, n_y}$ are the amplitude and phase of the field, $\{n_x, n_y\} \in \mathbb{N}^2$, $\textbf{r}_\perp = x \, \textbf{e}_x + y \, \textbf{e}_y$ and $\textbf{e}_P^{(p)}$ is the polarization vector standing in the $(xy)$ plane. Because of the symmetry of the problem, we will consider in the following two cases associated to two different polarizations of the Hermite-Gaussian modes, the parallel case where $\textbf{e}_P^{\parallel} = - \textbf{e}_y$ and the perpendicular case where $\textbf{e}_P^{\perp} = \textbf{e}_x$. In Eq.~\eqref{Eq:NW_in_cavity:General_Gaussian_beam_definition}, $H_n(x) = (-1)^n \, \e{x^2} d^n (\e{-x^2}) / dx^n$ is the Hermite polynomial of order $n$, $\Psi$ the Gouy phase of the beam, $R$ the curvature radius of the wavefronts, and $w$ characterizes the transverse spreading ($w_0$ being the waist). Their expressions are given by $w(z) = w_0 \sqrt{1 + \left( z / z_R \right)^2}$, $\Psi(z) = \arctan \left( z / z_R \right)$ and $R(z) = z_R^2 / z + z$,
%\begin{subequations}
%\label{Eq:NW_in_cavity:spreadin_Gouy_phase_curvature_radius}
%\begin{align}
%& w(z) = w_0 \sqrt{1 + \left( \frac{z}{z_R} \right)^2}, \\
%& \Psi(z) = \arctan \left( \frac{z}{z_R} \right), \\
%& R(z) = \frac{z_R^2}{z} + z,
%\end{align}
%\end{subequations}
where $z_R = \pi w_0^2 / \lambda$ is the Rayleigh length defining the distance over which the beam can be considered as non diverging.

%\sout{In the paraxial approximation, $\textbf{E}^{(\pm, p)}_{n_x, n_y}$ satisfies the vectorial Helmholtz equation. Additionally,} (?)
The set of Hermite-Gaussian modes forms a complete orthogonal basis of solution of the Helmholtz equation in the paraxial approximation~\cite{jones2015optical}, i.e. any two-dimensional complex light field can be obtained as a superposition of Hermite Gaussian beams. The scalar product associated to this basis for $\pm z$ propagation is
\begin{eqnarray}
\big \langle \textbf{E}^{(\pm, p)}_{p_x, p_y} \vert \textbf{E}^{(\pm, p')}_{n_x, n_y} \big \rangle &\equiv& \delta_{p, p'} \, \iint_{\mathcal{S}} \mathrm{d}^2 \textbf{r} \, E^{(\pm)}_{n_x, n_y} (\textbf{r}) \left. E^{(\pm)}_{p_x, p_y} \right.^\ast (\textbf{r}) \nonumber \\
	&=& \delta_{p, p'} \, \delta_{n_x, p_x} \, \delta_{n_y, p_y},
\label{Eq:NW_in_cavity:scalar_product_scalar_field}
\end{eqnarray}
where the integration surface $\mathcal{S}$ can be any surface having the $z$ axis as symmetry axis and a collection angle of $\pi$, i.e. collecting the whole field in the transverse direction. The eigenmodes's normalization factors $\mathcal{A}_0^{n_x, n_y}$ appearing in~\eqref{Eq:NW_in_cavity:General_Gaussian_beam_definition} are derived with the above scalar product. In the following, we will note $(\pm, n_x, n_y, p)$ a beam propagating along $\pm z$ in the $(n_x, n_y)$ Hermite-Gaussian mode with polarization $p$.

\subsubsection{Hermite-Gaussian modes propagation}
\label{sec:NW_in_cavity:Hermite_Gaussian_cavity_modes_propagation}

As mentioned before, in an optical cavity the geometrical parameters of the Hermite-Gaussian modes are fully constrained~\cite{siegman1986lasers}, the Rayleigh length being set to $z_R = (L_\mathrm{cav}/2) \sqrt{(1 + g)/(1 - g)}$ with $g = 1 - L_\mathrm{cav}/R_c$. As such, the knowledge of the field strength at one position can be used to deduce its value at any other position. This allows, for a beam in the $(\pm, n_x, n_y, p)$ mode, to introduced the \textit{reduced field} $F^{(\pm, p)}_{n_x, n_y}(z) = A_{\pm} \e{ \pm i \varphi_0^{n_x, n_y}(z)}$ defined as $\textbf{E}^{(\pm, p)}(\textbf{r}) = \rho_{n_x, n_y}(\textbf{r}_\perp, z) \e{\pm i k \frac{\textbf{r}_\perp^2}{2 R(z)}} \, F^{(\pm, p)}_{n_x, n_y}(z) \, \textbf{e}_P^{(p)}$ with $A_{\pm}$ the complex amplitude of the field and $\varphi_0^{n_x, n_y}(z) = k z - (1 + n_x + n_y) \Psi(z)$.

As a consequence, we restrict the field propagation analysis to the one of the reduced field.
In the slowly varying envelope approximation, the propagation (in vacuum) between two positions $z_1$ and $z_2$ is given by
\begin{eqnarray}
\VecTwoD{F^{(+, p)}_{n_x, n_y}(z_2)}{F^{(-, p)}_{n_x, n_y}(z_2)} = M_{z_1, z_2}^{n_x, n_y} \VecTwoD{F^{(+, p)}_{n_x, n_y}(z_1)}{F^{(-, p)}_{n_x, n_y}(z_1)},
\end{eqnarray}
where
\begin{eqnarray}
\hspace*{-0.5cm}
M_{z_1, z_2}^{n_x, n_y} {=} \MatTwoD{\e{i \left[ \varphi_0^{n_x, n_y}(z_2) {-} \varphi_0^{n_x, n_y}(z_1) \right]}}{0}{0}{\hspace*{-1.2cm} \e{-i \left[ \varphi_0^{n_x, n_y}(z_2) {-} \varphi_0^{n_x, n_y}(z_1) \right]}}. \nonumber \\
\label{Eq:NW_in_cavity:transfer_matrix_free_space}
\end{eqnarray}

\begin{figure*}[t]
\begin{center}
\includegraphics[width=0.98\linewidth]{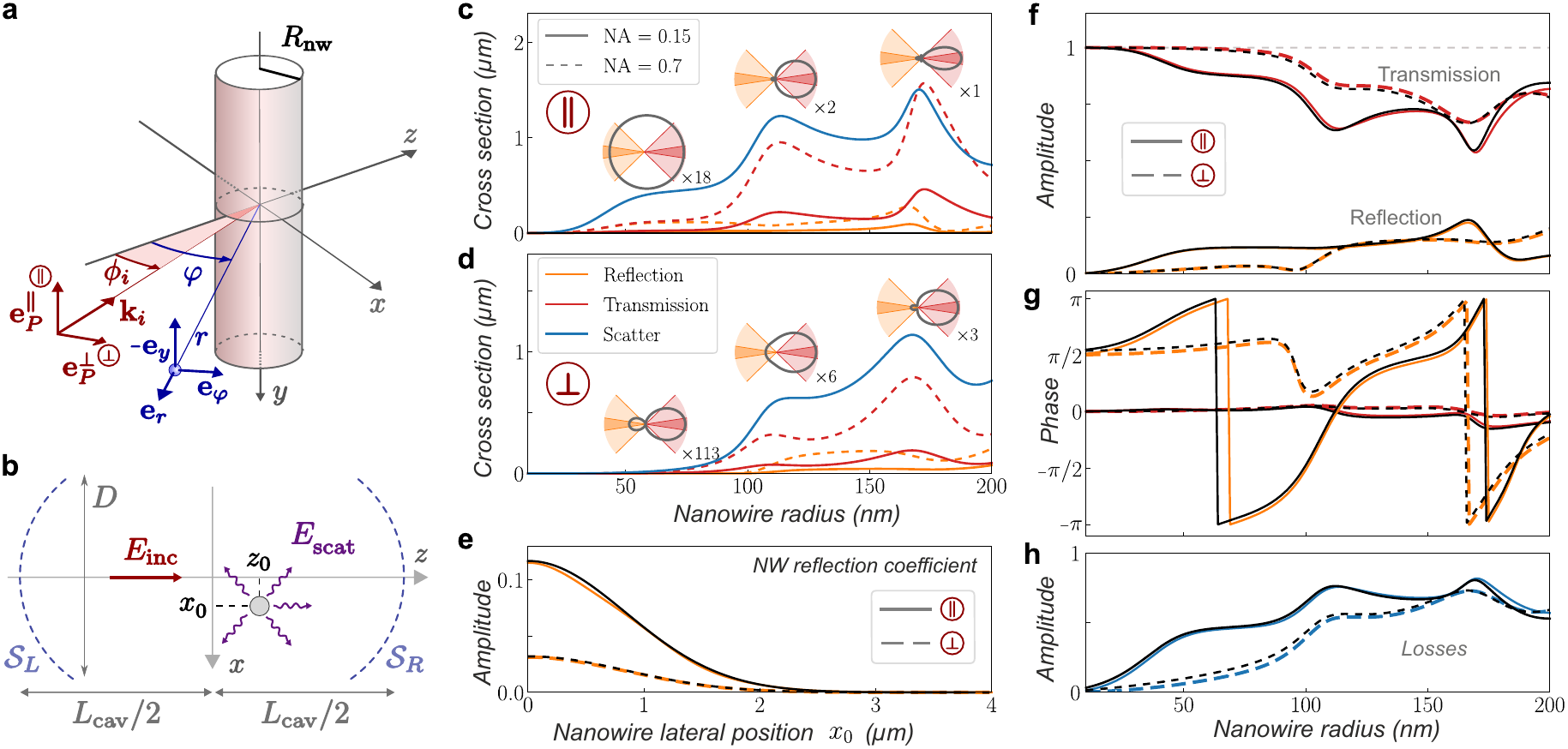}
\caption{\textbf{(a)} Infinite cylinder under single plane wave illumination. The incidence angle in the (xz) plane is noted $\phi_i$ and the polarization can be either parallel or perpendicular to the cylinder axis. \textbf{(b)} Scheme of the configuration used to compute the reflection and transmission coefficients of a Gaussian beam on a nanowire. An incident beam $\textbf{E}_\mathrm{inc}$ generates a scattered field $\textbf{E}_\mathrm{scat}$ evaluated on the cavity mirror surfaces $\mathcal{S}_{L, R}$ using the Mie formalism. The coefficients are obtained by considering that the effective reflected and transmitted fields in a given mode are the result of the scattered field projection on this mode. \textbf{(c-d)} Reflection, transmission and scattered 1D cross-sections as function of the cylinder radius for an excitation wavelength $\lambda = 770 \, \mathrm{nm}$. The solid lines are associated to a numerical aperture of 0.15 corresponding to the situation described in this paper while the dashed line is associated to a numerical aperture of 0.7 corresponding to the situation where a usual microscope objective is used to collect the scattered light. Inset: angular emission diagrams for different cylinder radii. The coloured areas highlight the collection angles associated to the two numerical apertures. \textbf{(e)} Transverse dependence of the reflection coefficient of an SiC nanowire in the fundamental Hermite-Gaussian mode ($R_\mathrm{NW} = 65 \, \mathrm{nm}$, $\lambda = 770 \, \mathrm{nm}$, $w_0 = 1.7 \, \mu \mathrm{m}$), see text. \textbf{(f-g)} Transmission and reflection coefficients for a nanowire located at $\textbf{r}_0 = \bf{0}$. \textbf{(h)} Amplitude of the nanowire induced cavity losses coefficients. In e, f, g and h, the black lines correspond to the complete description of the scattered field while the coloured lines are the result of the approximated method (see text).}
\label{Fig:NW_in_cavity:Mie_scattering}
\end{center}
\end{figure*}

\subsubsection{Field interaction with cavity mirrors}

We now consider a cavity mirror located at position $z_\mathrm{M}$ with real reflection and transmission coefficients $r$ and $t$ verifying $r^2 + t^2 = 1$. It is possible for any Hermite-Gaussian mode to express the reduced fields at the right of the mirror in terms of the reduced fields at its left,

\begin{eqnarray}
\VecTwoD{F_R^{(+)}(z_\mathrm{M})}{F_R^{(-)}(z_\mathrm{M})} = M_\mathrm{BS} \VecTwoD{F_L^{(+)}(z_\mathrm{M})}{F_L^{(-)}(z_\mathrm{M})},
\end{eqnarray}
with $M_\mathrm{BS}$ given by
\begin{eqnarray}
M_\mathrm{BS} = \frac{1}{t} \MatTwoD{1}{\eta r}{\eta r}{1}, \label{Eq:NW_in_cavity:transfer_matrix_BS}
\end{eqnarray}
where the index $R$ and $L$ indicate the side of the mirror on which the field is given. Here we have omitted the mode and polarization labels for simplicity. The existence of the parameters $\eta = \pm 1$ is due to a phase choice, consequence of the energy conservation in the beam splitter.

\subsection{Cavity - Nanowire interaction}

\subsubsection{Presentation of the formalism}

We now treat the interaction between the nanowire and the optical cavity modes, the aim being to obtain the transfer matrix of the nanowire inside the cavity. For that purpose, we first compute in section~\ref{sec:NW_in_cavity:R_T_coeff_NW_in_Gaussian_beam} the nanowire-induced reflection and transmission coefficients from one Hermite-Gaussian mode impinging on the nanowire to another one. The incoming field produces a scattered field which is expanded on the set of Hermite-Gaussian modes propagating along $\pm z$. Then, using the fact that the intra-cavity field is fully characterized by the value of the reduced field on the cavity axis (see~\ref{sec:NW_in_cavity:Hermite_Gaussian_cavity_modes_propagation}), the reflection and transmission coefficient are defined as the ratio of the outgoing and incoming reduced fields. Note that the scattered coefficients are computed between the propagative modes of the cavity since they are the relevant ones to describe how the nanowire impacts the intra-cavity field. However in the following when calculating optical forces, we will take into account the complete electromagnetic field surrounding the nanowire as given by the Mie theory.

When the nanowire is inserted in the cavity at position $\textbf{r}_0$, it is simultaneously illuminated by the forward and backward propagating fields, each producing a scattered field. In the matrix formalism, the nanowire will act as a scattering element which will cross-couple the different transverse modes propagating along the 2 directions and for the 2 polarizations, leading to a $4N \times 4N$ dimension scattering matrix if one restricts the analysis to the first $N^\mathrm{th}$ transverse modes. For the scattering from the modes $(+, n_x, n_y)$ and $(-, n_x', n_y')$ towards $(+, p_x, p_y)$ and $(-, p_x', p_y')$ for all possible polarizations, we have

\begin{eqnarray}
\label{Eq:NW_in_cavity:general_transfer_matrix}
\VecFourD{F^{(+, \parallel)}_{p_x, p_y}(z_0)}{F^{(-, \parallel)}_{p_x', p_y'}(z_0)}{F^{(+, \perp)}_{p_x, p_y}(z_0)}{F^{(-, \perp)}_{p_x', p_y'}(z_0)} = S_\mathrm{nw}(\textbf{r}_0) \VecFourD{F^{(+, \parallel)}_{n_x, n_y}(z_0)}{F^{(-, \parallel)}_{n_x', n_y'}(z_0)}{F^{(+, \perp)}_{n_x, n_y}(z_0)}{F^{(-, \perp)}_{n_x', n_y'}(z_0)},
\end{eqnarray}
with $S_\mathrm{nw}$ given by
\begin{eqnarray}
\label{Eq:NW_in_cavity:general_transfer_matrix_expression}
&& S_\mathrm{nw}(\textbf{r}_0) = \nonumber \\
&& \begin{pmatrix}
\left. C_t^{(+)} \right._{n_x, n_y \, \parallel}^{p_x, p_y \, \parallel}  &
\left. C_r^{(-)} \right._{n_x', n_y' \, \parallel}^{p_x, p_y \, \parallel}  &
\left. C_t^{(+)} \right._{n_x, n_y \, \perp}^{p_x, p_y \, \parallel}  &
\left. C_r^{(-)} \right._{n_x', n_y' \, \perp}^{p_x, p_y \, \parallel} \\
\left. C_r^{(+)} \right._{n_x, n_y \, \parallel}^{p_x', p_y' \, \parallel} &
\left. C_t^{(-)} \right._{n_x', n_y' \, \parallel}^{p_x', p_y' \, \parallel} &
\left. C_r^{(+)} \right._{n_x, n_y \, \perp}^{p_x', p_y' \, \parallel} &
\left. C_t^{(-)} \right._{n_x', n_y' \, \perp}^{p_x', p_y' \, \parallel} \\
\left. C_t^{(+)} \right._{n_x, n_y \, \parallel}^{p_x, p_y \, \perp}  &
\left. C_r^{(-)} \right._{n_x', n_y' \, \parallel}^{p_x, p_y \, \perp}  &
\left. C_t^{(+)} \right._{n_x, n_y \, \perp}^{p_x, p_y \, \perp}  &
\left. C_r^{(-)} \right._{n_x', n_y' \, \perp}^{p_x, p_y \, \perp} \\
\left. C_r^{(+)} \right._{n_x, n_y \, \parallel}^{p_x', p_y' \, \perp} &
\left. C_t^{(-)} \right._{n_x', n_y' \, \parallel}^{p_x', p_y' \, \perp} &
\left. C_r^{(+)} \right._{n_x, n_y \, \perp}^{p_x', p_y' \, \perp} &
\left. C_t^{(-)} \right._{n_x', n_y' \, \perp}^{p_x', p_y' \, \perp} \\
\end{pmatrix}, \nonumber \\
\end{eqnarray}
where $\left. C_{r, t}^{(\pm)} \right._{m_x, m_y \, (p)}^{q_x, q_y \, (p')}$ are the reflection and transmission coefficient from the mode $(m_x, m_y, p)$ towards $(q_x, q_y, p')$, the upper index $(\pm)$ indicating the direction of the incident beam on the nanowire. For simplification, we have omitted the different indexes in $S_\mathrm{nw}$ and the position dependence of the coefficients.% In section~\ref{sec:NW_in_cavity:Transfer_matrix_NW_in_Gaussian_beam} we will rewrite Eq.~\eqref{Eq:NW_in_cavity:general_transfer_matrix} in a form suited for the situation considered here.

Moreover, if one assumes (i) that the cavity modes are non-degenerated, so that one can pump the cavity close to a single optical mode, the other being far from resonance, (ii) the cross coupling induced by the nanowire does not bring a transverse mode close to being at resonance with the pump laser, (iii) the polarization eigenbasis of the cavity is aligned with the nanowire orientation so that there is no cross coupling between modes of different polarizations (see later), then one can restrict the analysis to the situation where a single cavity mode remains resonant inside the cavity. In that situation, the nanowire-induced cross coupling to the other transverse modes can be viewed as a loss channel, while the structure of the cavity propagation matrix can be restricted to a single family of transverse mode, of dimension 2 only for each polarizations. This will be the subject of section~\ref{sec:NW_in_cavity:Impact_on_TM00} where we will focus on the case of the fundamental Hermite-Gaussian mode.

\subsubsection{Reflection and transmission coefficients of the nanowire}
\label{sec:NW_in_cavity:R_T_coeff_NW_in_Gaussian_beam}

Here we describe the structure of the scattering coefficients, on the example of an incident field in the $(+, n_x, n_y, p)$ mode, $\textbf{E}_\mathrm{inc}(\textbf{r}) = A_{\mathrm{inc}} \, \textbf{E}^{(+, \, p)}_{n_x, n_y} (\textbf{r})$,
%\begin{eqnarray}
%\textbf{E}_\mathrm{inc}(\textbf{r}) = A_{\mathrm{inc}} \, \textbf{E}^{(+, \, p)}_{n_x, n_y} (\textbf{r}),
%\label{Eq:NW_in_cavity:general_incident_field}
%\end{eqnarray}
$A_{\mathrm{inc}}$ being the field's amplitude. When illuminating the nanowire, it generates a scattered field $\textbf{E}_{\mathrm{scat}} = A_{\mathrm{inc}} \, \textbf{E}_{n_x, n_y}^{\mathrm{scat} \, (p)}$ (see Fig.~\ref{Fig:NW_in_cavity:Mie_scattering}(b)) where $\textbf{E}_{n_x, n_y}^{\mathrm{scat} \, (p)}$ is the field scattered by the nanowire due to the incidence of $\textbf{E}^{(+, \, p)}_{n_x, n_y}$. It can be expanded on the Hermite-Gaussian modes propagating along $\pm z$ as
\begin{subequations}
\label{sec:NW_in_cavity:alpha_beta_definition}
\begin{align}
& \textbf{E}_{n_x, n_y}^{\mathrm{scat} \, (p)}(\textbf{r}) {=} \sum_{\substack{p_x, p_y \\ p' = \{\parallel, \perp\}}} \bigg( \alpha^{p_x, p_y \, (p')}_{n_x, n_y \, (p)} \, \textbf{E}^{(+, \, p')}_{p_x, p_y}(\textbf{r}) \nonumber \\
& \hspace*{2.5cm} + \beta^{p_x, p_y \, (p')}_{n_x, n_y \, (p)} \, \textbf{E}^{(-, \, p')}_{p_x, p_y}(\textbf{r}) \bigg) + \tilde{\textbf{E}}_{n_x, n_y}^{\mathrm{scat} \, (p)}(\textbf{r}) , \\
& \alpha^{p_x, p_y \, (p')}_{n_x, n_y \, (p)} {=} \iint_{\mathcal{S}_R} \mathrm{d}^2 \, \textbf{r} \, \textbf{E}_{n_x, n_y}^{\mathrm{scat} \, (p)}(\textbf{r}) \cdot \left. \textbf{E}^{(+, \, p')}_{p_x, p_y} \right.^\ast (\textbf{r}), \\
& \beta^{p_x, p_y \, (p')}_{n_x, n_y \, (p)} {=} \iint_{\mathcal{S}_L} \mathrm{d}^2 \, \textbf{r} \, \textbf{E}_{n_x, n_y}^{\mathrm{scat} \, (p)}(\textbf{r}) \cdot \left. \textbf{E}^{(-, \, p')}_{p_x, p_y} \right.^\ast (\textbf{r}),
\end{align}
\end{subequations}
where the coefficients $\alpha^{p_x, p_y \, (p')}_{n_x, n_y \, (p)}$ and $\beta^{p_x, p_y \, (p')}_{n_x, n_y \, (p)}$ depend on the laser wavelength and polarization, as well as on the nanowire position and properties (refractive index and radius). These two groups of coefficients correspond to the part of the scattered field propagating along $+\textbf{e}_z$ and $-\textbf{e}_z$ respectively while $\tilde{\textbf{E}}_{n_x, n_y}^{\mathrm{scat} \, (p)}$ is associated to the off axis contribution of the scattered field. This last term exists especially because of the finite lateral extension of the cavity mirrors. The integration surfaces $\mathcal{S}_{R/L}$ are chosen to be the surfaces of the spherical cavity mirrors located at $\pm L_\mathrm{cav}/2$, ensuring the orthogonality of the Hermite-Gaussian modes as discussed in~\ref{sec:NW_in_cavity:Hermite_Gaussian_cavity_modes}. The normalization is guaranteed by the fact that the transverse size of the cavity mirrors $D$ is assumed to be large compare to the transverse spreading of the Hermite-Gaussian modes which is the case in ref.~\cite{Fogliano2021} since $D = 12 \, \mu \mathrm{m}$ and $w(\pm L_\mathrm{cav}/2) = 2.1 \, \mu \mathrm{m}$.
%However, it is possible to go beyond this approximation in order to take into account the eventual clipping losses of the cavity mirrors.
%Their numerical aperture is $\mathrm{NA} = \sin\left[ \arctan(D/L_\mathrm{cav}) \right]$ where $D$ is the transverse size of the cavity mirrors and allows to eventually take into account the clipping losses due to the finite size of the mirrors.
Finally, in the Mie formalism used in the following to compute the scattered field (see Appendix~\ref{Appendix:Optical_prop_SiC_NW}), the total field due to the incidence of $\textbf{E}_\mathrm{inc}$ on the nanowire is $\textbf{E}_{\mathrm{tot}}(\textbf{r}) = \textbf{E}_\mathrm{inc}(\textbf{r}) + \textbf{E}_{\mathrm{scat}}(\textbf{r})$.
%\begin{eqnarray}
%\textbf{E}_{\mathrm{tot}}(\textbf{r}) &=& \textbf{E}_\mathrm{inc}(\textbf{r}) + \textbf{E}_{\mathrm{scat}}(\textbf{r}) \nonumber \\
%	&=& A_{\mathrm{inc}} \, \bigg[ \sum_{\substack{p_x, p_y \\ p' = \{\parallel, \perp\}}} \bigg( \bigg[\alpha^{p_x, p_y \, (p')}_{n_x, n_y \, (p)} \nonumber \\
%	&& \hspace*{1.2cm} + \delta_{n_x, p_x} \, \delta_{n_y, p_y} \, \delta_{p, p'} \bigg] \, \textbf{E}^{(+, \, p')}_{p_x, p_y}(\textbf{r}) \nonumber \\
%&& \hspace*{1.3cm} + \beta^{p_x, p_y \, (p')}_{n_x, n_y \, (p)} \, \textbf{E}^{(-, \, p')}_{p_x, p_y}(\textbf{r}) \bigg) + \tilde{\textbf{E}}_{n_x, n_y}^{\mathrm{scat} \, (p)}(\textbf{r}) \bigg]. \nonumber \\
%\end{eqnarray}

Since the intra-cavity field is fully characterized by the reduced fields, the reflection and transmission coefficients are defined as the ratio of the outgoing and incoming reduced fields. For the scattering from a $(+, n_x, n_y, p)$ Hermite-Gaussian mode to another $(\pm, p_x, p_y, p')$ mode, these coefficients reads as
%Since the intra-cavity field is fully characterized by its value on the cavity axis (the reduced field defined Eq.~\eqref{Eq:NW_in_cavity:reduced_field_definition}) we define the transmission and reflection coefficients from a $(+, n_x, n_y, p)$ Hermite-Gaussian mode to another $(\pm, p_x, p_y, p')$ modes as
\begin{subequations}
\label{sec:NW_in_cavity:Cr_Ct_NW_definition}
\begin{align}
& \left. C_r^{(+)} \right._{n_x, n_y \, (p)}^{p_x, p_y \, (p')}(\textbf{r}_0) = \frac{F^{\mathrm{tot} \ (-, p')}_{p_x, p_y}(z_0)}{F^{\mathrm{inc} \ (+, p)}_{n_x, n_y}(z_0)}, \\
& \left. C_t^{(+)} \right._{n_x, n_y \, (p)}^{p_x, p_y \, (p')}(\textbf{r}_0) = \frac{F^{\mathrm{tot} \ (+, p')}_{p_x, p_y}(z_0)}{F^{\mathrm{inc} \ (+, p)}_{n_x, n_y}(z_0)},
\end{align}
\end{subequations}
where
\begin{subequations}
\label{sec:NW_in_cavity:reduced_amplitude_for_Cr_Ct}
\begin{align}
& F^{\mathrm{inc} \ (+, p)}_{n_x, n_y}(z_0) = A_{\mathrm{inc}} \, \e{ i \varphi_0^{n_x, n_y}(z_0)}, \\
& F^{\mathrm{tot} \ (-, p')}_{p_x, p_y}(z_0) = A_{\mathrm{inc}} \beta^{p_x, p_y \, (p')}_{n_x, n_y \, (p)} \, \e{ -i \varphi_0^{p_x, p_y}(z_0)}, \\
& F^{\mathrm{tot} \ (+, p')}_{p_x, p_y}(z_0) = A_{\mathrm{inc}} \left[\alpha^{p_x, p_y \, (p')}_{n_x, n_y \, (p)} + \delta_{n_x, p_x} \, \delta_{n_y, p_y} \, \delta_{p, p'} \right] \nonumber \\
& \hspace*{5cm}  \times\e{ i \varphi_0^{p_x, p_y}(z_0)}.
\end{align}
\end{subequations}
In Eq.~\eqref{sec:NW_in_cavity:Cr_Ct_NW_definition}, the $(+)$ notation reminds that these coefficients have been obtained for a propagation of the incident beam along $+z$. Injecting Eq.~\eqref{sec:NW_in_cavity:reduced_amplitude_for_Cr_Ct} in Eq.~\eqref{sec:NW_in_cavity:Cr_Ct_NW_definition} leads to the reflection and transmission coefficients from the $(+, n_x, n_y, p)$ to the $(\pm, p_x, p_y, p')$ Hermite-Gaussian mode as
\begin{subequations}
\label{sec:NW_in_cavity:Cr_Ct_NW_final_expressions}
\begin{align}
& \left. C_r^{(+)} \right._{n_x, n_y \, (p)}^{p_x, p_y \, (p')}(\textbf{r}_0) = \beta^{p_x, p_y \, (p')}_{n_x, n_y \, (p)} \, \e{ -i \left[ \varphi_0^{p_x, p_y}(z_0) + \varphi_0^{n_x, n_y}(z_0) \right]} , \label{sec:NW_in_cavity:Cr_Ct_NW_final_expressions_Cr} \\
& \left. C_t^{(+)} \right._{n_x, n_y \, (p)}^{p_x, p_y \, (p')}(\textbf{r}_0) = \left( \alpha^{p_x, p_y \, (p')}_{n_x, n_y \, (p)} + \delta_{n_x, p_x} \, \delta_{n_y, p_y} \, \delta_{p, p'} \right) \nonumber \\
& \hspace*{4cm} \times \e{ i \left[ \varphi_0^{p_x, p_y}(z_0) - \varphi_0^{n_x, n_y}(z_0) \right]} \label{sec:NW_in_cavity:Cr_Ct_NW_final_expressions_Ct}
\end{align}
\end{subequations}
where $\alpha^{p_x, p_y \, (p')}_{n_x, n_y \, (p)}$ and $\beta^{p_x, p_y \, (p')}_{n_x, n_y \, (p)}$ are given by Eq.~\eqref{sec:NW_in_cavity:alpha_beta_definition} while $\varphi_0^{n_x, n_y}$ has been introduced in Section~\ref{sec:NW_in_cavity:Hermite_Gaussian_cavity_modes_propagation}. The second part of parenthesis in Eq.~\eqref{sec:NW_in_cavity:Cr_Ct_NW_final_expressions_Ct} is the remaining contribution from the incident field which should not be forgotten. These coefficients depend on the laser wavelength and polarization as well as on the nanowire position $\textbf{r}_0$ and properties ($R_\mathrm{nw}$ and $n$). Those calculations thus help connecting the $\alpha$ and $\beta$ coefficients, which account for the 3D vectorial structure of the problem, to the scattering coefficients of the nanowire in the transfer matrix formalism, which presents a 1D structure. In order to map the optomechanical coupling, they will be computed for any position of the nanowire in the cavity.

\subsubsection{Transfer matrix of a nanowire in a cavity}
\label{sec:NW_in_cavity:Transfer_matrix_NW_in_Gaussian_beam}

We now derive the transfer matrix associated to the nanowire which cross-couples the different transverse intra-cavity modes. In order to simplify the formalism, and come closer to the experimental configuration, we note that because of the geometry considered here (very short cavity), the intra-cavity modes are almost non diverging, leading to zero cross polarization reflection and transmission coefficients. In that case the transfer matrix $S_\mathrm{nw}(\textbf{r}_0)$ given in Eq.~\eqref{Eq:NW_in_cavity:general_transfer_matrix_expression} becomes a diagonal bloc matrix, each bloc corresponding to a given polarization. It allows to work in the following with 2 by 2 transfer matrices for each polarizations according to
%We consider two counter-propagating incident beams on the nanowire in the $(\pm, n_x, n_y, p)$ mode. The $S$-transfer matrix giving the reduced output fields in the $(\pm, p_x, p_y, p')$ modes in terms of the reduced input fields in the $(\pm, n_x, n_y, p)$ mode according to,
%\begin{subequations}
%\label{Eq:NW_in_cavity:S_transfer_matrix_NW}
%\begin{align}
%& \VecTwoD{F^{(+, p')}_{p_x, p_y}(z_0)}{F^{(-, p')}_{p_x, p_y}(z_0)} = \left. S_\mathrm{nw} \right._{n_x, n_y \, (p)}^{p_x, p_y \, (p')}(\textbf{r}_0) \VecTwoD{F^{(+, p)}_{n_x, n_y}(z_0)}{F^{(-, p)}_{n_x, n_y}(z_0)}, \\
%& \nonumber \\
%& \left. S_\mathrm{nw} \right._{n_x, n_y \, (p)}^{p_x, p_y \, (p')}(\textbf{r}_0) = \MatTwoD{\left. C_t^{(+)} \right._{n_x, n_y \, (p)}^{p_x, p_y \, (p')}(\textbf{r}_0)}{\left. C_r^{(-)} \right._{n_x, n_y \, (p)}^{p_x, p_y \, (p')}(\textbf{r}_0)}{\left. C_r^{(+)} \right._{n_x, n_y \, (p)}^{p_x, p_y \, (p')}(\textbf{r}_0)}{\left. C_t^{(-)} \right._{n_x, n_y \, (p)}^{p_x, p_y \, (p')}(\textbf{r}_0)},
%\end{align}
\begin{eqnarray}
\hspace*{-0.5cm}
\VecTwoD{F^{(+, p)}_{p_x, p_y}(z_0)}{F^{(-, p)}_{p_x', p_y'}(z_0)} = S_\mathrm{nw}^{(p)}(\textbf{r}_0) \VecTwoD{F^{(+, p)}_{n_x, n_y}(z_0)}{F^{(-, p)}_{n_x', n_y'}(z_0)},
\end{eqnarray}
with $S_\mathrm{nw}^{(p)}(\textbf{r}_0)$  (we omit the mode indices here again) reading as
\begin{eqnarray}
S_\mathrm{nw}^{(p)}(\textbf{r}_0) = \MatTwoD{\left. C_t^{(+, p)} \right._{n_x, n_y}^{p_x, p_y}(\textbf{r}_0)}{\left. C_r^{(-, p)} \right._{n_x', n_y'}^{p_x, p_y}(\textbf{r}_0)}{\left. C_r^{(+, p)} \right._{n_x, n_y}^{p_x', p_y'}(\textbf{r}_0)}{\left. C_t^{(-, p)} \right._{n_x', n_y'}^{p_x', p_y'}(\textbf{r}_0)}. \nonumber \\
\end{eqnarray}
%where $\left. C_{r/t}^{(-, p)} \right._{m_x, m_y}^{q_x, q_y}(\textbf{r}_0)$ are the reflection and transmission coefficients from the $(-, m_x, m_y, p)$ mode towards the $(\pm, q_x, q_y, p)$ modes.
Thus, the coefficients of the $M$-transfer matrix giving the reduced fields on the right of the nanowire in terms of the reduced fields on its left according to
%\begin{subequations}
%\label{Eq:NW_in_cavity:M_transfer_matrix_NW}
%\begin{align}
%& \VecTwoD{F^{(+, p')}_{p_x, p_y}(z_0)}{F^{(-, p)}_{n_x, n_y}(z_0)} = \left. M_\mathrm{nw} \right._{n_x, n_y \, (p)}^{p_x, p_y \, (p')}(\textbf{r}_0) \VecTwoD{F^{(+, p)}_{n_x, n_y}(z_0)}{F^{(-, p')}_{p_x, p_y}(z_0)}, \\
%& \nonumber \\
%& \hspace*{-0.2cm} \left. M_\mathrm{nw} \right._{n_x, n_y \, (p)}^{p_x, p_y \, (p')}(\textbf{r}_0) {=} \frac{1}{C_t^{(-)}} \MatTwoD{C_t^{(+)} C_t^{(-)} - C_r^{(+)} C_r^{(-)}}{C_r^{(-)}}{-C_r^{(+)}}{1}, \label{Eq:NW_in_cavity:M_transfer_matrix_NW_matrix}
%\end{align}
%\end{subequations}
\begin{eqnarray}
\label{Eq:NW_in_cavity:M_transfer_matrix_NW}
\hspace*{-0.5cm}
\VecTwoD{F^{(+, p)}_{p_x, p_y}(z_0)}{F^{(-, p)}_{n_x', n_y'}(z_0)} = M_\mathrm{nw}^{(p)}(\textbf{r}_0) \VecTwoD{F^{(+, p)}_{n_x, n_y}(z_0)}{F^{(-, p)}_{p_x', p_y'}(z_0)},
\end{eqnarray}
can be written in term of the S matrix coefficients as
\begin{subequations}
\label{Eq:NW_in_cavity:M_transfer_matrix_NW_coef}
\begin{align}
& \left( M_\mathrm{nw}^{(p)} \right)_{11} {=} \frac{1}{\left. C_t^{(-, p)} \right._{n_x', n_y'}^{p_x', p_y'}} \bigg( \left. C_t^{(+, p)} \right._{n_x, n_y}^{p_x, p_y} \, \left. C_t^{(-, p)} \right._{n_x', n_y'}^{p_x', p_y'} \nonumber \\
& \hspace*{3cm} - \left. C_r^{(+, p)} \right._{n_x, n_y}^{p_x', p_y'} \, \left. C_r^{(-, p)} \right._{n_x', n_y'}^{p_x, p_y} \bigg), \\
& \left( M_\mathrm{nw}^{(p)} \right)_{12} {=} \frac{\left. C_r^{(-, p)} \right._{n_x', n_y'}^{p_x, p_y}}{\left. C_t^{(-, p)} \right._{n_x', n_y'}^{p_x', p_y'}}, \\
& \left( M_\mathrm{nw}^{(p)} \right)_{21} {=} -\frac{\left. C_r^{(+, p)} \right._{n_x, n_y}^{p_x', p_y'}}{\left. C_t^{(-, p)} \right._{n_x', n_y'}^{p_x', p_y'}}, \\
& \left( M_\mathrm{nw}^{(p)} \right)_{22} {=} \frac{1}{\left. C_t^{(-, p)} \right._{n_x', n_y'}^{p_x', p_y'}}.
%\MatTwoD{C_t^{(+)} C_t^{(-)} - C_r^{(+)} C_r^{(-)}}{C_r^{(-)}}{-C_r^{(+)}}{1}, \label{Eq:NW_in_cavity:M_transfer_matrix_NW_matrix} \nonumber \\
\end{align}
\end{subequations}

In order to describe the optomechanical coupling for any position of the nanowire within the cavity, it is necessary to compute those coefficients for all possible positions of the nanowire. However, here we consider the ideal situation of a symmetric cavity, leading to
\begin{subequations}
\begin{align}
& \left. C_{r/t}^{(-)} \right._{m_x, m_y \, (p)}^{q_x, q_y \, (p')}(x_0, z_0) = \left. C_{r/t}^{(+)} \right._{m_x, m_y \, (p)}^{q_x, q_y \, (p')}(x_0, -z_0), \\
& \left. C_{r/t}^{(\pm)} \right._{m_x, m_y \, (p)}^{q_x, q_y \, (p')}(-x_0, z_0) {=} (-1)^{m_x + q_x} \left. C_{r/t}^{(\pm)} \right._{m_x, m_y \, (p)}^{q_x, q_y \, (p')}(x_0, z_0).
\end{align}
\end{subequations}
It is then sufficient to compute the reflection and transmission coefficients for an incident beam propagating along $+z$ in the half plane $x_0 \geq 0$ to obtain the $M$-transfer matrix appearing in Eq.~\eqref{Eq:NW_in_cavity:M_transfer_matrix_NW} for any position of the nanowire.

\begin{figure}[t!]
\begin{center}
\includegraphics[width=0.95\linewidth]{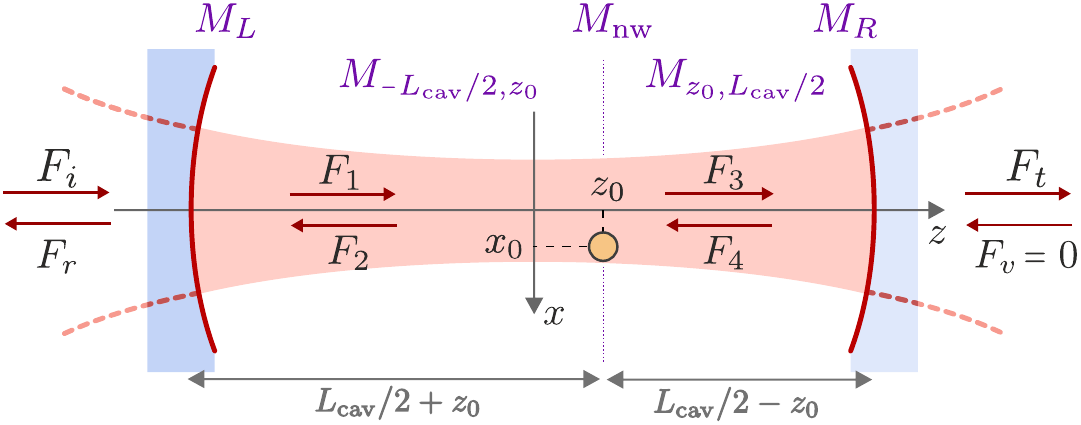}
\caption{Scheme of the intra-cavity field propagation for an optical cavity pumped from the left side. $M_{\mathrm{BS}, L/R}$ are the transfer matrices of the cavity mirrors, $M_{-L_\mathrm{cav}/2, z_0}$ and $M_{z_0, L_\mathrm{cav}/2}$ propagate the fields in the left and right sub-cavities respectively, and $M_\mathrm{nw}$ accounts for the propagation through the nanowire.}
 \label{Fig:NW_in_cavity:Schema_full_field_propagation}
\end{center}
\end{figure}

\subsubsection{Restriction to the fundamental Hermite-Gaussian mode}
\label{sec:NW_in_cavity:Impact_on_TM00}

We now focus on the optomechanical coupling between the nanowire and the fundamental cavity mode $n_x = n_y = 0$ and $p_x = p_y = 0$, and for readability we omit the indices in the following expressions. It corresponds to the cases where (i) the cavity length is locked at the resonance of the fundamental cavity mode; (ii) the cavity length is scanned around the resonance of the fundamental mode, when no other resonant mode is present in the scanning range. It allows to simplify the formalism presented above but also to stick to the experimental configuration studied in ref.~\cite{Fogliano2021}. Nonetheless, the formalism exposed is suitable to treat the case of any higher order intra-cavity optical modes or even to study the interaction between several cavity modes possibly mediated by the nanowire scattering. The reflection and transmission coefficients from the fundamental cavity mode ($+, p$) towards ($\pm, p$) are obtained combining Eq.~\eqref{sec:NW_in_cavity:Cr_Ct_NW_final_expressions} and~\eqref{sec:NW_in_cavity:alpha_beta_definition},
\begin{subequations}
\label{sec:NW_in_cavity:Cr_Ct_NW_fundamental_mode}
\begin{align}
& C_r^{(+, p)}(\textbf{r}_0) {=} \e{ -2i \varphi_0(z_0)} \iint_{\mathcal{S}_L} \mathrm{d}^2 \, \textbf{r} \, \textbf{E}_{\mathrm{scat}}^{(p)}(\textbf{r}) \cdot \left. \textbf{E}^{(-, \, p)}_{0, 0} \right.^\ast (\textbf{r}) , \\
& C_t^{(+, p)}(\textbf{r}_0) = \delta_{p, p'} + \iint_{\mathcal{S}_R} \mathrm{d}^2 \, \textbf{r} \, \textbf{E}_{\mathrm{scat}}^{(p)}(\textbf{r}) \cdot \left. \textbf{E}^{(+, \, p)}_{0, 0} \right.^\ast (\textbf{r}), \label{sec:NW_in_cavity:Cr_Ct_NW_fundamental_mode_transmission}
\end{align}
\end{subequations}
where the scattered field $\textbf{E}_{\mathrm{scat}}^{(p)}$ is generated by the incident field $\textbf{E}^{(+, \, p)}_{0, 0}$.

A first way to obtain the scattered field $\textbf{E}_{\mathrm{scat}}^{(p)}$ is to expand the incident Gaussian beam onto the plane wave spectrum (see Appendix~\ref{Appendix:Gaussian_beam}) and to compute the total scattered field as the sum of all the contributions due to the different incidences. The surface integrals in~\eqref{sec:NW_in_cavity:Cr_Ct_NW_fundamental_mode} can then be calculated numerically (see Appendix~\ref{Appendix:numerical_considerations} for details) for different nanowire radius and positions in the cavity using the experimental parameters of ref~\cite{Fogliano2021} ($L_\mathrm{cav} = 12 \, \mu \mathrm{m}$, $R_c = 28 \, \mu \mathrm{m}$, $D = 12 \, \mu \mathrm{m}$, $\lambda = 770 \, \mathrm{nm}$ and $n = 2.61$). Fig.~\ref{Fig:NW_in_cavity:Mie_scattering}(e) shows in black the transverse dependence of $\abs{C_{r}^{(+, \, p)}}$ for a nanowire radius $R_\mathrm{nw} = 65 \, \mathrm{nm}$ at $z_0 = 0$. As expected, the amplitude is maximum at the center of the beam and decreases with a Gaussian profile laterally. Additionally, Fig.~\ref{Fig:NW_in_cavity:Mie_scattering}(fg) exhibits (black lines) the radius dependence of the amplitude and phase of the reflection and transmission coefficients for a nanowire located at $\textbf{r}_0 = \bf{0}$. We observe a strong dependence of these coefficients with respect to the nanowire radius due to the Mie resonances, as observed in the variation of the 1D cross section presented Fig.~\ref{Fig:NW_in_cavity:Mie_scattering}(cd). Finally, the energy conservation allows to obtain the nanowire induced cavity losses, defined in modulus as
\begin{eqnarray}
\abs{C_\mathrm{losses}^{(+, p)}} = \sqrt{1 - \left( \abs{C_{r}^{(+, p)}}^2 + \abs{C_{t}^{(+, p)}}^2 \right)},
\end{eqnarray}
and shown (black lines) in Fig.~\ref{Fig:NW_in_cavity:Mie_scattering}(h). This coefficient then characterizes the amount of light scattered out of the cavity axis and towards other modes than the fundamental cavity mode. With no surprise, one notice that the nanowire can scatter an important fraction of the light. It will thus be important to carefully position it in the standing wave structure of the cavity mode, in order to maximize the optomechanical coupling. Finally, the reflection and transmission coefficients from one polarization to the other have been numerically shown to be zero justifying the fact to consider independent polarization states in section~\ref{sec:NW_in_cavity:Transfer_matrix_NW_in_Gaussian_beam}.

In order to reduce the numerical calculation time we developed a second method to compute the scattered field $\textbf{E}_{\mathrm{scat}}^{(p)}$. It consists in making the assumption that the incoming optical field wave front is almost flat on the extent of the sub-wavelength sized nanowire. In that case, we assume that the incident field can be approximated by a single plane wave incidence with a wavevector $\textbf{k}_i$ orthogonal to the phase curvature of the beam at the nanowire position, with an amplitude $E_{0,0}^{(+)}(\textbf{r}_0)$ evaluated at the nanowire position, and a polarization vector $\textbf{e}_P$ which can be parallel ($\textbf{e}_P^\parallel = -\textbf{e}_y$) or perpendicular ($\textbf{e}_P^\perp = \textbf{e}_x$) to the nanowire axis. Since we consider a nanowire positions close to the center of the cavity in the following ($z_0 \ll z_R$), the incident wavevector takes the form $\textbf{k}_i = k \, \textbf{e}_z$. It generates a scattered field in the $(xz)$ plane given by Eq.~\eqref{sec:NW_in_cavity:scattered_fields_single_orthogonal_incidence} for $\phi_i = 0$. The vertical $y$-axis dependence can be accounted phenomenologically by considering that the scattered field intensity follows the Gaussian distribution of the incident beam, $f(y, z) = \e{-y^2/w(z)^2}$ where $w(z)$ is the waist of the fundamental mode at position $z$. Under these assumptions, the scattered fields for both polarizations are
\begin{subequations}
\label{sec:NW_in_cavity:scattered_fields_TM00}
\begin{align}
& \textbf{E}_{\mathrm{scat}}^{\parallel}(\textbf{r}) {=} E_{0,0}^{(+)}(\textbf{r}_0) \, \e{- \frac{y^2}{w(z)^2} } \sum_{l = - \infty}^{+ \infty} (-i)^l b_l^\parallel H_l^{(1)}(kr) \, \e{i l \varphi} \, \textbf{e}_y, \\
& \textbf{E}_{\mathrm{scat}}^{\perp}(\textbf{r}) {=} E_{0,0}^{(+)}(\textbf{r}_0) \, i \, \e{- \frac{y^2}{w(z)^2} } \nonumber \\
& \hspace*{1.1cm} \times \sum_{l = - \infty}^{+ \infty} (-i)^l a_l^\perp \bigg[ il \frac{H_l^{(1)}(kr)}{kr} \, \textbf{e}_r {-} H_l^{(1)'}(kr) \, \textbf{e}_\varphi \bigg] \, \e{i l \varphi},
\end{align}
\end{subequations}
where the coefficients $a_l^\perp$ and $b_l^\parallel$ are given in Appendix~\ref{Appendix:Optical_prop_SiC_NW} and where the Cartesian coordinate $z$ is expressed in terms of the cylindrical coordinates centred at the nanowire position as $z = z_0 - r \cos \varphi$. In that case, it is clear from the expressions of the scattered fields Eq.~\eqref{sec:NW_in_cavity:scattered_fields_TM00}, that an incident field polarized along $\textbf{e}_P^\parallel$ ($\textbf{e}_P^\perp$) will generate a scattered field with no contribution along $\textbf{e}_P^\perp$ ($\textbf{e}_P^\parallel$), leading to zero coefficients for cross polarizations. The result of this procedure is shown Fig.~\ref{Fig:NW_in_cavity:Mie_scattering}(e-h) in color lines and we observe a very good agreement with the method presented before, the second method being two orders of magnitude faster.

\section{Optomechanical coupling with the fundamental cavity mode}
\label{sec:optomecha_coupling_TM00}

\subsection{Formalization of the problem}
\label{sec:optomecha_coupling_TM00:formalization}

The optomechanical interaction  between an optical cavity mode and a single mechanical mode vibrating along $z$ (pulsations $\omega_\mathrm{cav} \, \vert \, \Omega_\mathrm{m}$; ladder operators $\hat{a}, \hat{a}^\dagger \, \vert \vert \, \hat{b}, \hat{b}^\dagger$) arises from the parametric dependence of the optical cavity resonance pulsation $\omega_\mathrm{cav}(z_0)$ on the oscillator position $z_0$. It is described in second quantization by the coupling Hamiltonian $\hat{H}_\mathrm{int} = \hbar g_0 \, \hat{a}^\dagger \hat{a} (\hat{b} + \hat{b}^\dagger)$ where $g_0 = G \, \delta z^\mathrm{zpf}$ is the single photon coupling strength with $G = \partial \omega_\mathrm{cav} / \partial z_0$ and $\delta z^\mathrm{zpf} = \sqrt{\hbar / 2 M_\mathrm{eff} \Omega_\mathrm{m}}$ the spatial spreading of the oscillator zero point fluctuations (effective mass $M_\mathrm{eff}$~\cite{pinard1999effective}). A single photon in the cavity ($\langle \hat{a}^\dagger \hat{a} \rangle = 1$) generates an optical force on the oscillator $F^{(1)} = - \hbar g_0 / \delta z^\mathrm{zpf}$ leading to a static displacement $\delta z^{(1)} = F^{(1)}/M_\mathrm{eff} \Omega_m^2 = 2 (g_0 / \Omega_m) \delta z^\mathrm{zpf}$. The static effect of a single photon in the cavity will then be observable only if $\delta z^{(1)}$ is larger than $\delta z^\mathrm{zpf}$, a criteria equivalent to $2g_0 / \Omega_m \geq 1$. Moreover, $\delta z^{(1)}$ also have to be larger than the Brownian spreading of the oscillator position $\Delta z^\mathrm{th} = \sqrt{k_B T / M_\mathrm{eff} \Omega_m^2}$ due to thermal fluctuation, which makes the temperature a relevant parameter of the problem. Finally, the static deformation of the oscillator leads to a parametric shift of the cavity resonance, $\delta \omega_\mathrm{cav}^{(1)} = g_0 \, \delta z^{(1)} / \delta z^\mathrm{zpf}$. If it exceeds the cavity linewidth, $\delta \omega_\mathrm{cav}^{(1)} > \kappa_\mathrm{cav}$, the system presents a static optomechanical non-linearity at the single intra-cavity photon level. This exotic regime is achievable if the \textit{single photon parametric cooperativity} $\mathcal{C}^{(1)} = 2 g_0^2 / \kappa_\mathrm{cav} \Omega_m$ is larger than one.

In this Section we study the optomechanical interaction between the nanowire and the fundamental cavity mode which acquires a vectorial character, $\textbf{G} = \bm{\nabla} \omega_\mathrm{cav} \vert_{\textbf{r}_0}$, the nanowire being able to oscillate in both transverse $(xz)$ directions. However, due to the faster variations of the intra-cavity field intensity along the cavity axis (over $\lambda/4$ along $z$ compared to $w_0$ along $x$), we will focus in the following on the optomechanical interaction along that direction since it provides the larger coupling strength ($G_z / G_x \sim 10$). We will show that the NIM configuration open the road to the study of the intrinsic optomechanical non-linearity at the single photon scale, a long standing goal of cavity optomechanics.

\subsection{Optomechanical coupling strength}
\label{sec:optomecha_coupling_TM00:coupling_strength}

To obtain the coupling strength $G_z = \partial \omega_\mathrm{cav} / \partial z_0$ it is essential to know how the presence of the nanowire in the optical cavity shifts the cavity resonance. Because in the system depicted here the cavity shifts are of the order of several hundreds of GHz which is very large compared to the fast tunability of most available lasers, it has been chosen in the experimental work of ref.~\cite{Fogliano2021} to adjust the cavity length instead of the laser frequency and we follow the same approach here. For small relative shifts of the cavity frequency, the coupling strength is given by
\begin{eqnarray}
G_z \approx \frac{4 \pi c}{N \lambda^2} \, \frac{\partial L_{cav}}{\partial z_0}, \label{Eq:optomecha_coupling:G_fct_dL/dz}
\end{eqnarray}
where $N$ is the longitudinal mode order of the pumped cavity mode ($N = 32$ in this work, meaning that a node is located at the center of the cavity at $z = 0$).
\begin{figure*}[t]
\begin{center}
\includegraphics[width=0.99\linewidth]{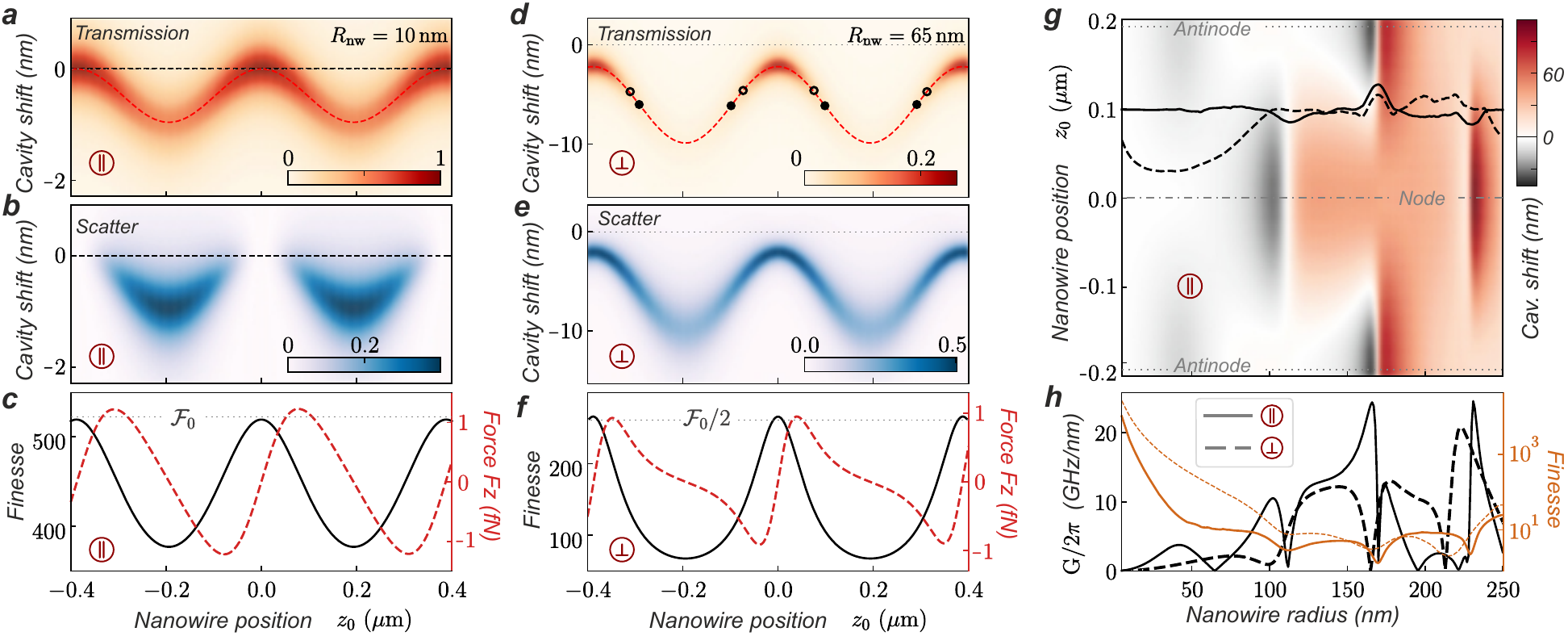}
\caption{Cavity transmission and losses coefficients for different positions of the nanowire on the cavity axis while scanning the cavity length: (ab) $R_\mathrm{nw} = 10 \, \mathrm{nm}$ and parallel polarization of the light, (de) $R_\mathrm{nw} = 65 \, \mathrm{nm}$ and perpendicular polarization. (c) and (f) show the nanowire position dependence of the cavity finesse (black solid lines) and of the optical force applied by the intra-cavity field at resonance on the nanowire (red dashed lines) obtain in~\ref{sec:2D_carac:optical_force}. In (c), the cavity finesse remains larger than $\mathcal{F}_0 / 2$ meaning that the nanowire only probe the intra-cavity field. In (f) on the contrary, the nanowire strongly impacts the intra-cavity field, leading to an important decrease of the cavity finesse at an anti-node of the field and then to a drop of the nanowire induced cavity losses. In (d) $\bullet$ and $\circ$ indicate nanowire positions maximizing the coupling strength $G_z$ and the single photon parametric cooperativity $\mathcal{C}^{(1)}$ respectively. (g) Resonant cavity shift maps for varying nanowire radii and positions on the cavity axis, for a parallel polarization of the light. The black solid and dashed lines indicate positions of maximum coupling strength $G_z$ and maximum ratio $G_z^2/\kappa_\mathrm{cav}$ involved in the single photon static cooperativity, $\kappa_\mathrm{cav}$ being the linewidth of the cavity. (h) Radius dependence of the maximum optomechanical coupling strength along z for both light polarizations (thick black lines). The thin orange lines represent the cavity finesse at the position of the maximum coupling strength. (g) and (h) have been obtained for an empty cavity finesse of $\mathcal{F}_0 \approx 50000$ which is the highest value one can reach using the experimental configuration of ref.~\cite{Fogliano2021}.}
\label{Fig:NW_in_cavity:LZ_map_fig}
\end{center}
\end{figure*}

The cavity shift $\partial L_{cav} / \partial z_0$ is obtained by computing the cavity resonant length for different nanowire position. It is determined from the transmission of the cavity, which is  the channel used in the experiment to lock the cavity at resonance. To do so, we propagate the field through the optical cavity (see Fig.~\ref{Fig:NW_in_cavity:Schema_full_field_propagation}) using the transfer matrix formalism presented in~\ref{sec:NW_in_cavity:optical_cavity_field}. The reduced fields on the right side of the cavity are expressed in terms of the reduced fields at its left as
\begin{subequations}
\label{Eq:optomecha_coupling:general_propagation_equations}
\begin{align}
& \VecTwoD{F_t^{(p)}}{F_v^{(p)}} = M^{(p)} \VecTwoD{F_i^{(p)}}{F_r^{(p)}}, \\
\nonumber \\
& M^{(p)} = M_{\mathrm{BS}, R} \, M_{z_0, \frac{L_\mathrm{cav}}{2}} \, M_{\mathrm{nw}}^{(p)} \, M_{-\frac{L_\mathrm{cav}}{2}, z_0} \, M_{\mathrm{BS}, L} \label{Eq:optomecha_coupling:general_propagation_equations_transfer_matrix},
\end{align}
\end{subequations}
where $(p)$ stands for the polarization of the incident light which can be either parallel or perpendicular to the nanowire axis. In this expression, $F_r^{(p)}$ and $F_t^{(p)}$ correspond to the reflected and transmitted reduced fields while $F_i^{(p)}$ and $F_v^{(p)}$ are the incoming reduced fields on the cavity, from the left and right respectively. In the following, we will consider a cavity pumped only from the left side such as $F_v^{(p)} = 0$. In~\eqref{Eq:optomecha_coupling:general_propagation_equations_transfer_matrix}, $M_{\mathrm{BS}, L/R}$ are the transfer matrices of the cavity mirrors given in Eq.~\eqref{Eq:NW_in_cavity:transfer_matrix_BS} where we have chosen $\eta_L = -1$ and $\eta_R = 1$ to ensure the symmetry of the reflected intra-cavity fields on both cavity mirrors, and $R_L = R_R = 0.994$. The propagation of the fields in the left and right sub-cavities is taken into account through $M_{-L_\mathrm{cav}/2, z_0}$ and $M_{z_0, L_\mathrm{cav}/2}$ respectively, these matrices being obtained using Eq.~\eqref{Eq:NW_in_cavity:transfer_matrix_free_space} in the case of the fundamental cavity mode ($n_x = n_y = 0$). The transfer matrix of the nanowire $M_{\mathrm{nw}}^{(p)}$ which depends on the laser wavelength and polarization, as well as on the nanowire position and geometry, is obtained using Eq.~\eqref{Eq:NW_in_cavity:M_transfer_matrix_NW_coef} where the reflection and transmission coefficients have been obtained in~\ref{sec:NW_in_cavity:Impact_on_TM00}. As already discussed, the nanowire does not couple the two polarizations which can then be treated independently. Finally, we define the reflection and transmission coefficients of the cavity for a given polarization as
\begin{subequations}
\label{Eq:optomecha_coupling:definition_Cr_Ct}
\begin{align}
& C_r^{(p)} = F_r^{(p)}/F_i^{(p)}, \\
& C_t^{(p)} = F_t^{(p)}/F_i^{(p)}.
\end{align}
\end{subequations}
They allow to define the intensity coefficients as $C_R^{(p)} = \abs{C_r^{(p)}}^2$ and $C_T^{(p)} = \abs{C_t^{(p)}}^2$ corresponding to the ratio of the intensity of the reflected or transmitted field divided by the intensity of the incident field on the cavity. Moreover, using the energy conservation, it is possible to evaluate the amount of light scattered out of the cavity mode due to the nanowire, $C_L^{(p)} = 1 - C_R^{(p)} - C_T^{(p)}$.

The evaluation of~\eqref{Eq:optomecha_coupling:general_propagation_equations} and~\eqref{Eq:optomecha_coupling:definition_Cr_Ct} for different nanowire positions on the cavity axis and for different cavity lengths around the resonance of the fundamental mode leads to the so called LZ maps, from which the resonant cavity length $L_\mathrm{cav}(z_0)$ and the quantitative evaluation of $\partial L_{cav} / \partial z_0$ can be derived. Fig.~\ref{Fig:NW_in_cavity:LZ_map_fig} represents the LZ maps of the transmission and losses coefficients as well as the position dependence of the cavity finesse $\mathcal{F}$ for two nanowire radii and different incident polarization. Note that the finesse $\mathcal{F}$ takes into account the losses due to the finite reflectivity of the cavity mirrors as well as the ones due to the scattering of the nanowire. On these plots, the origin of the cavity shift matches the resonant cavity length of the symmetric $32\mathrm{th}$ longitudinal cavity mode when no nanowire is present in the cavity ($L_\mathrm{cav}^0 \approx 12.440 \, \mu \mathrm{m}$, $\mathcal{F}_0 \approx 522$), meaning that a node of the intra-cavity field is present at the center of the cavity ($z = 0$). The left column is obtained for a small nanowire radius ($R_\mathrm{nw} = 10 \, \mathrm{nm}$, parallel polarization) and corresponds to the dipole-like case where the nanowire hardly disrupts the intra-cavity field. We observe the $\lambda / 2$ periodic oscillation of the cavity resonant length in agreement with the standing wave profile of the intra-cavity field. At the nodes of the intra-cavity field, there is no cavity shift or losses induced by the nanowire, while they are maximized at an anti-node.

The middle column of Fig.~\ref{Fig:NW_in_cavity:LZ_map_fig} is obtained for a nanowire radius $R_\mathrm{nw} = 65 \, \mathrm{nm}$ (perpendicular polarization) and corresponds to the experimental situation of ref.~\cite{Fogliano2021}. We observe cavity shifts of the order of $10 \, \mathrm{nm}$ in a good quantitative agreement with the experimental results. It is worth noticing that because of the large enough radius of the nanowire, residual cavity shifts and losses are also observable at the nodes of the intra-cavity filed. Moreover, contrary to the previous case where the nanowire induced cavity losses remained small compared to the intrinsic losses of the cavity mirrors here they become larger ($\mathcal{F} < \mathcal{F}_0/2$, see Fig.~\ref{Fig:NW_in_cavity:LZ_map_fig}(f)). This finesse reduction is then responsible for a reduction of the intra-cavity field, so that the total resonant losses get reduced when the nanowire is largely inserted in the optical mode. We will see in~\ref{sec:2D_carac:2D_carac_intra_cavity_field} that this effect is also at the origin of the ring shapes appearing in the XZ maps of the scattered field when the cavity length is locked at resonance.

Using the same procedure for different nanowires radii gives access to the cavity shifts maps presented Fig.~\ref{Fig:NW_in_cavity:LZ_map_fig}(g) for $z_0 \in [-\lambda/4, \lambda/4]$, where we plot the resonant cavity shifts obtained for a parallel polarizations of the light (see Appendix~\ref{Appendix:cavity_shifts_perp_polar} for the perpendicular case). The first interesting result consists in the existence of radius ranges where the cavity shift is more important at a node of the intra-cavity field than at an anti-node, for example for a nanowire radius between $65 \, \mathrm{nm}$ and $112 \, \mathrm{nm}$. This effect, also observable in \textit{membrane in the middle} systems, is due to internal resonances and depends on the number of field oscillations storable inside the nanowire, which structures its reflection and transmission coefficients, demonstrating the radius dependence of this phenomena. More surprisingly, we observe \textit{positive shifts} of the resonant cavity length as soon as the diameter of the nanowire becomes large enough. This effect was unexpected since the insertion of a dielectric of refractive index larger than one could at first sight only be expected to increase the cavity optical path length, then leading to a reduction of its resonant length. Contrary to the previous situation, it is not observable in MIM systems and is a specificity of the nanowire in the middle configuration due to the dimensionality of the system. Indeed, for 1D objects inserted in a 2D optical field, the input/output formalism leads to new relations between the reflection and transmission coefficients of the scatterer which enriches the phenomenology of the system. The experimental investigation of this effect will be the subject of a future work.

From the above results it is clear that the position of the nanowire in the cavity can be used to tune the optomechanical interaction. It is null at a node or an anti-node of the intra-cavity field since the resonant cavity length shift is extremum, and maximum in-between as shown by the black solid line in Fig.~\ref{Fig:NW_in_cavity:LZ_map_fig}(g) which represents the positions of the maximum coupling strength shown in (h). As expected from the existence of internal Mie resonances which structure the light-nanowire interaction, we observe  a strong variation of the maximum coupling strength with the nanowire radius and light polarization. These results can be used to maximize the optomechanical coupling of the system, so as to reach original regimes as discussed below. %Because induced induced nanowire cavity losses also plays a fundamental role
\begin{figure}[t!]
\begin{center}
\includegraphics[width=0.95\linewidth]{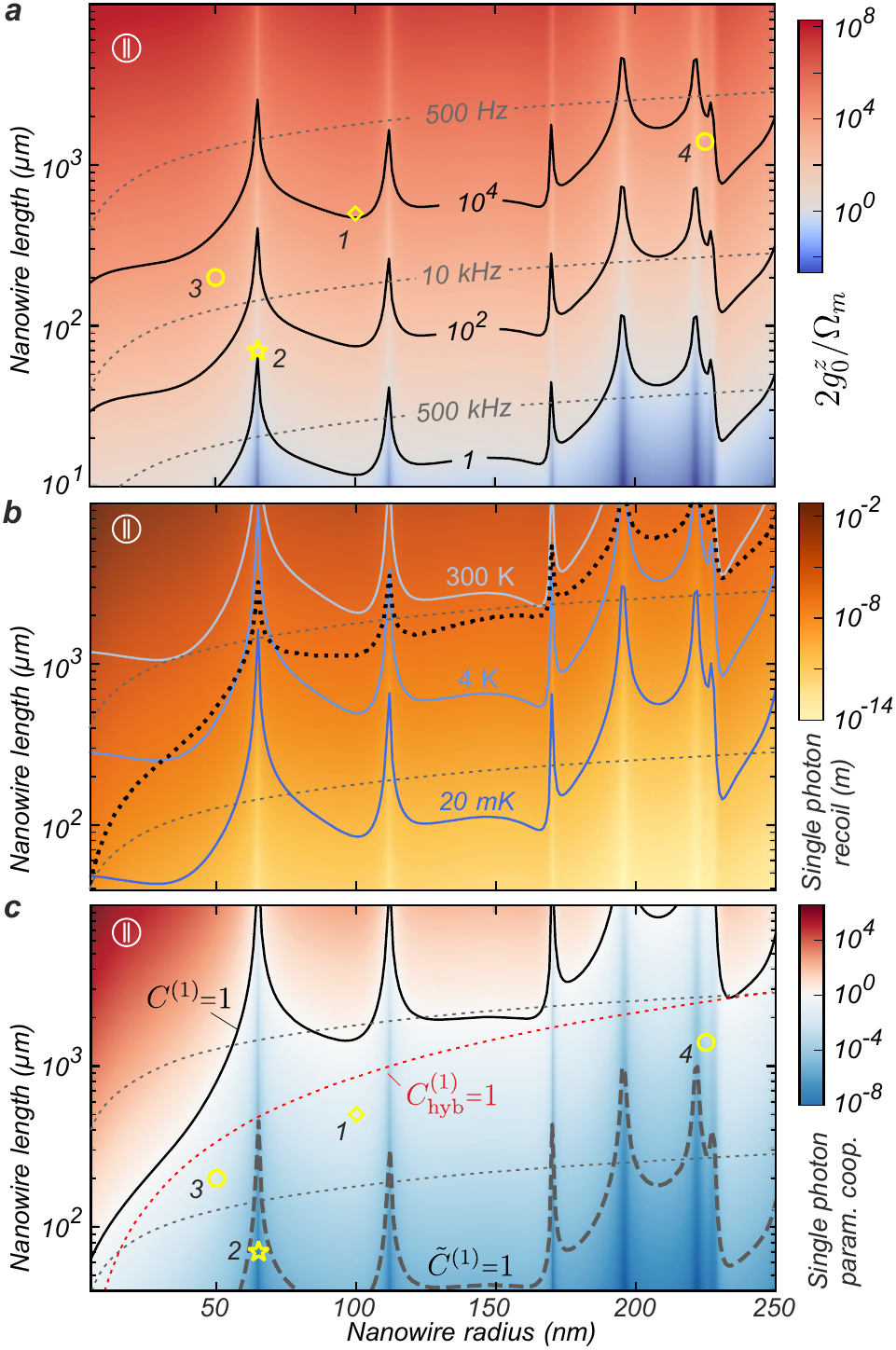}
\caption{Dependence on the nanowire dimensions of $\delta z ^{(1)} / \delta z^\mathrm{zpf} = 2 g_0^z / \Omega_\mathrm{m}$ (a), of the single photon recoil $\delta z^{(1)}$ (b), and of the static single photon parametric cooperativity $\mathcal{C}^{(1)} = 2 \left. g_0^z \right. ^2 / \kappa_\mathrm{cav} \Omega_m$ (c) for parallel polarization of the intra-cavity field (see Appendix~\ref{Appendix:single_photon_optomecha_perp_polar} for the perpendicular case). Note that a nanowire length beyond 1\,mm is within experimental reach. The grey dashed lines are the fundamental vibrational mode iso-frequencies. In (b), the blue lines show the nanowire dimensions ensuring $\delta z^{(1)} =\Delta z^{\mathrm{th}}$ for different bath temperatures while the black dotted line indicates nanowire dimensions for which $\delta z^{(1)} = R_\mathrm{nw}$. In those extreme regimes of giant light-induced nanowire deformations, one cannot expect the linear modelization of \ref{sec:optomecha_coupling_TM00:formalization} to remain pertinent. The diamond and star markers correspond to a relevant nanowire for cryogenic temperature experiment (NW\,1: $R_\mathrm{nw} = 100 \, \mathrm{nm}$ and  $L_\mathrm{nw} = 500 \, \mu \mathrm{m}$) and to the nanowire used in ref.~\cite{Fogliano2021} (NW\,2: $65 \, \mathrm{nm}, 70 \, \mu \mathrm{m}$) respectively. The two circles indicate two nanowires recently studied in our group: NW\,3 ($50 \, \mathrm{nm}, 200 \, \mu \mathrm{m}$) and NW\,4 ($ 225 \, \mathrm{nm},  1400 \, \mu \mathrm{m}$). In (c) we also show nanowire dimensions corresponding to a static single photon parametric cooperativity of one (solid black line) and to a dynamical single photon parametric cooperativity of one (dashed black line). The dashed red line indicates a static single photon parametric cooperativity of one when the nanowire extremity is functionalized to maximize the optical interaction while minimizing the optical losses (see text).}
\label{Fig:NW_in_cavity:Optomechanical_coupling}
\end{center}
\end{figure}

\subsection{Optomechanics at the single photon level}
\label{sec:optomecha_coupling_TM00:strong_coupling_and_cooperativity}

We have seen that the optomechanical coupling strength strongly depends on the nanowire diameter.  We will now explore the geometrical parameters of the nanowire (length and diameter) in order to identify a regime of parameters which maximize the optomechanical interaction. In the framework of~\ref{sec:optomecha_coupling_TM00:formalization}, the static force exerted on the nanowire by a single intra-cavity photon is potentially observable if $\delta z ^{(1)} / \delta z^\mathrm{zpf} = 2 g_0^z / \Omega_m > 1$, this ratio involving the optomechanical coupling strength $G_z$ as well as the effective mass $M_\mathrm{eff}$ and frequency $\Omega_\mathrm{m} / 2 \pi$ of the fundamental vibrational mode. From the beam theory~\cite{cleland2013foundations} we have $\Omega_\mathrm{m} / 2 \pi = \kappa_\Omega R_\mathrm{nw} / L_\mathrm{nw}^2 $ where $\kappa_\Omega = 3126 \, \mathrm{Hz.m}$ is obtained from the Euler-Lagrange equation using a Young modulus $E = 400 \, \mathrm{GPa}$ and density $\rho = 3210 \, \mathrm{kg.m}^{-3}$ of SiC (iso-frequency are plotted as grey dashed lines in Fig.~\ref{Fig:NW_in_cavity:Optomechanical_coupling}). The effective mass of the fundamental vibrational mode represents a fourth of the nanowire mass, leading to $M_\mathrm{eff} = \rho \pi R_\mathrm{nw}^2 L_\mathrm{nw} / 4$. Using the radius dependence of the parametric coupling strength $G_z$ exposed in section~\ref{sec:optomecha_coupling_TM00:coupling_strength}, we compute the maximum value of $2 g_0^z / \Omega_m$ as a function of the nanowire dimensions for a parallel polarization of the intra-cavity field (Fig.~\ref{Fig:NW_in_cavity:Optomechanical_coupling}(a)). This ratio is largely above unity for a wide range of accessible nanowire geometries, which demonstrates the possibility to largely enter in the regime where $\delta z ^{(1)} / \delta z^\mathrm{zpf} > 1$. It reaches $10^4$ for a nanowire of radius $R_\mathrm{nw} = 100 \, \mathrm{nm}$ and length $L_\mathrm{nw} = 500 \, \mu \mathrm{m}$ (NW 1, diamond marker in Fig.~\ref{Fig:NW_in_cavity:Optomechanical_coupling}) relevant for cryogenic temperature experiments~\cite{Fogliano2021a}. Additionally, we indicate by a star in Fig.~\ref{Fig:NW_in_cavity:Optomechanical_coupling} the nanowire used in ref.~\cite{Fogliano2021} (NW 2: $R_\mathrm{nw} = 65 \, \mathrm{nm}$ and $L_\mathrm{nw} = 70 \, \mu \mathrm{m}$) and we show in Appendix~\ref{Appendix:single_photon_optomecha_perp_polar} the ratio $2 g_0^z / \Omega_m$ for a perpendicular polarization of the light (situation of ref.~\cite{Fogliano2021}). In order to highlight where our experiment stands compared to the theory, we also show (black circles) two nanowires recently studied in our group: NW~3 ($R_\mathrm{nw} = 50 \, \mathrm{nm}$, $L_\mathrm{nw} = 200 \, \mu \mathrm{m}$) and NW~4 ($R_\mathrm{nw} = 225 \, \mathrm{nm}$, $L_\mathrm{nw} = 1400 \, \mu \mathrm{m}$).

However, satisfying the condition $2 g_0^z / \Omega_m > 1$ is not sufficient alone in a real experiment to directly measure the static impact of a single intra-cavity photon on the nanowire. The single photon deformation $\delta z^{(1)}$ has also to be larger than the Brownian spreading of the nanowire position $\Delta z^{\mathrm{th}}$ associated to its coupling to the thermal bath. The experimental work of ref.~\cite{Fogliano2021} was conducted at room temperature where $\delta z^{(1)} \sim 20 \, \mathrm{pm}$ while $\Delta z^{\mathrm{th}} \sim 10 \, \mathrm{nm}$, which underlines the importance to operate at low temperature where the thermal spreading can be reduced by several orders of magnitude. Fig.~\ref{Fig:NW_in_cavity:Optomechanical_coupling}(b) shows the value of $\delta z^{(1)}$ in terms of the nanowire dimensions for a parallel polarization of the light (see Appendix~\ref{Appendix:single_photon_optomecha_perp_polar} for the perpendicular case). The blue lines indicates the nanowire dimensions ensuring $\delta z^{(1)} = \Delta z^{\mathrm{th}}$ for different bath temperatures. Interestingly, NW~1 presents a single photon static displacement larger than its Brownian spreading as soon as the temperature is reduced below $4 \, \mathrm{K}$. For this nanowire, the ratio $\delta z^{(1)} / \Delta z^{\mathrm{th}}$ even reach a value of the order of $300$ at $20 \, \mathrm{mK}$, demonstrating the possibility of largely entering such a regime experimentally. It is one of the reason why a specific effort has been put recently to develop an experimental set-up at cryogenic temperature, already showing promising results~\cite{Fogliano2021a}. Note that the measurements realized in ref.~\cite{Fogliano2021} have allowed to measure the action of the intra-cavity field on the nanowire at unitary photon numbers. First, by using a pump-probe technique and a temporal averaging, we manage to detect optical forces generated by mean photon number changes smaller than one. Secondly, we demonstrated that the impact of the gradients of such optomechanical force field (produced by approximatively one photon) was observed and does govern the mechanical properties of the nanowire (its vibration frequency in particular). In comparison, the temperature criteria discussed above leads to the desirable regime where the nanowire dynamics and its position fluctuations are totally dominated by intra-cavity light quantum fluctuations, which could then be observable at the single photon level.

We have thus shown that the optomechanical interaction is extremely large in such a configuration and that a single intra-cavity photon can have a measurable impact on the nanowire dynamics. We now investigate a second order optomechanical effect, the possibility to reach the regime where one can observe a static bistability of the cavity, at the single photon level. The static bistability is observed when the intra-cavity field generates a static deformation of the nanowire, which in turn shifts the cavity resonance by more than its optical linewidth. This highly non-linear regime is achieved when the static single photon parametric cooperativity $\mathcal{C}^{(1)} = 2 \left. g_0^z \right. ^2 / \kappa_\mathrm{cav} \Omega_m$ becomes larger than 1. The expression of $\mathcal{C}^{(1)}$ highlights the fact that a large cooperativity requires an appropriate balance between the dispersive and dissipative coupling: the cavity shift has to be as large as possible to ensure a large value of $g_0^z$ while preserving a small cavity loss rate $\kappa_\mathrm{cav}$, or in other words a sufficiently large cavity finesse. As a consequence, there exists optimum positions in the standing wave which maximize the single photon parametric cooperativity: an example of this compromise is shown in Fig.~\ref{Fig:NW_in_cavity:LZ_map_fig}(d) where the positions along $z$ maximizing $\mathcal{C}^{(1)} $ (indicated by $\circ$) do not coincide with the positions associated to the maximum coupling strength (indicated by $\bullet$) since the nanowire modifies the cavity finesse has shown in Fig.~\ref{Fig:NW_in_cavity:LZ_map_fig}(f). Fig.~\ref{Fig:NW_in_cavity:Optomechanical_coupling}(c) presents the maximum value of $\mathcal{C}^{(1)}$ as a function of the nanowire dimensions for an empty cavity finesse $\mathcal{F}_0 \approx 50000$ (typical values one can expect with the fiber micro-cavity of ref.~\cite{Fogliano2021}) and for a parallel polarization of the light (see Appendix~\ref{Appendix:single_photon_optomecha_perp_polar} for the perpendicular case). The values obtained with existing nanowires are smaller than one but can approach unity. It means that it is necessary to put $1/\mathcal{C}^{(1)}$ photons inside the cavity to observe a static bistability, namely 25 (7) photons in the situation of NW~1 (NW~4) for a parallel (perpendicular) polarization of the light. However operating those ultra-soft nanowires in the middle of a fiber microcavity remains a true experimental challenge. There are possible avenues for improving the single photon parametric coupling strength, such as optimizing the optical mode geometry to minimize the nanowire induced photon losses, or operating with thin and long nanowires (which maximizes their mechanical susceptibility) functionalized  at their extremity with a sub-micron sphere to maximize the optical interaction while minimizing the optical losses. To give an order of magnitude of the latter possibility, we show in Fig.~\ref{Fig:NW_in_cavity:Optomechanical_coupling}(c) (black dotted line) a situation where the nanowire has been functionalized by a second nanowire maximizing the ratio $G_z^2 / \kappa$.

Finally, the above discussions were connected to the static deformation and the static bistability. One can also evaluate the dependence on the nanowire geometry of the more common dynamical optomechanical single photon cooperativity, $\tilde{\mathcal{C}}^{(1)} = 2 g_0^2/\kappa_\mathrm{cav} \Gamma_{\mathrm{m}}$. It makes use of the resonant mechanical susceptibility instead of the static one which is thus enhanced by the mechanical quality factor, and $\tilde{\mathcal{C}}^{(1)} \, N_\mathrm{cav}$ generally sets the strength of the second order optomechanical effects, such as optomechanical cooling or parametric instability \cite{Aspelmeyer2014,Arcizet2006a}.
In particular, large values of $\tilde{\mathcal{C}}^{(1)}$ corresponds to a strong squeezing of the outgoing light at the mechanical frequency~\cite{Purdy2013a, safavi2013squeezed, nielsen2017multimode} while $\mathcal{C}^{(1)} N_\mathrm{cav} \gg 1 $ leads to a broadband squeezing in the cavity bandwidth in the adiabatic regime~\cite{Fabre1994} which is the situation of this work ($\Omega_\mathrm{m} \ll \kappa_\mathrm{cav}$). The black dashed line in Fig.~\ref{Fig:NW_in_cavity:Optomechanical_coupling}(c) indicates $\tilde{\mathcal{C}}^{(1)} = 1$ for a nanowire quality factor $Q = 10^5$  (as observed at $20 \, \mathrm{mK}$~\cite{Fogliano2021a}) demonstrating the high potential of the NIM configuration. It opens the road to applications in quantum optics still operating at low photon numbers, where non Gaussian states of the light \cite{Reynaud1989} should be reachable with a mean photon number close to one.

\section{Two-dimensional characterization of a nanowire in the middle system}
\label{sec:2D_charac}

The above analyses were restricted to the situation where the nanowire is positioned on the optical axis. In this section we will explore the 2D character of the system, and in particular the optomechanical force field experienced by the nanowire.

\subsection{Two-dimensional characterization of the intra-cavity field}
\label{sec:2D_carac:2D_carac_intra_cavity_field}

\begin{figure}[t!]
\begin{center}
\includegraphics[width=0.99 \linewidth]{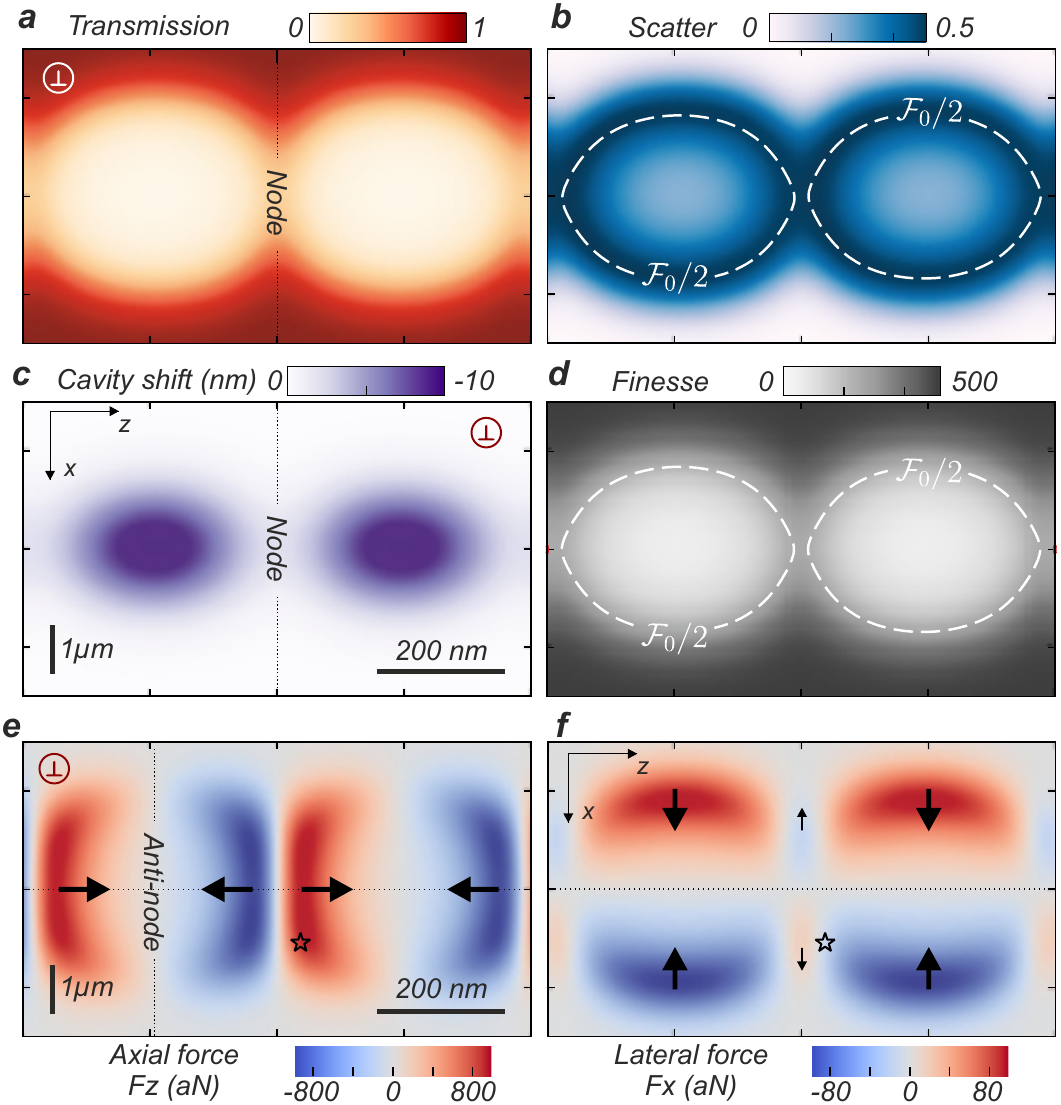}
\caption{Transmission (a), cavity shift (b), scattered coefficient (c) and cavity finesse (d) maps obtained when scanning a nanowire of radius $R_\mathrm{nw} = 65 \, \mathrm{nm}$ in the $(xz)$ plane while locking the cavity at resonance for a perpendicular polarization of the light. The dashed lines in (c) and (d) indicates positions where $\mathcal{F} = \mathcal{F}_0 / 2$ and coincide with the rings in the scattered map. (ef) Optomechanical force maps applied by the intra-cavity field on the nanowire for an input power of $1 \, \mu \mathrm{W}$ and a perpendicular polarization of the light. The $\star$ symbol indicates the position of maximum force curl ($1 \, \mathrm{nN.m}^{-1}$), see text.}
\label{Fig:NW_in_cavity:XZ_map_65nm}
\end{center}
\end{figure}

Due to its sub-wavelength-sized diameter, the nanowire can be efficiently used as a scanning probe to map and explore the structure of the intra-cavity field, \textit{i.e.} the wave function of the confined photonic mode. This capacity is largely used experimentally in the alignment and optimisation sequences. In the following, we will "lock" the cavity length at resonance, exploiting the cavity transmission signal as in the experimental case, to ensure that for any position of the nanowire in the cavity the resonance condition of the fundamental cavity mode remains satisfied. In practice, for a given nanowire position, we numerically evaluate Eq.\eqref{Eq:optomecha_coupling:general_propagation_equations}, compute the intensity transmission coefficient while scanning the cavity length around the resonance (which is equivalent to a vertical cut in the transmission map of Figure~\ref{Fig:NW_in_cavity:LZ_map_fig}), and fit it to extract the resonant cavity length as well as the cavity finesse. Then, reproducing such a procedure for different positions of the nanowire in the $(xz)$ plane, we obtain two dimensional maps of the resonant cavity shift, cavity finesse, transmission and scatter coefficients as shown Figure~\ref{Fig:NW_in_cavity:XZ_map_65nm} for NW~2 (perpendicular polarization). It allows a direct visualization of the intra-cavity standing waves, the nodes (anti-nodes) of the field appearing here as region of large (low) transmission and small (large) cavity shifts.

For this nanowire, the scatter map presents ring shapes due to the large dispersive coupling achieved when the nanowire is positioned in the middle of an antinode. Outside of the rings, the optical losses induced by the nanowire are smaller than the intrinsic losses of the cavity mirrors and the nanowire locally probe the intra-cavity field without too much alteration. Inside the rings, the dispersive coupling becomes so large that the nanowire induced losses overpass the cavity mirrors losses, thus lowering the intra-cavity field and leading to a decrease of the amount of scattered light. Actually, the rings correspond to nanowire positions where the losses due to the nanowire and to the finite mirror reflectivity are equivalent, thus reducing the cavity finesse by a factor two ($\mathcal{F} = \mathcal{F}_0 / 2$).

If the reduction of the intra-cavity field strength is usually seen as a limitation preventing the observation of large cooperativities or optomechanical forces, it also provides a new dispersive measurement channel featuring large  variations with the nanowire position (stronger than the one observed in the  transmission or reflection channels), a key ingredient to realize efficient optical readout of the nanowire position~\cite{gloppe2014bidimensional, de2017universal, de2018eigenmode}. Experimentally, the signal to noise ratio obtained on the laterally scattered signals was always significantly larger than the one obtained on the usual measurement channels. As a comparison, we give in Appendix~\ref{Appendix:Simu_Rnw_10nm} the XZ maps obtained for a dipole-like behaving nanowire of radius $R_\mathrm{nw} = 10 \, \mathrm{nm}$ and for a parallel polarization of the light. In this case, the nanowire hardly disturbs the intra-cavity field (see Fig.~\ref{Fig:NW_in_cavity:LZ_map_fig}(c)) and no ring is observable in the scattered map.

\subsection{Optomechanical force exerted on the nanowire}
\label{sec:2D_carac:optical_force}

The previous scanning probe analysis allows to understand how the nanowire impacts the intra-cavity field, which is a first aspect of the optomechanical coupling. To fully characterize the NIM system, it is crucial to also study the reverse interaction, namely the optomechanical force applied by the intra-cavity field on the oscillator. Contrary to usual cavity optomechanical setup, this dual investigation is of first importance in NIM configurations because the spatial profile and the pumping efficiency of the cavity mode vary with the nanowire position. As such, the photon number present in the optical mode will also depend on the nanowire localization, which prevents to estimate the optical force from the single knowledge of the parametric coupling strength $g_0$. In that view, the reasoning of section~\ref{sec:optomecha_coupling_TM00:strong_coupling_and_cooperativity} should be seen as a qualitative evaluation of the the optomechanical interaction expected in the NIM system.

The time-average of the optical force experienced by the nanowire is computed by integrating the Maxwell stress tensor on the nanowire surface when inserted inside the cavity. It is given by
\begin{eqnarray}
&& \textbf{F} = \frac{\varepsilon_0}{2} \oint_\mathcal{S} \mathrm{d}^2 S \bigg[ \mathrm{Re} \left[ (\textbf{E} \cdot \textbf{n}) \textbf{E}^\ast \right] + c^2 \mathrm{Re} \left[(\textbf{B} \cdot \textbf{n}) \textbf{B}^\ast \right] \nonumber \\
&& \hspace*{2.5cm} - \frac{1}{2} \left( (\textbf{E} \cdot \textbf{E}^\ast) \textbf{n} + c^2 (\textbf{B} \cdot \textbf{B}^\ast) \textbf{n} \right) \bigg],
\label{Eq:2D_carac:general_expression_optical_force}
\end{eqnarray}
where $\ast$ denotes the complex conjugate quantity, $\mathrm{Re}[.]$ is the real part operator, $\textbf{n}$ is the outward normal vector to $\mathcal{S}$, and $\textbf{E}$ and $\textbf{B}$ are the complex electric and magnetic fields evaluated on the closed surface $\mathcal{S}$ enclosing the nanowire. In practice, these fields will be evaluated at the nanowire outer periphery, making use of the Mie expressions given above. %The fields appearing in Eq.~\eqref{Eq:2D_carac:general_expression_optical_force} will be calculated without restriction to their projections on the cavity mode, which will allow to compute the optical force vector for any position of the nanowire inside the cavity, giving access to a mechanical exploration of the intra-cavity photon wave function.
They will be calculated by taking into account the vectorial structure of the Gaussian modes, which will allow to compute the optical force vector for any position of the nanowire inside the cavity. This procedure will lead to a correct description of the forces as required for a mechanical exploration of the intra-cavity photon wave function. Additionally, we consider a spatially limited Gaussian cavity mode of waist $w_0$ and we assume that there is no efficient wave-guiding mechanism along the nanowire, in such a way that the surfacic integration can be restricted to the one of the infinite cylinder surrounding the nanowire, leading to

$ \oint_\mathcal{S} \mathrm{d}^2 S \ \ast \rightarrow R_\mathrm{nw} \int_{-\infty}^{+\infty} \mathrm{d}y \int_{0}^{2 \pi} \mathrm{d} \varphi \ \ast
$.

When inserted inside the cavity, the nanowire is simultaneously illuminated by the left and right incoming fields propagating towards the nanowire in each sub-cavity. The total incident field then reads as $\textbf{E}_\mathrm{inc}^{(p)}(\textbf{r}) {=} \sqrt{P_\mathrm{inc} / \varepsilon_0 c} \left( A^{(+, p)} \, \textbf{E}_{0,0}^{(+, p)}(\textbf{r}) + B^{(-, p)} \, \textbf{E}_{0,0}^{(-, p)}(\textbf{r}) \right)$, where $P_\mathrm{inc}$ is the incident power, $\textbf{E}_{0,0}^{(\pm, p)}$ are the cavity modes propagating along $\pm z$, and $A^{(+, p)}$ and $B^{(-, p)}$ are the dimensionless amplitudes of the corresponding fields associated to the $z = 0$ plane. These amplitudes are obtained from the transfer matrix formalism as
\begin{subequations}
\label{Eq:2D_carac:intra_cavity_amplitudes}
\begin{align}
& \VecTwoD{A^{(+, p)}}{A^{(-, p)}} = M_{-L/2, 0} \, M_{\mathrm{BS}, L} \VecTwoD{1}{C_r^{(p)}}, \\
& \VecTwoD{B^{(+, p)}}{B^{(-, p)}} = M_{z_0, 0} \, M_{\mathrm{nw}}^{(p)} \, M_{-L/2, z_0} \, M_{\mathrm{BS}, L} \VecTwoD{1}{C_r^{(p)}},
\end{align}
\end{subequations}
where $C_r^{(p)}$ is calculated using Eq.~\eqref{Eq:optomecha_coupling:general_propagation_equations} and~\eqref{Eq:optomecha_coupling:definition_Cr_Ct}.
%
%The electric and magnetic field appearing in~\eqref{Eq:2D_carac:general_expression_optical_force_2} are computed using the Mie formalism presented in Appendix~\ref{Appendix:Optical_prop_SiC_NW}.
Because we are interested in the vectorial aspect of the optical force in the $(xz)$ plane, it is necessary to take into account the transverse structure of the Gaussian cavity mode (the local approximation by a single incident plane wave would not be sufficient here to account for the transverse force along $x$). This is done by expanding the incident field on the plane wave basis, for which the scattered field takes an analytical form so that we can operate with analytical expressions for the electromagnetic field surrounding the nanowire. Furthermore, since we consider the case of an infinite nanowire inserted in a Gaussian beam, the force along $y$ cancels for symmetry reasons and it is a good approximation to restrict the plane wave expansion to the $(xz)$ plane for a weakly diverging cavity mode. Physically, this approximation is equivalent to neglect the beam divergence in the $y$ direction, so this calculation will be reliable in the Rayleigh volume of the cavity mode which is the position where the experiments of ref.~\cite{Fogliano2021} are conducted (field curvature small over the entire cavity length). Under these considerations, the fundamental Gaussian modes propagating along $\pm \textbf{e}_z$ can be approximated (see Appendix~\ref{Appendix:Gaussian_beam:Vectorial_Gaussian_beam_expansion_force_calculation}) by
%
%\begin{eqnarray}
%\textbf{E}_{0,0}^{(\pm, p)}(\textbf{r}) \approx f(y) \sum_{j \, / \, \abs{\abs{\bm{\kappa}_j}} = k} \mathcal{E}_0^j \ \e{i \bm{\kappa}_j^{(\pm)} \cdot \bm{\rho}} \ \textbf{e}_{P, j}^{(\pm, p)},
%\label{Eq:2D_carac:vectorial_gaussian_beam_2D_discretization}
%\end{eqnarray}
$\textbf{E}_{0,0}^{(\pm, p)}(\textbf{r}) \approx f(y) \sum_{j \, / \, \abs{\abs{\bm{\kappa}_j}} = k} \mathcal{E}_0^j \ \e{i \bm{\kappa}_j^{(\pm)} \cdot \bm{\rho}} \ \textbf{e}_{P, j}^{(\pm, p)}$
where $f(y) = \left( \frac{2}{\pi w(z)^2} \right)^{1/4} \, \e{-y^2/w(z)^2}$ has been normalized such as $\int_{-\infty}^{\infty} \mathrm{d} y \abs{f(y)}^2 = 1$ ($p$ characterizes the polarization of the field). In this expression, $\bm{\rho} = x \, \textbf{e}_x + z \, \textbf{e}_z$ is the 2D position vector in the $(xz)$ plane, and $\mathcal{E}_0^j$, $\bm{\kappa}_j^{(\pm)}$ and $\textbf{e}_{P, j}^{(\pm, p)}$ are the amplitude, wave vector and polarization vector of each plane wave contribution (given in Appendix~\ref{Appendix:Gaussian_beam:Vectorial_Gaussian_beam_expansion_force_calculation}). The total incident field is then expressed as a sum of plane waves,% $\textbf{E}_\mathrm{inc}^{(p)}(\textbf{r}) = \sqrt{P_\mathrm{inc} / \varepsilon_0 c} \, f(y) \, \sum_{j} E_0^{j (p)} \, \e{i \bm{\kappa_j \cdot \bm{\rho}}} \, \textbf{e}_P^{j (p)}$
\begin{eqnarray}
&& \textbf{E}_\mathrm{inc}^{(p)}(\textbf{r}) = \sqrt{ \frac{P_\mathrm{inc}}{\varepsilon_0 c}} \, f(y) \, \Big[ A^{(+, p)} \, \sum_{j} \mathcal{E}_0^j \, \e{i \bm{\kappa_j^{(+)} \cdot \bm{\rho}}} \, \textbf{e}_{P, j}^{(+, p)} \nonumber \\
&& \hspace*{2.8cm} + B^{(-, p)} \, \sum_{j} \mathcal{E}_0^j \, \e{i \bm{\kappa_j^{(-)} \cdot \bm{\rho}}} \, \textbf{e}_{P, j}^{(-, p)} \Big],
\end{eqnarray}
with $\abs{\kappa_j} = 2 \pi / \lambda$, which is the required form for optical force calculation.% and $E_0^{j (p)} = \mathcal{E}_0^j \, A^{(+, p)}$ or $E_0^{j (p)} = \mathcal{E}_0^j \, B^{(-, p)}$ depending of the propagation direction of the $j^\mathrm{th}$ plane wave along $z$.

%Each incident plane wave $\textbf{E}_\mathrm{inc}^{j (p)} = E_0^{j (p)} \, \e{i \bm{\kappa_j \cdot \bm{\rho}}} \, \textbf{e}_P^{j (p)}$
Each incident plane wave of this expansion $\textbf{E}_\mathrm{inc}^{j (p)}$ generates a scattered field $\textbf{E}_\mathrm{scat}^{j (p)}$ which can be calculated using the Mie formalism presented in Appendix~\ref{Appendix:Optical_prop_SiC_NW:Mie_formalism_infinite_cylinder}, leading to a total field of the form $\textbf{E}^{(p)}(\textbf{r}) = \sqrt{P_\mathrm{inc} / \varepsilon_0 c} \, f(y) \sum_j \textbf{E}_j^{(p)}(\bm{\rho})$ with $\textbf{E}_j^{(p)} = \textbf{E}_\mathrm{inc}^{j (p)} + \textbf{E}_\mathrm{scat}^{j (p)}$. Injecting it in the expression of the optical force given above, we get
\begin{subequations}
\begin{align}
& \textbf{F}^{(p)} = \sum_{j_1, j_2} \textbf{F}_{j_1 \, j_2}^{(p)}, \\
& \textbf{F}_{j_1 \, j_2}^{(p)} {=} \frac{P_\mathrm{inc} R_\mathrm{nw}}{2 c} \int_{0}^{2 \pi} \mathrm{d} \varphi \bigg[ \mathrm{Re} \left[ (\textbf{E}_{j_1}^{(p)} \cdot \textbf{n}) \left. \textbf{E}_{j_2}^{(p)} \right. ^\ast \right]  \nonumber \\
& \hspace*{4.5cm} + c^2 \mathrm{Re} \left[(\textbf{B}_{j_1}^{(p)} \cdot \textbf{n}) \left. \textbf{B}_{j_2}^{(p)} \right.^\ast \right] \nonumber \\
& \hspace*{1.7cm} - \frac{1}{2} \left( (\textbf{E}_{j_1}^{(p)} \cdot \left. \textbf{E}_{j_2}^{(p)} \right.^\ast) \textbf{n} + c^2 (\textbf{B}_{j_1}^{(p)} \cdot \left. \textbf{B}_{j_2}^{(p)} \right.^\ast) \textbf{n} \right) \bigg], \nonumber \\
\end{align}
\end{subequations}
$\textbf{F}_{j_1 \, j_2}$ being the force contribution due to the incidence of the two plane waves $j_1$ and $j_2$, which can propagate either along similar or opposite directions. Following the methodology proposed by Grzegorczyk and Kong~\cite{grzegorczyk2007analyticalTM, grzegorczyk2007analyticalTE} we derive its expression for two incident plane waves with different complex amplitudes, leading to $\textbf{F}_{j_1 \, j_2}^{(p)} = \mathrm{Im}\left[ \tilde{F}_{j_1 \, j_2}^{(p)} \right] \, \textbf{e}_x + \mathrm{Re}\left[ \tilde{F}_{j_1 \, j_2}^{(p)} \right] \, \textbf{e}_z$ with
\begin{eqnarray}
&& \tilde{F}_{j_1 \, j_2}^{(p)} = P_\mathrm{inc} \, K \,\abs{ E_0^{j_1 (p)}} \, \abs{E_0^{j_2 (p)}} \, \mathcal{P}_{j_1, j_2}^{(p)} \, \e{- i \Phi_{j_1 j_2}} \nonumber \\
&& \hspace*{1.7cm} \times \sum_{l=0}^\infty \Lambda_l^{(p)} \, \mathrm{Im}\left[ D_l^{(p) \, \ast} D_{l+1}^{(p)} \, \e{-i (l + 1/2)(\phi_{j_1} - \phi_{j_2})} \right], \nonumber \\
\label{Eq:2D_carac:F_j1_j2}
\end{eqnarray}
where $K = 4(n^2 - 1)/\pi c R_\mathrm{nw}$ depends on the nanowire properties. The amplitudes $E_0^{j (p)} = \mathcal{E}_0^j \, A^{(+, p)}$ and $E_0^{j (p)} = \mathcal{E}_0^j \, B^{(-, p)}$ depend on the plane wave expansion of the Gaussian beam through $\mathcal{E}_0^j$ and on the intra-cavity amplitudes field through $A^{(+, p)}$ and $B^{(-, p)}$. The coefficient $\mathcal{P}_{j_1, j_2}^{(p)} = \pm 1$ is a factor depending on the polarization and on the incidence directions of the two plane waves (see Appendix~\ref{Appendix:Optical_force} for details). The angle $\Phi_{j_1 j_2}$ characterizes the spatial orientation of each $\textbf{F}_{j_1 \, j_2}$ force term, it is given by
\begin{eqnarray}
\Phi_{j_1 j_2} = \Psi^{(j_1)} - \Psi^{(j_2)} + (\bm{\kappa}_{j_1} - \bm{\kappa}_{j_2}) \cdot \bm{\rho}_0 + \frac{\phi_{j_1} + \phi_{j_2}}{2}, \nonumber \\
\end{eqnarray}
where $\phi_j$ gives the orientation of the wave vector $\bm{\kappa}_j$ in the $(xz)$ plane (see Figure~\ref{Fig:Intro:NW_and_cavity_intro}), $\Psi^{(j)}$ is the phase of the complex amplitude $E_0^{j (p)}$, and $\bm{\rho}_0$ is the nanowire position in the $(xz)$ plane. In this expression, the propagation of the field in the two sub-cavities is taken into account through $(\Psi^{(j_1)} + \bm{\kappa}_{j_1} \cdot \bm{\rho}_0) - ( \Psi^{(j_2)} + + \bm{\kappa}_{j_2} \cdot \bm{\rho}_0)$. In Eq.~\eqref{Eq:2D_carac:F_j1_j2}, $\Lambda_l^{(p)}$ and $D_l^{(p)}$ involve Bessel and Hankle functions of first kind and only depend on the nanowire diameter and refractive index and on the operating wavelength (see Appendix~\ref{Appendix:Optical_force} for details).

Fig.~\ref{Fig:NW_in_cavity:XZ_map_65nm}(ef) shows the optomechanical force field, computed along both transversal ($x$) and longitudinal ($z$) directions. Here we considered a nanowire radius $R_\mathrm{nw} = 65 \, \mathrm{nm}$, a perpendicular polarization of the injected and intra-cavity light fields, and an incoming optical power of $1 \mu \mathrm{W}$. Each incident Gaussian field is expanded onto 9 plane wave components while the  Mie expansion is limited to the $5^{\mathrm{th}}$ order term, leading to a calculation time of the order of 10 minutes on a regular computer for Fig.~\ref{Fig:NW_in_cavity:XZ_map_65nm} (additional details on the simulation procedure and parameters can be found in Appendix~\ref{Appendix:numerical_considerations}). Along the cavity axis, we observe the typical $\lambda/2$ periodicity due to the intra-cavity field structure with a repulsive (attractive) character close to the nodes (antinodes). A cut of the force field along the $z$ cavity axis (dashed line in Fig.~\ref{Fig:NW_in_cavity:XZ_map_65nm}(e)) is shown in Fig.~\ref{Fig:NW_in_cavity:LZ_map_fig}(f) and corresponds to the force at resonance for the LZ map shown in Fig.~\ref{Fig:NW_in_cavity:LZ_map_fig}(d). We find a very good qualitative and quantitative agreement with the experimental work of ref.~\cite{Fogliano2021}, the optical force along the cavity axis being of the order of the fN for an input power of $1 \, \mu \mathrm{W}$. On an ascending (descending) branch of the LZ map, the transfer matrix formalism shows that the intra-cavity field is mainly localized in the associated right (left) sub-cavity, leading to a negative (positive) optical force. The slight asymmetry between the maximum and minimum values of $F_z$ is here due to the fact that the cavity is optically pumped from the left side, leading to a better coupling to the resonance of the left sub-cavity. We note that this effect is not the only reason for the experimental asymmetry found in~\cite{Fogliano2021} which is also due to a slight asymmetry in the reflectivities of the cavity mirrors ($R_L \neq R_R$), as demonstrated in complementary simulations based on the same formalism. Concerning the force $F_x$ along the transverse direction (one order of magnitude smaller than $F_z$), we observe attractive forces toward the antinodes and smaller lateral repulsive forces from the nodes. The lateral optomechanical force field  thus presents a significant shear character with a curl reaching $1 \, \mathrm{nN.m}^{-1}$ in the situation of Figure~\ref{Fig:NW_in_cavity:XZ_map_65nm} (at the location indicated by $\star$ symbol). Note that shear force fields are known to break the nanowire eigenmodes orthogonality, leading to the violation of the fluctuation dissipation relation~\cite{de2018eigenmode} and generating a topological instability~\cite{gloppe2014bidimensional}. We believe that the prolongation of those studies in such a cavity nano-optomechanical configuration, and at very low photon numbers, is certainly of great interest.
%Due to the large versatility of the nanowire as mechanical resonator~\cite{de2018eigenmode}, it opens the possibility to study non reciprocal forces in NIM configuration.
Additional results in the case of a smaller nanowire ($R_\mathrm{nw} = 10 \, \mathrm{nm}$) and parallel polarization of the light can be found in Appendix~\ref{Appendix:Simu_Rnw_10nm}, showing attractive forces toward intensity maxima in both directions, as expected from a dipole-like approximation.
Another important aspect is the sawtooth-like profile of $F_z$ which is due to the spatial dependence of the dissipative coupling strength. At an antinode, the intra-cavity field is strongly affected by the presence of the nanowire (see~\ref{sec:2D_carac:2D_carac_intra_cavity_field}), leading to a strong decrease of the intra-cavity field and then of the optical force. Complementary to Section~\ref{sec:optomecha_coupling_TM00:strong_coupling_and_cooperativity}, it highlights the essential balance between the dispersive and dissipative coupling required to reach new regimes where a small intra-cavity photon number would have a significant impact on the nanowire. This observation also underlines that the simple knowledge of $g_0$ is not sufficient to characterize the optomechanical force field experienced by the nanowire (which would have otherwise followed a sinusoidal profile).

\begin{figure}[t!]
\begin{center}
\includegraphics[width=0.99\linewidth]{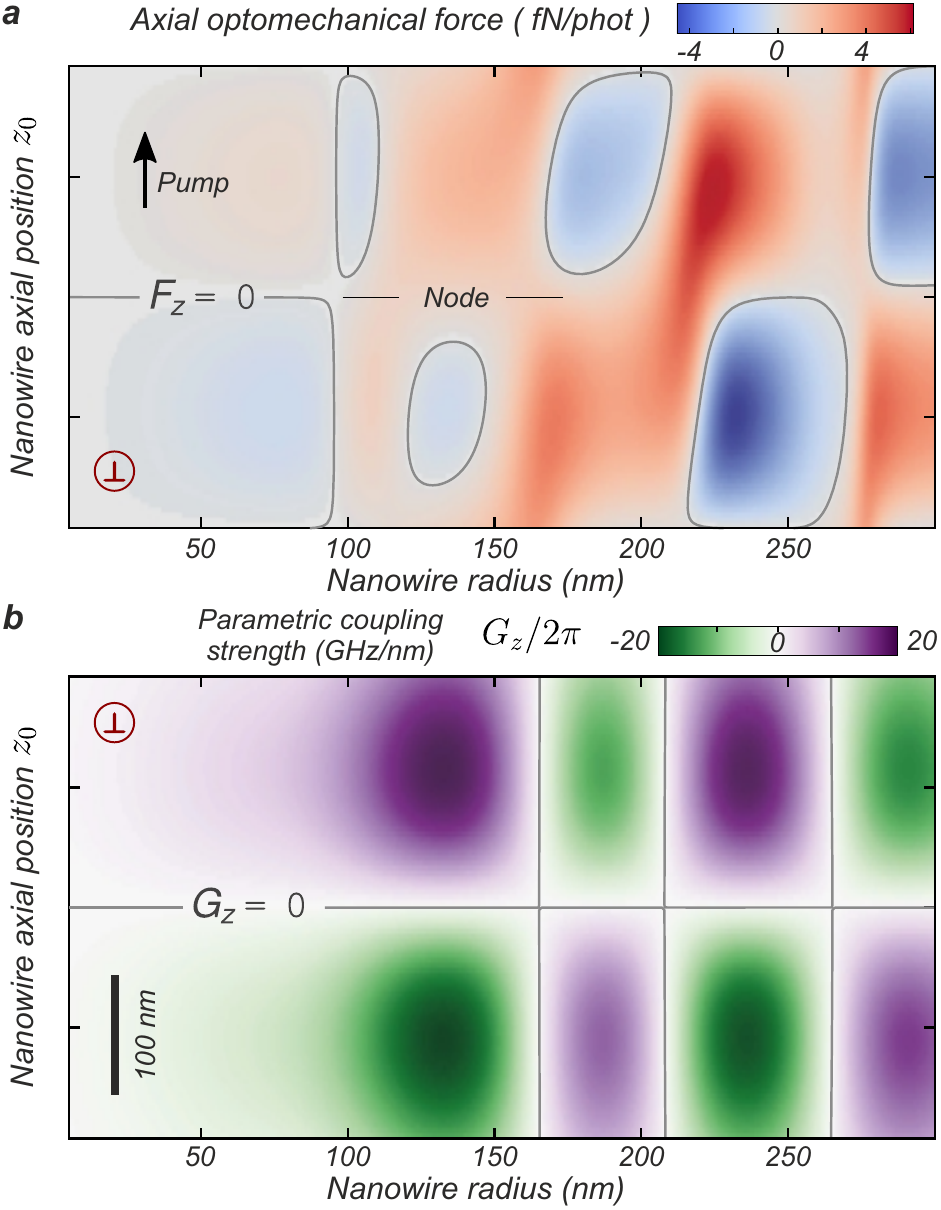}
\caption{Position and nanowire radius dependence of the optical force per intra-cavity photon (a) (perpendicular light polarization, $1 \, \mu \mathrm{W}$ input power) and of the parametric coupling strength along $z$ (b). These simulations have been done for a nanowire on the cavity axis ($x_0 = 0$). The cavity central node is located at $z=0$ and the pump direction propagates along $+ \textbf{e}_z$. The grey lines indicate positions where $F_z = 0$ and $G_z = 0$ illustrating that the knowledge of the coupling strength $G_z$ is not sufficient in NIM system to infer the optomechanical force exerted on the resonator. Results for a parallel light polarization can be found in Appendix~\ref{Appendix:Simu_Rnw_10nm}.}
\label{Fig:2D_charac:Fz_Gz_Rnw_scan}
\end{center}
\end{figure}

Additionally, it is worth mentioning the major impact of the nanowire radius on the structure of the optical force, as expected from Mie resonances. This opens the door to a large variety of configurations and a new phenomenology which cannot be observed in other 1D-like optomechanical systems such as membrane in the middle experiments. Fig.~\ref{Fig:2D_charac:Fz_Gz_Rnw_scan}(a) presents the $z$-position ($x_0 = 0$) and nanowire radius dependence of the optical force per intra-cavity photon number for a perpendicular polarization of the light. It presents a maximum for a nanowire radius close to $230 \, \mathrm{nm}$, in agreement with the relatively large static single photon cooperativity observed for NW~4 ($R_\mathrm{nw} = 225 \, \mathrm{nm}$, $\mathcal{C}^{(1)} = 0.14$) in the perpendicular case (see Fig.~\ref{Fig:appendix:Optomechanical_coupling_perp_appendix} in Appendix~\ref{Appendix:single_photon_optomecha_perp_polar}).

Finally, we show in Fig.~\ref{Fig:2D_charac:Fz_Gz_Rnw_scan}(b) the radius and position dependence of the optomechanical coupling strength $G_z$, while the grey lines indicates locations where $G_z = 0$ (and $F_z = 0$ in (a)). The large discrepancies observed at sufficiently large diameters further underlines the importance of calculating the optical force independently from the optomechanical coupling strength for nanowire in the middle systems as already discussed above. Indeed, we clearly observe no agreement between the sign of $F_z$ and $G_z$, breaking down the naive Hamiltonian approach where $F_z = -\hbar G_z N_\mathrm{cav}$, $N_\mathrm{cav}$ being the intra-cavity photon number. It means that for \textit{in the middle} configuration, the knowledge of $g_0$ is not sufficient to infer the optical force applied on the resonator and then to evaluate the impact of a single intra-cavity photon. This shows first that a proper Hamiltonian approach of such systems is nowadays still lacking, while highlighting at the same time the importance of performing direct force measurements to fully characterize both facets of the optomechanical coupling as previously stressed in ref.~\cite{Fogliano2021}.

\section{Conclusion and perspectives}
\label{sec:conclu_and_perpectives}

In this paper we have studied the optomechanical system made of a sub-wavelength-sized nanowire inserted in a small mode volume optical cavity. Using the Mie formalism and the fact that the set of Hermite-Gaussian mode forms a complete orthogonal basis of the two-dimensional propagating fields, we derived the reflection, transmission and scattering coefficients between different cavity modes, and how they depend on the nanowire position within the cavity. This lays the foundations for a complete study of the \textit{nanowire in the middle} system based on higher order and multiple optical modes, which is an essential step prior to investigate the quantum light field fluctuations in such a complex system, which lies beyond the scope of this paper.

In a second time we restricted our study to the situation where only the fundamental Gaussian cavity mode is addressed, the scattering towards others mode being considered as a loss channel. We examined how the presence of the nanowire shifts the cavity resonance, which allows to evaluate the vectorial optomechanical coupling strength $\textbf{g}_0$, demonstrating for this system a ratio $\delta z^{(1)} / \delta z^{\mathrm{zpf}} = 2 g_0^z/\Omega_\mathrm{m}$ of the order of $10^4$ for existing nanowires. This highlights the possibility to achieve the adiabatic single intra-cavity photon regime of cavity optomechanics, which was shown to become within experimental reach at modest cryogenic temperature ($4 \, \mathrm{K}$) despite the residual thermal position fluctuation of the nanowire, then motivating the development of such an experimental configuration~\cite{Fogliano2021a}. This analysis has also surprisingly revealed the existence of positive cavity shifts, a situation where the cavity length has to be increased in order to match the optical resonance condition. Preliminary work shows it is a consequence of the geometry of the system where a one dimensional scatterer is inserted inside the cavity. This could lead to original Casimir forces that should be observable making use of the very high force sensitivity of the nanowire~\cite{Fogliano2021, Fogliano2021a}.

Next, we have demonstrated the possibility to observe a static bistability close to the single photon scale, showing that it can be reached for only 7 intra-cavity photons in the optimal configuration. We suggest possibilities to reduce this number, for example by ingenering the optical mode geometry in order to minimize the induced nanowire losses which are responsible of a decrease of the optomechanical back action. Another promissing approach consists in functionalizing the nanowires at their extremity with a sub-micron sphere to maximize the optical interaction while minimizing the losses. Moreover, in this configuration which largely operates in the adiabatic regime where the cavity field instantaneously reacts on mechanical time-scales to a displacement of the nanowire ($\Omega_\mathrm{m} \ll \kappa_\mathrm{cav}$), the large cooperativity achieved leads to interesting applications in quantum optics. Indeed, it opens the road to a broadband squeezing of the outgoing cavity field in the cavity bandwidth~\cite{Fabre1994}, and then to the possibility to generate non Gaussian state of the light~\cite{Reynaud1989} for small photon numbers.

Finally, we investigated the two-dimensional specificities of the \textit{nanowire in the middle} configuration. We first demonstrated the possibility to obtain a direct picture of the intra-cavity field by locking the cavity length on the transmission signal while scanning the nanowire in the cavity mode volume. Secondly, we gave an analytic expression of the vectorial optomechanical force applied by the intra-cavity field on the nanowire, which was verified to lead to a qualitative and quantitative agreement with the experimental work of ref.~\cite{Fogliano2021}. It also highlights the limitations of traditional 1D like parametric analysis to describe the reverse optomechanical interaction for \textit{in the middle} systems, due to the dependence of the nanowire-dressed optical modes spatial profiles and pumping efficiency with the scatterer position. Contrary to historical 1D-like optomechanical systems where one of the cavity mirror oscillates around its equilibrium position, the optical force can not be obtained from the single knowledge of the parametric coupling strength but requires a proper experimental investigation as performed in ref.~\cite{Fogliano2021} or a specific calculation as done here.

\section{Acknowledgements}

We warmly thank G. Bachelier for the helpful discussions on the numerical approach of the Mie formalism and S. Reynaud for the interaction on the general context and applications of this work.

F.F. acknowledges funding from the LANEF (ANR-10-LABX-51-01). P.H. acknowledges funding from the European Union H2020 programme (Marie Sklodowska-Curie grant 754303). This project is supported by the French National Research Agency (SinPhoCOM project), and by the European Research Council under the EU's Horizon 2020 research and innovation programme, grant agreements No 671133 (EQUEMI project), 767579 (CARTOFF) and 820033 (AttoZepto).\\

\section*{Appendix}

\appendix

\section{Optical properties of SiC nanowire}
\label{Appendix:Optical_prop_SiC_NW}

In this Appendix we briefly present the Mie formalism used to described the scattering of an electromagnetic field on an infinite nanowire. Following the procedure developed by C. Bohren~\cite{Bohren1998}, we give the expressions of the scattered field due to the normal incidence of a single plane wave (incident wave vector orthogonal to the nanowire axis). In a second time, we define the one dimensional cross section of a nanowire characterizing the angular emission diagrams of the scattered field.

\subsection{Mie formalism for an infinite cylinder}
\label{Appendix:Optical_prop_SiC_NW:Mie_formalism_infinite_cylinder}

The nanowire is modeled as an infinite lossless dielectric (non magnetic) cylinder collinear to the $y$ axis and located at $\textbf{r}_0 = x_0 \, \textbf{e}_x + z_0 \, \textbf{e}_z$. The incident field onto the cylinder is assumed to be a single plane wave  with a wave vector $\textbf{k}_i = 2 \pi / \lambda \, \textbf{e}_i$ belonging to the $(xz)$ plane ($k_{i, y} = 0$), its direction being characterized by the incident angle $\phi_i$. The polarization of the electric field can be either parallel or orthogonal to the cylinder axis, leading to the so-called parallel and perpendicular cases. To describe this system we make use of the cylindrical coordinates $(r, \varphi, y)$ centred at the nanowire position $\textbf{r}_0$, as shown Fig.~\ref{Fig:NW_in_cavity:Mie_scattering}(a). It is important to note that in order to match the coordinates system used in previous experimental papers \cite{gloppe2014bidimensional, pigeau2015observation, de2017universal, de2018eigenmode}, $(\textbf{e}_r, \textbf{e}_\varphi, \textbf{e}_y)$ is an indirect orthonormal system, which leads to an additional minus sign in the expression of the rotational operator. This explains the differences between the results presented here and what can be found in the literature \cite{Bohren1998, grzegorczyk2007analyticalTE} when $(\textbf{e}_r, \textbf{e}_\varphi, \textbf{e}_y)$ is direct.

The Mie formalism consists in solving the vectorial Helmholtz equation to obtain the expressions of the scattered field as well as the field inside the cylinder. First one obtains the general form of the solutions in terms of the cylindrical vector harmonics. In a second time, expanding the incident plane wave on these vector harmonics and using the continuity equations of the electromagnetic field at the cylinder interface leads to the expressions of the incident, scattered and internal fields. In the case of a parallel polarization of the incident field along $- \, \textbf{e}_y$, the electric fields read as
\begin{subequations}
\label{Eq:Optical_prop_SiC_NW:E_field_para}
\begin{align}
\textbf{E}_\mathrm{i}^{\parallel}(\textbf{r}) &= E_0 \, \e{i \textbf{k}_i \cdot (\textbf{r}_0 + \textbf{r})} (- \textbf{e}_y) \nonumber \\
											  &= \sum_{l=-\infty}^{+\infty} \mathcal{E}_l \, \textbf{N}_l^{(1)}(\textbf{r}, k), \\
\textbf{E}_\mathrm{1}^{\parallel}(\textbf{r}) &= \sum_{l=-\infty}^{+\infty} \mathcal{E}_l \, f_l^{\parallel} \, \bm{N}_l^{(1)}(\textbf{r}, nk), \\
\textbf{E}_\mathrm{s}^{\parallel}(\textbf{r}) &= -\sum_{l=-\infty}^{+\infty} \mathcal{E}_l \, b_l^{\parallel} \, \textbf{N}_l^{(3)}(\textbf{r}, k), \label{Eq:Optical_prop_SiC_NW:E_field_para_scatrtered}
\end{align}
\end{subequations}
while in the case of a perpendicular polarization along $- \, \textbf{e}_{\phi_i}$, the electric fields is given by
\begin{subequations}
\label{Eq:Optical_prop_SiC_NW:E_field_perp}
\begin{align}
\textbf{E}_\mathrm{i}^{\perp}(\textbf{r}) &= E_0 \, \e{i \textbf{k}_i \cdot (\textbf{r}_0 + \textbf{r})} (- \textbf{e}_{\phi_i}) \nonumber \\
								 &= -i \sum_{l=-\infty}^{+\infty} \mathcal{E}_l \ \textbf{M}_l^{(1)}(\textbf{r}, k), \\
\textbf{E}_\mathrm{1}^{\perp}(\textbf{r}) &= -i \sum_{l=-\infty}^{+\infty} \mathcal{E}_l \ g_l^{\perp} \ \bm{M}_l^{(1)}(\textbf{r}, nk), \\
\textbf{E}_\mathrm{s}^{\perp}(\textbf{r}) &= i \sum_{l=-\infty}^{+\infty} \mathcal{E}_l \, a_l^{\perp} \ \textbf{M}_l^{(3)}(\textbf{r}, k), \label{Eq:Optical_prop_SiC_NW:E_field_perp_scatrtered}
\end{align}
\end{subequations}
where $\textbf{k}$ and $n \textbf{k}$ are the wave vectors in the background (the vacuum here) and in the cylinder respectively. The cylindrical vector harmonics are obtained using the solution of the scalar Helmholtz equation and $- \textbf{e}_y$ as the pilot vector,
\begin{subequations}
\label{Eq:Optical_prop_SiC_NW:vector_harmonics}
\begin{align}
& \textbf{M}_l^{(\eta)}(\textbf{r}, k) = k \left( il \frac{Z_l(k r)}{k r} \textbf{e}_r - Z'_l(k r) \textbf{e}_\varphi \right) \e{i l \varphi }, \\
& \textbf{N}_l^{(\eta)}(\textbf{r}, k) = -k Z_l(k r) \e{i l \varphi} \textbf{e}_y,
\end{align}
\end{subequations}
where the prime denotes a derivative with respect to the argument. In Eq.~\eqref{Eq:Optical_prop_SiC_NW:vector_harmonics}, $Z_l$ is the Bessel function of the first kind $J_l$ if $\eta = 1$ while it is the Hankel function of the first kind $H_l^{(1)}$ if $\eta = 3$. The $\mathcal{E}_l$ coefficient appearing in Eq.~\eqref{Eq:Optical_prop_SiC_NW:E_field_para} and~\eqref{Eq:Optical_prop_SiC_NW:E_field_perp} is obtained from the projection of the incident field onto the cylindrical vector harmonics given above  and reads as
\begin{eqnarray}
\mathcal{E}_l = \frac{E_0(\textbf{r}_0) \, (-i)^l \e{- i l \phi_i}}{k}, \label{Eq:Optical_prop_SiC_NW:El_ortho_incidence}
\end{eqnarray}
where $E_0(\textbf{r}_0)$ is the complex amplitude of the incident plane wave at the cylinder position. The other coefficients present in Eq.~\eqref{Eq:Optical_prop_SiC_NW:E_field_para} and~\eqref{Eq:Optical_prop_SiC_NW:E_field_perp} are obtained from the continuity relations at the cylinder interface and read as
\begin{subequations}
\label{Eq:Optical_prop_SiC_NW:a_b_f_g_coef_ortho_incidence}
\begin{align}
& b_l^\parallel = \frac{J_l(n \rho_R)J'_l(\rho_R) - nJ'_l(n\rho_R)J_l(\rho_R)}{J_l(n\rho_R)H_l^{(1) '}(\rho_R) - nJ'_l(n\rho_R)H_l^{(1)}(\rho_R)}, \\
& f_l^\parallel = \frac{J_l(\rho_R) - \beta_l^\parallel H_l^{(1)}(\rho_R)}{n J_l(n \rho_R)}, \\
& a_l^\perp = \frac{nJ_l(n\rho_R)J'_l(\rho_R) - J'_l(n\rho_R)J_l(\rho_R)}{nJ_l(n\rho_R)H_l^{(1) '}(\rho_R) - J'_l(n\rho_R)H_l^{(1)}(\rho_R)}, \\
& g_l^\perp = \frac{J_l(\rho_R) - \alpha_l^\perp H_l^{(1)}(\rho_R)}{n^2 J_l(n \rho_R)},
\end{align}
\end{subequations}
where $\rho_R = k R_\mathrm{nw}$. Note that when using the continuity relations, we consider $\textbf{E}_1$ and $\textbf{E}_2 = \textbf{E}_\mathrm{i} + \textbf{E}_\mathrm{s}$ as the internal and external field respectively, meaning that the total field outside the nanowire is the sum of the incident and scattered field.

\subsection{Expression of the magnetic fields}
\label{Appendix:additional_result_mie_scattering:B_field_ortho_incidence}

The expressions of the magnetic field for both polarizations can be directly obtained from the Maxwell-Faraday equation and from the vector harmonics identities $\bm{\nabla} \times \textbf{N}_l = k \, \textbf{M}_l$ and $\bm{\nabla} \times \textbf{N}_l = k \, \textbf{M}_l$. For a parallel polarization of the incident field along $- \, \textbf{e}_y$, we have
\begin{subequations}
\begin{align}
& \textbf{B}_\mathrm{i}^{\parallel}(\textbf{r}) = - \frac{i}{c} \sum_{l=-\infty}^{+\infty} \mathcal{E}_l \, \textbf{M}_l^{(1)}(\textbf{r}, k), \\
& \textbf{B}_\mathrm{1}^{\parallel}(\textbf{r}) = - \frac{i}{c} n \sum_{l=-\infty}^{+\infty} \mathcal{E}_l \, f_l^{\parallel} \, \bm{M}_l^{(1)}(\textbf{r}, nk), \\
& \textbf{B}_\mathrm{s}^{\parallel}(\textbf{r}) = \frac{i}{c} \sum_{l=-\infty}^{+\infty} \mathcal{E}_l \, b_l^{\parallel} \, \textbf{M}_l^{(3)}(\textbf{r}, k),
\end{align}
\end{subequations}
while for a perpendicular polarization along $- \, \textbf{e}_{\phi_i}$, the magnetic fields are given by
\begin{subequations}
\begin{align}
& \textbf{B}_\mathrm{i}^{\perp}(\textbf{r}) = - \frac{1}{c} \sum_{l=-\infty}^{+\infty} \mathcal{E}_l \ \textbf{N}_l^{(1)}(\textbf{r}, k), \\
& \textbf{B}_\mathrm{1}^{\perp}(\textbf{r}) = - \frac{1}{c} n \sum_{l=-\infty}^{+\infty} \mathcal{E}_l \ g_l^{\perp} \bm{N}_l^{(1)}(\textbf{r}, nk), \\
& \textbf{B}_\mathrm{s}^{\perp}(\textbf{r}) = \frac{1}{c} \sum_{l=-\infty}^{+\infty} \mathcal{E}_l \, a_l^{\perp} \ \textbf{N}_l^{(3)}(\textbf{r}, k),
\end{align}
\end{subequations}
where the vector harmonics are given Eq.~\eqref{Eq:Optical_prop_SiC_NW:vector_harmonics} and the other coefficients Eq.~\eqref{Eq:Optical_prop_SiC_NW:El_ortho_incidence} and~\eqref{Eq:Optical_prop_SiC_NW:a_b_f_g_coef_ortho_incidence}.

\subsection{Emission diagrams of an infinite cylinder}
\label{Appendix:additional_result_mie_scattering:emission_diagram}

We now define the angular one dimensional cross section $\mathrm{d} \sigma_{\mathrm{1D}} / \mathrm{d} \varphi$ characterizing the angular dependence of the scattered light. It is given, at a distance $r$, by the ratio between the outgoing scattered power per unit of angle and length (along y), and the incident flux (the intensity of the incident Poynting vector),
\begin{eqnarray}
\frac{\mathrm{d} \sigma_{\mathrm{1D}}}{\mathrm{d} \varphi} = \frac{r \, \bf{\Pi}_\mathrm{s} \cdot \textbf{e}_r}{\abs{\bf{\Pi}_\mathrm{i}}},
\end{eqnarray}
where $\bm{\Pi}_\mathrm{s}$ and $\bm{\Pi}_\mathrm{i}$ are the scattered and incident Poynting vectors given by $\bm{\Pi} = \mathrm{Re}\left[ \textbf{E} \wedge \left. \textbf{B} \right.^\ast \right] / 2 \mu_0$. Using the expressions of the field given previously, we obtain for each polarizations
\begin{subequations}
\label{Eq:Optical_prop_SiC_NW:angular_cross_sections}
\begin{align}
& \frac{\mathrm{d} \sigma_{\mathrm{1D}}^\parallel}{\mathrm{d} \varphi} = r \, \mathrm{Re} \Bigg[ i \sum_{l, m} b_l^\parallel \left. b_m^\parallel \right.^\ast H_l^{(1)}(kr) \left. H_m^{(1) '}\right.^\ast(kr) \nonumber \\
& \hspace*{4cm} \times \e{i (l - m) (\varphi - \phi_i - \frac{\pi}{2})} \Bigg], \\
& \frac{\mathrm{d} \sigma_{\mathrm{1D}}^\perp}{\mathrm{d} \varphi} = r \, \mathrm{Re} \Bigg[ i \sum_{l, m} a_l^\perp \left. a_m^\perp \right.^\ast H_l^{(1)}(kr) \left. H_m^{(1) '}\right.^\ast(kr) \nonumber \\
& \hspace*{4cm} \times \e{i (l - m) (\varphi - \phi_i - \frac{\pi}{2})} \Bigg],
\end{align}
\end{subequations}
where the sum over $l$ and $m$ extend from $-\infty$ to $+\infty$.
Fig.~\ref{Fig:NW_in_cavity:Mie_scattering}(cd) in the main text shows the angular emission diagrams resulting from Eq.~\eqref{Eq:Optical_prop_SiC_NW:angular_cross_sections} for an excitation wavelength $\lambda = 770 \, \mathrm{nm}$, an incident angle $\phi_i = 0$ and for three different cylinder radius. Here, like in the rest of this paper, we only consider 11 terms in each sum over the cylindrical harmonics ($\{ l, m \} \in \{ -5, 5 \}^2$) which is sufficient considering the working wavelength and cylinder radius smaller than $250 \, \mathrm{nm}$.

The reflected, transmitted and scattered one dimensional cross-section defined in the main text are given by
\begin{subequations}
\label{Eq:Optical_prop_SiC_NW:R_T_Scat_cross_sections}
\begin{align}
& \sigma_\mathrm{R} = \int_{-\theta_\mathrm{col}/2}^{\theta_\mathrm{col}/2} \mathrm{d} \varphi \, \frac{\mathrm{d} \sigma_{\mathrm{1D}}}{\mathrm{d} \varphi}, \\
& \sigma_\mathrm{T} = \int_{\pi - \theta_\mathrm{col}/2}^{\pi + \theta_\mathrm{col}/2} \mathrm{d} \varphi \, \frac{\mathrm{d} \sigma_{\mathrm{1D}}}{\mathrm{d} \varphi}, \\
& \sigma_\mathrm{scat} = \int_{\theta_\mathrm{col}/2}^{\pi - \theta_\mathrm{col}/2} \mathrm{d} \varphi \, \frac{\mathrm{d} \sigma_{\mathrm{1D}}}{\mathrm{d} \varphi} + \int_{\pi + \theta_\mathrm{col}/2}^{-\theta_\mathrm{col}/2} \mathrm{d} \varphi \, \frac{\mathrm{d} \sigma_{\mathrm{1D}}}{\mathrm{d} \varphi}.
\end{align}
\end{subequations}
where $\theta_\mathrm{col}$ is the collection angle. Fig.~\ref{Fig:NW_in_cavity:Mie_scattering}(cd) in the main text shows these three quantities as function of the cylinder radius for both polarizations of the incident plane wave and for numerical apertures $\mathrm{NA} = 0.15$ ($\theta_\mathrm{col} \approx 17\char23$) and $\mathrm{NA} = 0.7$ ($\theta_\mathrm{col} \approx 90\char23$).

\section{Numerical considerations}
\label{Appendix:numerical_considerations}

The work presented in this paper is based on the numerical evaluation (using Python 3 and usual packages) of the outgoing (transmission, reflection and scatter) and intra-cavity fields (used to compute the optical force). We present here the idea of the procedure used to perform the simulations and give additional information about the numerical parameters used.

\paragraph{Unperturbed intra-cavity field}
The intra cavity field is assume to be in the fundamental Gaussian Hermite mode given Eq.~\eqref{Eq:Annex:TM00} and we focus on the 32$^{\mathrm{th}}$ mode corresponding to an unperturbed cavity length $L_\mathrm{cav}^0 \approx 12.4 \, \mu \mathrm{m}$ with a node of the field at the cavity center ($z = 0$). The cavity length (always close to $L_\mathrm{cav}^0$), the curvature radius of the mirrors $R_c = 28 \, \mu \mathrm{m}$ and the optical wavelength $\lambda = 770 \, \mathrm{nm}$, fully constrains the mode profile leading to a beam waist $w_0 = 1.7 \, \mu \mathrm{m}$ at $z=0$ and a Rayleigh length $z_R = 11.5 \mu \mathrm{m}$.

\paragraph{Reflection and transmission coefficient of a nanowire in a Gaussian beam}
The calculation of the reflection and transmission coefficients given Eq.~\eqref{sec:NW_in_cavity:Cr_Ct_NW_fundamental_mode} allows to obtain the transfer matrix of the nanowire used to propagate the intra-cavity field through the whole system.

In the approximated method, we assume that the nanowire stays in a spatial area where the Gaussian beam does not diverge (in the waist of the beam). We then consider that the incident field is equivalent to a single plane wave incidence with a wave vector $\textbf{k}_i = k \, \textbf{e}_z$, an amplitude $E_{0,0}^{(+)}(\textbf{r}_0)$ at the nanowire position obtained from Eq.~\eqref{Eq:NW_in_cavity:General_Gaussian_beam_definition} (with $A_0 = \sqrt{2/\pi w_0^2}$), and a polarization vector $\textbf{e}_P$ which can be parallel ($\textbf{e}_P^\parallel = -\textbf{e}_y$) or perpendicular ($\textbf{e}_P^\perp = \textbf{e}_x$) to the nanowire axis. It generates a scattered field calculated on the cavity mirrors surface (located at $\pm L_\mathrm{cav}/2$ with transverse size $D$) using the Mie formalism presented in Appendix~\ref{Appendix:Optical_prop_SiC_NW}. The overlaps between the scattered field and the Gaussian cavity modes on the cavity mirrors surface, i. e. the surface integrals appearing Eq.~\eqref{sec:NW_in_cavity:Cr_Ct_NW_fundamental_mode}, are calculated numerically using the spherical coordinates associated to each cavity mirrors. For a function $f$ defined on a spherical surface $\mathcal{S}$ of curvature radius $R_c$, we have
\begin{eqnarray}
& \iint_{\mathcal{S}} \mathrm{d}^2 \textbf{r} \, f(\textbf{r}) = R_c^2 \int_{\Phi_i}^{\Phi_f} \mathrm{d} \Phi \int_{\theta_i}^{\theta_f} \mathrm{d} \theta \ f\left[ \textbf{r}(\Phi, \theta) \right],
\end{eqnarray}
where in our case, the azimuthal angle $\Phi$ varies between $\Phi_i = 0$ and $\Phi_f = 2\pi$. The polar angle range depends on the considered integration surfaces and is characterizes by the curvature radius of the mirrors and by their transverse size. For the right integration surface $\mathcal{S}_R$, it varies between $\theta_{i, R} = 0$ and $\theta_{f, R} = \arccos \left( \sqrt{1 - D^2/4 R_c^2} \right)$ while for the left integration surface $\mathcal{S}_L$ we have $\theta_{i, L} = \pi - \theta_{f, R}$ and $\theta_{f, L} = \pi$. Thus, to perform the numerical integration, we only need to express the Cartesian coordinates $(x, y, z)$ and cylindrical coordinates $(r, \varphi, y)$ in terms of the spherical angles $(\Phi, \theta)$. The mesh is chosen sufficiently small to resolve the phase oscillation of the fields overlap. We then construct a database of the reflection and transmission coefficients for different position of the nanowire in the beam and for different nanowire radius, using the experimental parameters used in ref~\cite{Fogliano2021} ($L_\mathrm{cav} = 12 \, \mu \mathrm{m}$, $R_c = 28 \, \mu \mathrm{m}$, $D = 12 \, \mu \mathrm{m}$, $\lambda = 770 \, \mathrm{nm}$ and $n = 2.61$). Moreover, since it is essential to consider a finite number of terms in the cylindrical expansion of the scattered field for a numerical evaluation, we restrict the sum appearing in Eq.~\eqref{sec:NW_in_cavity:scattered_fields_TM00} to the 11 "first" terms, $l \in \{ -5, 5 \}$, which is sufficient when considering nanowires of radius smaller than $300 \, \mathrm{nm}$ for an excitation wavelength at $770 \, \mathrm{nm}$.

The second and exact way to compute these coefficients consists in expanding the Gaussian beam onto the plane wave spectrum (see Appendix~\ref{Appendix:Gaussian_beam}), then evaluating the total scattered field as the sum of all the scattered fields due to the different incidences. For that purpose it is essential to be able to compute the scattered field due to a glancing incidence as detailed in~\cite{Bohren1998}, paying attention to the axis choice performed here. The discretization of the Gaussian beam is done following the procedure given in ~\ref{Appendix:Gaussian_beam:Vectorial_Gaussian_beam_expansion}. In practice it is sufficient to keep only 9 terms in the expansion along each directions (with a discretization step $\Delta_{x/y} = 0.75 \, \mu \mathrm{m}^{-1}$) leading to a total of 81 terms. As a consequence, this method require much longer calculation time, and was simply used to ensure the validity of the above approximation.

\paragraph{Reflected, transmitted and scattered fields}
For an SiC nanowire located at $\textbf{r}_0 = x_0 \, \textbf{e}_x + z_0 \, \textbf{e}_z$ in an optical cavity of length $L_\mathrm{cav}$, the cavity reflection and transmission coefficients are obtained using the transfer matrix formalism by solving Eq.~\eqref{Eq:optomecha_coupling:general_propagation_equations}. We chose symmetric cavity mirrors described as lossless beam splitters with intensity reflection coefficients $R_L = R_R = 0.994$. The amount of light scattered out of the cavity is then evaluated using the energy conservation.

\paragraph{LZ maps}
The so called LZ maps are obtained scanning the nanowire position along the cavity axis while scanning at the same time the cavity length around the resonance. The cavity finesse, which depends on the nanowire position along $z$ due to the spatial dependence of the dissipative coupling, is evaluated by fitting the cavity transmission with a Lorentzian profile. As long as the nanowire induced cavity shift remains small compare the free spectral range of the cavity ($L_\mathrm{FSR} = \lambda/2 = 385 \, \mathrm{nm}$ equivalent to $\omega_\mathrm{FSR} \approx 12 \, \mathrm{THz}$), we observe no deviation from the Lorentzian profile.

\paragraph{Locked XZ map}
For a given nanowire position, the cavity lock on the transmission signal is done by scanning the cavity length around the optical resonance, then fitting the transmission with a Lorentzian profile to extract the resonant length and finesse. The cavity length is adjusted for every nanowire position producing the XZ locked maps shown Figure~\ref{Fig:NW_in_cavity:XZ_map_65nm} and ~\ref{Fig:appendix:XZ_map_10nm}.

\paragraph{Optical force evaluation}

The optical force is calculated from the intra-cavity field using the Maxwell stress tensor\cite{jackson1999classical_chap6, grzegorczyk2007analyticalTE, grzegorczyk2007analyticalTM}. The amplitudes of the intra-cavity fields propagating along $\pm z$ are first obtain using the transfer matrix formalism Eq.~\eqref{Eq:2D_carac:intra_cavity_amplitudes}. We then consider two Gaussian beams with the amplitudes obtained previously incoming on the nanowire. The total scattered field on the nanowire surface is evaluated by expanding the Gaussian beams on the plane wave spectrum and summing the scattered fields due to each contributions. The optical force is finally obtained using the results of~\ref{sec:2D_carac:optical_force}. The Gaussian beam discretization is done following the procedure described in~\ref{Appendix:Gaussian_beam:Vectorial_Gaussian_beam_expansion_force_calculation} where we keep only 9 terms in the expansion. The incoming optical power on the cavity $P_\mathrm{inc}$ appearing in Eq.~\eqref{Eq:2D_carac:F_j1_j2} depends on the total injected power $P_\mathrm{inc}^0 = 1 \mu \mathrm{W}$, on the coupling coefficient between the laser and the optical fiber ($\eta_\mathrm{fiber} = 0.8$) and on the coupling coefficient between the optical mode in the fiber and the cavity mode ($T_\mathrm{in} = 0.5$), $P_\mathrm{inc} = \eta_\mathrm{fiber} \, T_\mathrm{in} \, P_\mathrm{inc}^0$. Similarly to all the other parameters, $\eta_\mathrm{fiber}$ and $T_\mathrm{in}$ have been chosen to be close to the experimental parameters of ref.~\cite{Fogliano2021}.

\section{Plane wave expansion of the fundamental Hermite-Gaussian mode}
\label{Appendix:Gaussian_beam}

\subsection{Expansion of a scalar Gaussian beam on the plane wave spectrum in the paraxial approximation}
\label{Appendix:Gaussian_beam:Scalar_Gaussian_beam_expansion}

We consider the fundamental Hermite-Gaussian mode simply refereed in the following as the \textit{Gaussian beam} for simplicity. Its expression given Eq.~\eqref{Eq:NW_in_cavity:General_Gaussian_beam_definition} in the paraxial approximation becomes for the fundamental mode ($n_x = n_y = 0$) propagating along $+z$,
\begin{eqnarray}
\label{Eq:Annex:TM00}
E_{0,0}^{(+)}(\textbf{r}) = \sqrt{\frac{2}{\pi w_0^2}} \frac{w_0}{w(z)} \, \e{-\frac{\textbf{r}_\perp^2}{w^2(z)}} \, \e{i \left[ kz - \Psi(z) + k \frac{\textbf{r}_\perp}{2R(z)} \right]}, \nonumber \\
\end{eqnarray}
where $k = 2 \pi / \lambda$, $\textbf{r}_\perp = x \, \textbf{e}_x + y \, \textbf{e}_y$, $w_0$ is the waist of the beam and where the transverse spreading $w(z)$, the Gouy phase $\Psi(z)$ and the curvature radius of the beam are given in the main text (\ref{sec:NW_in_cavity:Hermite_Gaussian_cavity_modes}). The normalization factor $\mathcal{A}_0^{0,0} = \sqrt{2/\pi w_0^2}$ is obtained considering a planar integration surface in the scalar product since $w(\pm L_\mathrm{cav}/2) \ll R_c$.

In the paraxial approximation, valid for a weakly diverging beam, the plane wave expansion of the field $E_{0,0}^{(+)}(\textbf{r})$ given Eq.~\eqref{Eq:Annex:TM00} reads as
\begin{eqnarray}
\label{Eq:Annex:TM00_plane_wave_expansion}
E_{0,0}^{(+)}(\textbf{r}) = \iint_{\abs{\abs{\textbf{k}}} = k} \mathrm{d} k_x \mathrm{d} k_y \, \mathcal{E}_0(k_x, k_y) \, \e{i \textbf{k}(k_x, k_y) \cdot \textbf{r}}, \nonumber \\
\end{eqnarray}
where $\mathcal{E}_0(k_x, k_y) = \frac{w_0}{(2\pi)^{3/2}} \e{-\frac{w_0^2}{4} \left( k_x^2 + k_y^2 \right)}$ is the amplitude associated to the wave vector $\textbf{k}(k_x, k_y) = k_x \, \textbf{e}_x + k_y \, \textbf{e}_y + \sqrt{k^2 - (k_{x}^2 + k_{y}^2)} \, \textbf{e}_z$.

It is worth mentioning that a vectorial field of the form $\textbf{E}_{0,0}^{(+)} = E_{0,0}^{(+)} \, \textbf{e}_P$, where $\textbf{e}_P$ is the polarization vector standing in the $(xy)$ plane, still satisfies the vectorial Helmholtz equation but is not an exact solution of the Maxwell-Gauss equation (for example for $\textbf{e}_P = \textbf{e}_x$ or $\textbf{e}_P = -\textbf{e}_y$).

\subsection{Expansion of a vectorial Gaussian beam on the plane wave spectrum}
\label{Appendix:Gaussian_beam:Vectorial_Gaussian_beam_expansion}

\begin{figure}[h!]
\begin{center}
\includegraphics[width=0.6\linewidth]{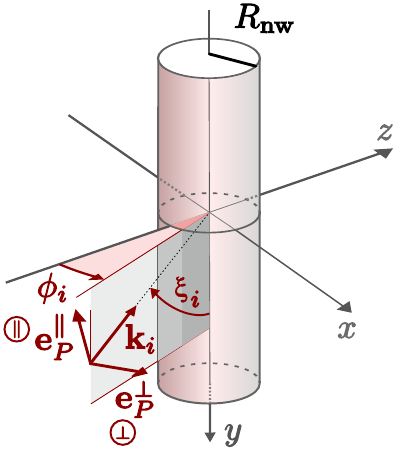}
\caption{Infinite cylinder under plane wave illumination. The incidence angles $\phi_i$ and $\xi_i$ defined the incident wavevector $\textbf{k}_i$. Considering additionally a parallel or perpendicular polarization, the polarization vector $\textbf{e}_P$ is also fully determined.}
\label{Fig:Annex:Mie_scattering_glancing_incidence}
\end{center}
\end{figure}

To introduce polarization aspects we follow the procedure developed by Igelsias and S\`aenz in ref.~\cite{iglesias2011scattering}. The idea is to consider an independent polarization vector for each plane waves in~\eqref{Eq:Annex:TM00_plane_wave_expansion} such as
\begin{eqnarray}
\label{Eq:Annex:TM00_plane_wave_expansion_polar}
\textbf{E}_{0,0}^{(+, p)}(\textbf{r}) &=& \iint_{\abs{\abs{\textbf{k}}} = k} \mathrm{d} k_x \mathrm{d} k_y \, \mathcal{E}_0(k_x, k_y) \, \e{i \textbf{k}(k_x, k_y) \cdot \textbf{r}} \nonumber \\
&& \hspace*{3cm} \times \, \textbf{e}_P^{(p)}(k_x, k_y),
\end{eqnarray}
where the polarization vector $\textbf{e}_P^{(p)}(k_x, k_y)$ is fully determined by the orientation of the wave vector $\textbf{k}(k_x, k_y)$ and by the polarization choice (parallel or perpendicular). Defining the incidence angles $\phi$ and $\xi$ as in Figure~\ref{Fig:Annex:Mie_scattering_glancing_incidence}, the incident wave vector and polarizations vectors read as
\begin{subequations}
\label{Eq:Annex:TM00_plane_wave_expansion_wavevector_polar_vector}
\begin{align}
& \textbf{k}(\phi, \xi) = k \left( -\sin \xi \sin \phi \, \textbf{e}_x + \sin \xi \cos \phi \, \textbf{e}_z - \cos \xi \, \textbf{e}_y \right), \\
& \textbf{e}_P^{\parallel}(\phi, \xi) = \cos \xi \sin \phi \, \textbf{e}_x - \cos \xi \cos \phi \, \textbf{e}_z - \sin \xi \, \textbf{e}_y, \\
& \textbf{e}_P^{\perp}(\phi, \xi) = \cos \phi \, \textbf{e}_x + \sin \phi \, \textbf{e}_z.
\end{align}
\end{subequations}
It is easy to show that each component of the vectorial field $\textbf{E}_{0,0}^{(+, p)}$ satisfy the scalar Helmholtz equation and then that the total field satisfies the vectorial Helmholtz equation $\bm{\nabla^2} \textbf{E}_{0,0}^{(+, p)} + k^2 \textbf{E}_{0,0}^{(+, p)} = 0$. Moreover, due to the relation between the wave vectors and the polarizations vectors, it also satisfies the Maxwell-Gauss equation, $\bm{\nabla} \cdot \textbf{E}_{0,0}^{(+, p)} = 0$.

The orthogonality and normalization of the two eigenmodes $\textbf{E}_{0,0}^{(+, \parallel)}$ and $\textbf{E}_{0,0}^{(+, \perp)}$ can be demonstrated similarly to Eq.~\eqref{Eq:NW_in_cavity:scalar_product_scalar_field} calculating
\begin{eqnarray}
\big \langle \textbf{E}_{0,0}^{(+, p)} \vert \textbf{E}_{0,0}^{(+, p')} \big \rangle &=& \iint_{\mathcal{S}} \mathrm{d}^2 \, \textbf{r} \textbf{E}_{0,0}^{(+, p)} (\textbf{r}) \cdot \left. \textbf{E}_{0,0}^{(+, p')} \right.^\ast (\textbf{r}) \nonumber \\
	&=& \delta_{p, p'}.
\end{eqnarray}
The last equality has been obtained by performing the change of variable $(k_x, k_y) \rightarrow (\phi, \xi)$ in~\eqref{Eq:Annex:TM00_plane_wave_expansion_polar} and considering the integration surface $\mathcal{S}$ to be be an infinite plane orthogonal to the $z$ axis.

Finally, the integral representation~\eqref{Eq:Annex:TM00_plane_wave_expansion_polar} can be discretized using the rectangle rule leading to
\begin{eqnarray}
\label{Eq:Annex:TM00_plane_wave_expansion_polar_discretized}
\textbf{E}_{0,0}^{(+, p)}(\textbf{r}) \approx \sum_{j_x, j_y \, / \, \abs{\abs{\textbf{k}_{j_x, j_y}}} = k} \mathcal{E}_0^{j_x, j_y} \, \e{i \textbf{k}_{j_x, j_y} \cdot \textbf{r}} \, \textbf{e}^{(p)}_{P, j_x, j_y},
\end{eqnarray}
where the wave vector of each plane wave contribution is given by $\textbf{k}_{j_x, j_y} = k_{x,j_x} \textbf{e}_x +  k_{y,j_y} \textbf{e}_y + \sqrt{ k^2 - (k_{x,j_x}^2 + k_{y,j_y}^2)} \textbf{e}_z$ with $k_{x,j_x} = j_x \Delta_{k_x}$ and $k_{y,j_y} = j_y \Delta_{k_y}$, $\Delta_{k_{x/y}}$ being the discretization step used in the rectangle rule along each axis while $j_{x/y}$ are integers. The plane wave amplitudes are $\mathcal{E}_0^{j_x, j_y} = (w_{0} / (2\pi)^{3/2}) \Delta_{k_x} \Delta_{k_y} \, \e{-\frac{w_0^2}{4} \left( k_{x, j_x}^2 + k_{y, j_y}^2 \right)}$ and the polarization vectors for each polarization are given Eq.~\eqref{Eq:Annex:TM00_plane_wave_expansion_wavevector_polar_vector} where the incidence angles $\phi^{(j_x, j_y)}$ and $\xi^{(j_x, j_y)}$  can be expressed in terms of the wave vector $\textbf{k}_{j_x, j_y}$ components. In practice, because of the weak divergence of the intra-cavity field, it is sufficient to consider 9 terms in each sum appearing in~\eqref{Eq:Annex:TM00_plane_wave_expansion_polar_discretized} with a discretization step $\Delta_{x/y} = 0.75 \, \mu \mathrm{m}^{-1}$ leading to a total of 81 terms.

\subsection{Vectorial Gaussian beam expansion for force calculation}
\label{Appendix:Gaussian_beam:Vectorial_Gaussian_beam_expansion_force_calculation}

We now realize the plane wave expansion of a fundamental vectorial Gaussian beam propagating along $\pm z$ suitable for optical force calculation in an NIM system. As already mentioned in the main text, because their is no optical force on the nanowire in the $y$ direction for symmetry reasons, it is correct to restrict the plane wave expansion to the $(xz)$ plane. Physically, this approximation is equivalent to neglect the contribution of the incident field polarized along $z$ due to the beam divergence in the $y$ direction. Mathematically, it means that the beam waist dimension along $y$ (noted $w_{0 y}$) can be chosen arbitrary large. In that case, the plane wave expansion of a scalar Gaussian beam in the paraxial approximation leads to
\begin{eqnarray}
E_{0,0}^{(\pm)}(\textbf{r}) \approx f(y) \ E_{(2D) \, 0,0}^{(\pm)}(x, z),
\end{eqnarray}
where $f(y) = (2 / \pi w_{0 y}^2)^{1/4} \, \e{-(y / w_{0 y})^2}$ has been defined such as $\int_{-\infty}^{\infty} \mathrm{d} y \abs{f(y)}^2 = 1$ and
\begin{eqnarray}
\label{Eq:Annex:Gaussian_beam_expansion_2D_0}
\hspace*{-0.5cm}
E_{(2D) \, 0,0}^{(\pm)}(x, z) = \frac{\sqrt{w_{0}}}{(2 \pi)^{3/4}} \int_0^{k} \mathrm{d} k_x \, \e{- \left( \frac{w_0 k_x}{2} \right)^2} \, \e{\pm i \left( k_x x + \sqrt{k^2 - k_x^2} z \right)}, \nonumber \\
\end{eqnarray}
with $k = 2 \pi / \lambda$.

The polarization of the beam is introduced following the same procedure as in Appendix~\ref{Appendix:Gaussian_beam:Vectorial_Gaussian_beam_expansion}. After discretization of the integral appearing in~\eqref{Eq:Annex:Gaussian_beam_expansion_2D_0} we get
\begin{eqnarray}
\label{Eq:Annex:Gaussian_beam_expansion_2D}
\textbf{E}_{0,0}^{(\pm)}(\textbf{r}) \approx f(y) \sum_{j \, / \, \abs{\abs{\bm{\kappa}_j}} = k} \mathcal{E}_0^j \ \e{i \bm{\kappa}_j^{(\pm)} \cdot \bm{\rho}} \ \textbf{e}_{P, j}^{(\pm, p)}.
\end{eqnarray}
where for each plane wave contribution the amplitude, wave vector and polarization vectors are given by
\begin{subequations}
\label{Eq:Annex:vectorial_beam_discretization_2D_amplitude_wavevector_polar_vectors}
\begin{align}
& \mathcal{E}_0^j = \frac{\sqrt{w_{0}}}{(2 \pi)^{3/4}} \Delta k_x \, \e{- \left( \frac{w_0 k_{x, j}}{2} \right)^2}, \\
& \bm{\kappa}_j^{(\pm)} = k_{x, j}^{(\pm)} \, \textbf{e}_x \pm \sqrt{k^2 - \left.k_{x, j}^{(\pm)}\right.^2} \, \textbf{e}_z, \\
& \textbf{e}_{P, j}^{(\pm, \parallel)} = -\textbf{e}_y, \label{Eq:Annex:vectorial_beam_discretization_2D_amplitude_wavevector_polar_vectors_polar_para} \\
& \textbf{e}_{P, j}^{(\pm, \perp)} = \pm \left( \cos \phi_j \, \textbf{e}_x + \sin \phi_j \, \textbf{e}_z \right). \label{Eq:Annex:vectorial_beam_discretization_2D_amplitude_wavevector_polar_vectors_polar_perp}
\end{align}
\end{subequations}
with $k_{x, j}^{(\pm)} = \pm j \Delta k_x$ ($j \in \mathbb{Z}$), $\Delta k_x$ the step used during the integral discretization and $\bm{\rho} = x \, \textbf{e}_x + z \, \textbf{e}_z$. The angle $\phi_j$ characterises the incidence angle of each plane wave and is given, depending of the propagation direction of the beam, by
\begin{subequations}
\begin{align}
& \phi_j^{(+)} = - \arctan\left( \frac{k_{x, j}^{(+)}}{\sqrt{k^2 - \left.k_{x, j}^{(+)}\right.^2}} \right) , \\
& \phi_j^{(-)} = \pi + \arctan\left( \frac{k_{x, j}^{(-)}}{\sqrt{k^2 - \left.k_{x, j}^{(-)}\right.^2}} \right).
\end{align}
\end{subequations}
In Eq.~\eqref{Eq:Annex:vectorial_beam_discretization_2D_amplitude_wavevector_polar_vectors_polar_perp}, the $\pm$ factor is due to the angle dependence of the polarization vector and accounts for the fact that in cavity, two beams propagating along opposite directions must have the same polarization. For instance, in the case $\phi^{(+)} = 0$ and $\phi^{(-)} = \pi$, we have $\textbf{e}_{P}^{(+, \perp)} = \textbf{e}_{P}^{(-, \perp)} = \textbf{e}_x$ which is valid. It is at the origin of the $\mathcal{P}_{j_1, j_2}^{(p)}$ factor appearing in the force expression Eq.~\eqref{Eq:2D_carac:F_j1_j2} and detailed in Appendix~\ref{Appendix:Optical_force}.

\section{Additional simulation results}
\label{Appendix:Simu_Rnw_10nm}

\subsection{Resonant cavity shifts for perpendicular polarization of the light}
\label{Appendix:cavity_shifts_perp_polar}

\begin{figure}[h!]
\begin{center}
\includegraphics[width=0.9\linewidth]{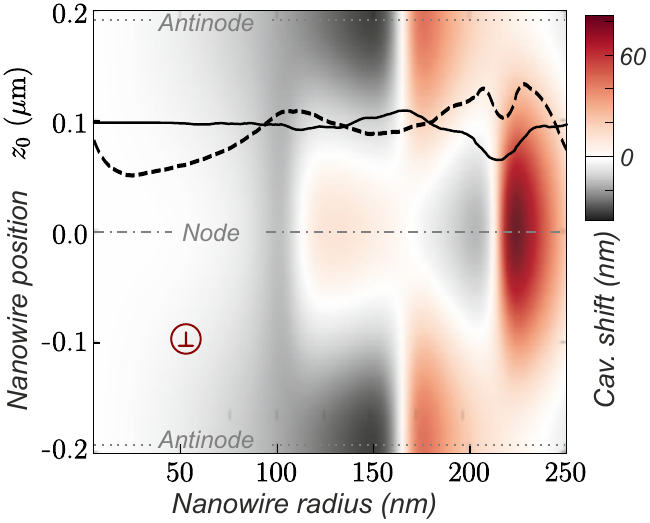}
\caption{Resonant cavity shift maps for different nanowire radius, positions on the cavity axis and perpendicular polarization of the light. The black solid and dashed lines indicate positions of maximum coupling strength $G_z$ and maximum ratio $G_z^2/\kappa$ characterizing the single photon static cooperativity, $\kappa$ being the linewidth of the cavity. Here we consider a cavity finesse without nanowire $\mathcal{F}_0 \approx 50000$ which is the highest value one can reach using the experimental configuration of ref.~\cite{Fogliano2021}.}
\label{Fig:appendix:cavity_shifts_perp_polar}
\end{center}
\end{figure}

We show Fig.~\ref{Fig:appendix:cavity_shifts_perp_polar} the resonant cavity shift as function of the nanowire radius and position along $z$ for a perpendicular polarization of the light. Similarly to the parallel case, we observe the existence of positive resonant cavity shifts.

\subsection{Single photon optomechanics for perpendicular polarization of the light}
\label{Appendix:single_photon_optomecha_perp_polar}

\begin{figure}[t!]
\begin{center}
\includegraphics[width=0.9\linewidth]{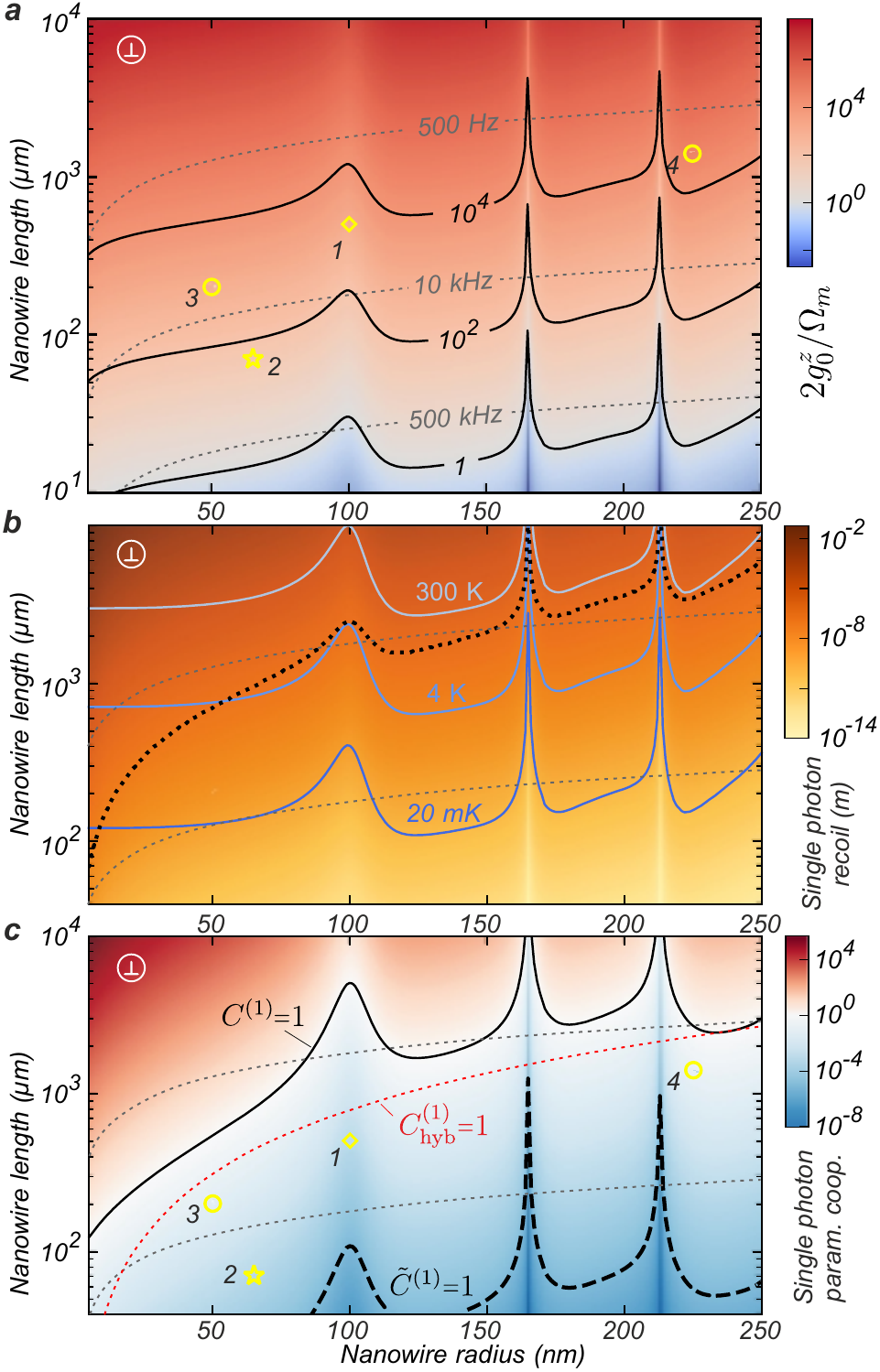}
\caption{Dependence on the nanowire dimensions of $\delta z ^{(1)} / \delta z^\mathrm{zpf} = 2 g_0^z / \Omega_\mathrm{m}$ (a), of the single photon deformation $\delta z^{(1)}$ (b), and of the static single photon parametric cooperativity $\mathcal{C}^{(1)} = 2 \left. g_0^z \right. ^2 / \kappa_\mathrm{cav} \Omega_m$ (c) for a perpendicular polarization of the light. The grey dashed lines indicate the fundamental vibrational mode iso-frequencies. In (b), the blue lines show the nanowire dimensions ensuring $\delta z^{(1)} =\Delta z^{(\mathrm{th})}$ for different bath temperatures while the black dotted line indicate nanowire dimensions for which $\delta z^{(1)} = R_\mathrm{nw}$. The diamond and star markers correspond to a relevant nanowire for cryogenic temperature experiment (NW 1: $R_\mathrm{nw} = 100 \, \mathrm{nm}$ and  $L_\mathrm{nw} = 500 \, \mu \mathrm{m}$) and to the nanowire used in ref.~\cite{Fogliano2021} (NW 2: $R_\mathrm{nw} = 65 \, \mathrm{nm}$ and $L_\mathrm{nw} = 70 \, \mu \mathrm{m}$) respectively. The two black circles indicate two nanowires recently studied in our group: NW~3 ($R_\mathrm{nw} = 50 \, \mathrm{nm}$, $L_\mathrm{nw} = 200 \, \mu \mathrm{m}$) and NW~4 ($R_\mathrm{nw} = 225 \, \mathrm{nm}$, $L_\mathrm{nw} = 1400 \, \mu \mathrm{m}$). In (c) we also show nanowire dimensions corresponding to a static single photon parametric cooperativity of one (solid black line) and to a dynamical single photon parametric cooperativity of one (dashed black line). The dotted red line indicates a static single photon parametric cooperativity of one when the nanowire has been functionalized at its extremity in order to maximize the optical interaction while minimizing the optical losses.}
\label{Fig:appendix:Optomechanical_coupling_perp_appendix}
\end{center}
\end{figure}

Here we present the results of Section~\ref{sec:optomecha_coupling_TM00:strong_coupling_and_cooperativity} discussing optomechanical effects at the single photon scale in the case of a perpendicular polarization of the light. Fig.~\ref{Fig:appendix:Optomechanical_coupling_perp_appendix} shows in terms of the nanowire dimensions the maximum value of the ratio $2 g_0^z / \Omega_m$ (a), the value of $\delta z^{(1)}$ (b) the blue lines indicating the nanowire dimensions ensuring $\delta z^{(1)} = \Delta z^{(\mathrm{th})}$ for different bath temperatures, and the maximum value of the single photon parametric cooperativity $\mathcal{C}^{(1)}$. On these three plots, the grey dashed lines indicate the iso-frequency of the nanowire fundamental vibrational mode.

\subsection{2D simulation results on a small nanowire}
\label{Appendix:Simu_Rnw_10nm}

\begin{figure}[t!]
\begin{center}
\includegraphics[width=0.95\linewidth]{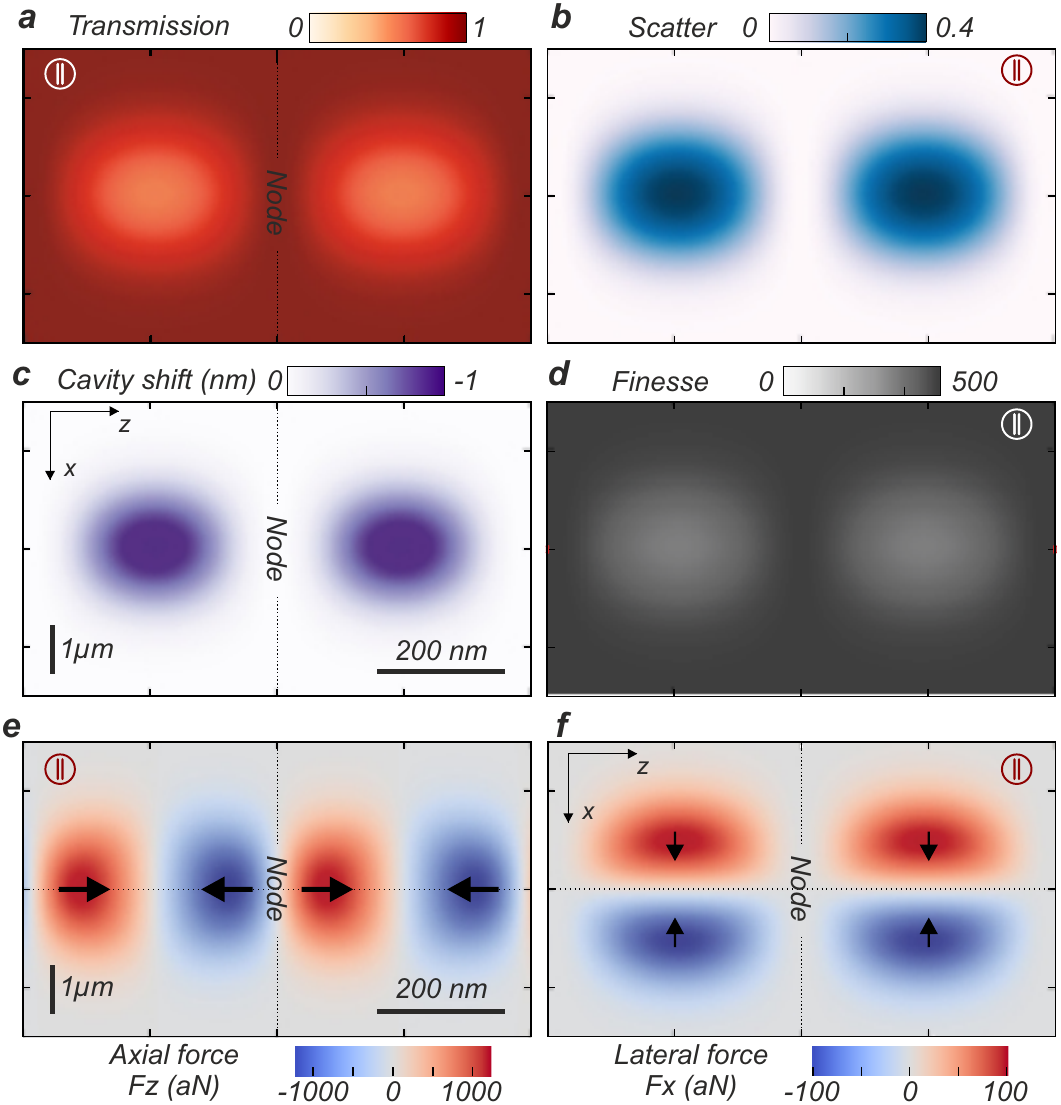}
\caption{Transmission (a), cavity shift (b), scattered coefficient (c) and cavity finesse (d) maps obtained when scanning a nanowire of radius $R_\mathrm{nw} = 10 \, \mathrm{nm}$ in the $(xz)$ plane while locking the cavity at resonance for a parallel polarization of the light. (ef) Optical force maps along $z$ (e) and $x$ (f) applied by the intra-cavity field on the nanowire for an input power of $1 \, \mu \mathrm{W}$.}
\label{Fig:appendix:XZ_map_10nm}
\end{center}
\end{figure}

We show Fig.~\ref{Fig:appendix:XZ_map_10nm} the XZ maps obtained for a cavity length locked at resonance, a nanowire of radius $R_\mathrm{nw} = 10 \, \mathrm{nm}$, and a parallel polarization of the light. It corresponds to the case of a small dispersive and dissipative coupling between the oscillator and the light, the nanowire only locally probing the intra-cavity field structure. At a node of the unperturbed cavity shift we observe no transmission drop, no cavity shift and no scattered light. On the opposite, the coupling is maximum at an anti-node where we have a drop of the cavity transmission associated to a maximum of the cavity shift as well as a maximum of the scattered light. We note a weak decrease of the cavity finesse which remains however larger than $\mathcal{F}_0/2$ ($\mathcal{F}_0$ being the finesse of the unperturbed cavity) explaining the absence of rings in the scattered map, as discussed in the main text.

The optical force field is similar to what one would obtain if scanning a dipole in the optical cavity. The nanowire is attracted towards position of high intensity, reflecting the dominance of gradient force for small nanowire radius and parallel polarization of the light.

\subsection{Optical force and coupling strength for parallel polarization}
\label{Appendix:Simu_Rnw_10nm}

\begin{figure}[t!]
\begin{center}
\includegraphics[width=0.95\linewidth]{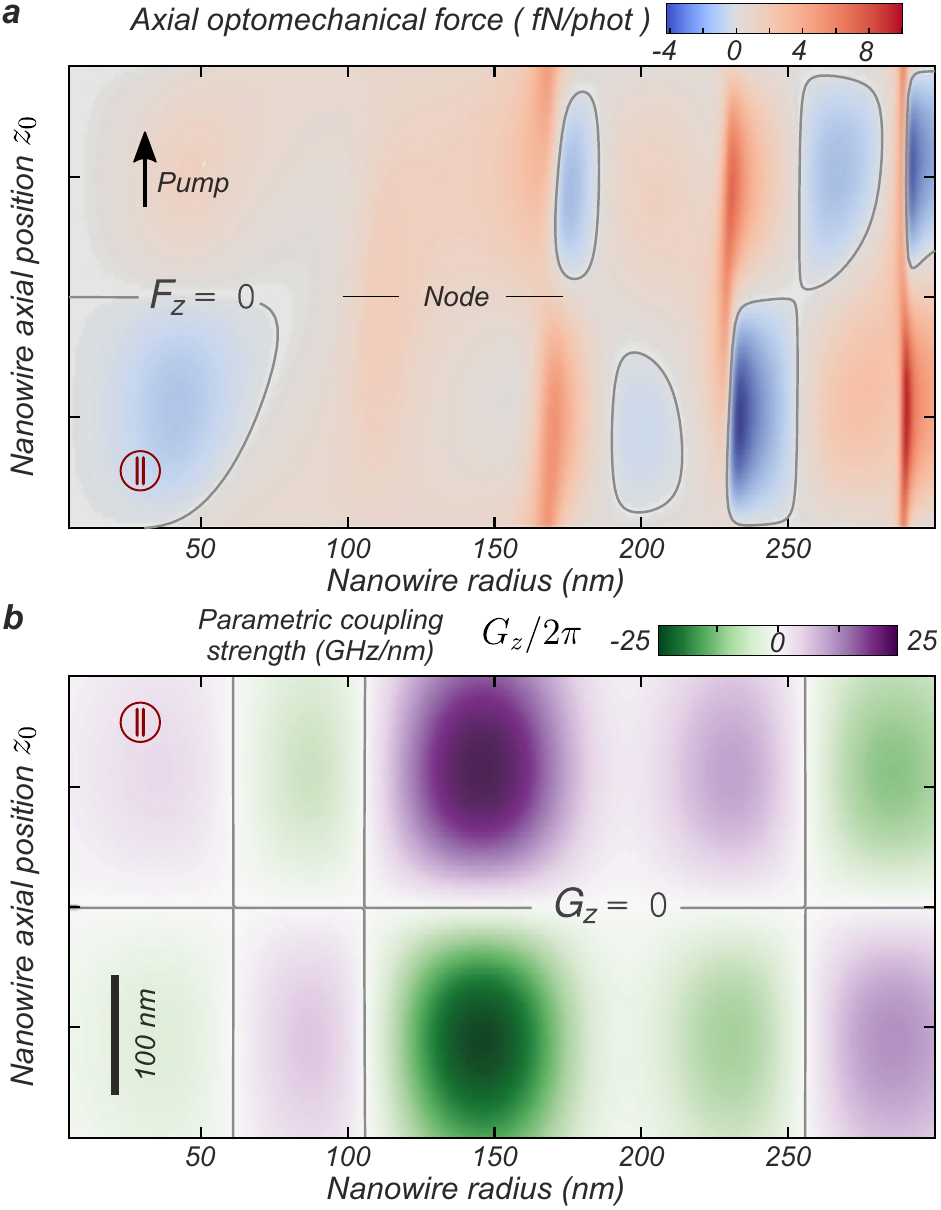}
\caption{Position and nanowire radius dependence of the optical force for a parallel polarization of the light and an input power of $1 \, \mu\mathrm{W}$ (a) and of the coupling strength along $z$ (b). The cavity central node is located at $z=0$ and the pump direction propagates along $+z$. The grey lines indicate positions where $F_z = 0$ and $G_z = 0$.}
\label{Fig:appendix:Fz_Gz_Rnw_scan_para}
\end{center}
\end{figure}

We show Fig.~\ref{Fig:appendix:Fz_Gz_Rnw_scan_para} the radius and position dependence of the optomechanical coupling strength $G_z$, where the grey lines indicates locations where $G_z = 0$ (and $F_z = 0$ in (a)). Here again it underlines the importance of calculating the optical force independently from the optomechanical coupling strength for nanowire in the middle system as already discussed in the main text.

\section{Optical force on a nanowire}
\label{Appendix:Optical_force}

We give below the explicit formulas of the different coefficients appearing in the expression of the optical force Eq.~\eqref{Eq:2D_carac:F_j1_j2} due to the incidence of two plane waves of amplitudes $E_0^{j_1 (p)}$ and $E_0^{j_2 (p)}$.

The coefficient $\mathcal{P}_{j_1, j_2}^{(p)}$ depends on the polarization and on the incidence directions of the two plane waves. It is given by
\begin{eqnarray}
\mathcal{P}_{j_1, j_2}^{(p)} = \sigma_{j_1}^{(p)} \, \sigma_{j_2}^{(p)},
\end{eqnarray}
where $\sigma_{j}^{(p)}$ depends on the polarization choice and on the propagation direction of the plane wave $j$. For parallel polarization, $\sigma_{j}^{(\parallel)} = 1$, reflecting the fact that the polarization vector of the Gaussian beam plane wave expansion given Eq.~\eqref{Eq:Annex:vectorial_beam_discretization_2D_amplitude_wavevector_polar_vectors_polar_para} does not depend on the propagation direction. On the opposite, for perpendicular polarization, the polarization vector of the plane wave expansion given Eq.~\eqref{Eq:Annex:vectorial_beam_discretization_2D_amplitude_wavevector_polar_vectors_polar_perp} depends on the propagation direction, leading to
\begin{subequations}
\begin{align}
& \sigma_{j}^{(\perp)} = 1 \quad \hspace*{0.26cm} \mathrm{if} \quad \bm{\kappa}_j \cdot \textbf{e}_z > 0, \\
& \sigma_{j}^{(\perp)} = -1 \quad \mathrm{if} \quad \bm{\kappa}_j \cdot \textbf{e}_z < 0.
\end{align}
\end{subequations}
In other words, it is 1 (-1) if the plane wave $j$ comes from the plane wave expansion of the Gaussian beam propagating along $+z$ ($-z$).

The other coefficients appearing in Eq.~\eqref{Eq:2D_carac:F_j1_j2} are given by
\begin{subequations}
\begin{align}
& \Lambda_l^{(\parallel)} = \frac{J_l(n \rho_R) J_{l+1}(n \rho_R)}{\abs{D_{l+1}^{\parallel}}^2 \abs{D_{l}^{\parallel}}^2}, \\
& \Lambda_l^{(\perp)} = \frac{\left( \frac{l(l + 1)}{\rho_R^2} \right) J_l(n \rho_R) J_{l+1}(n \rho_R) + J'_l(n \rho_R) J'_{l+1}(n \rho_R)}{\abs{D_{l+1}^{\perp}}^2 \abs{D_{l}^{\perp}}^2}, \\
& D_{l}^{\parallel} = k \left[ J_l(n\rho_R) H_l^{(1)'}(\rho_R) - n J_l'(n\rho_R) H_l^{(1)}(\rho_R) \right], \\
& D_l^\perp = k \left[ J'_l(n \rho_R) H_l^{(1)}(\rho_R) - n J_l(n \rho_R) H_l^{(1) '}(\rho_R) \right],
\end{align}
\end{subequations}
where $n$ is the refractive index of the nanowire, $k = 2 \pi / \lambda$, $\rho_R = k R_\mathrm{nw}$, and where $J_l$ and $H_l^{(1)}$ are the Bessel and Hankle functions of first kind of order $l$. In these equations, the prime denotes a derivative with respect to the argument.

Finally, the minus sign in the exponential characterizing the force orientation in~\eqref{Eq:2D_carac:F_j1_j2} differs from what can be found in~\cite{grzegorczyk2007analyticalTM, grzegorczyk2007analyticalTE}. This is simply due to the axis choice performed here.

%\bibliography{bibliocavNW}

\end{document}